\documentclass[sigconf]{acmart}

\renewcommand\footnotetextcopyrightpermission[1]{} 
\usepackage{fancyhdr}
\usepackage{wrapfig}
\usepackage{fancyhdr}
\usepackage{flushend}
\usepackage{algorithmic}
\usepackage{graphicx}
\usepackage{textcomp}
\usepackage{xspace}
\usepackage{xcolor}
\usepackage{array}
\usepackage{tikz}
\usepackage{soul}
\usepackage{booktabs} 
\usepackage{multirow}
\usepackage[bottom]{footmisc}
\usepackage{setspace}
\usepackage{colortbl}
\usepackage{amsmath}

\AtBeginDocument{%
  \providecommand\BibTeX{{%
    \normalfont B\kern-0.5em{\scshape i\kern-0.25em b}\kern-0.8em\TeX}}}

\copyrightyear{2021}
\acmYear{2021}
\setcopyright{acmcopyright}
\acmPrice{15.00}
\acmDOI{10.1145/3466752.3480085}
\acmISBN{978-1-4503-8557-2/21/10}

\definecolor{schrift}{RGB}{0,73,174}
\definecolor{newred}{rgb}{0.76, 0.13, 0.28}

\newcommand{\rpp}[2][pink]{{%
{#2}}%
}

 \newcommand{\hgg}[2][green]{{%
{#2}}%
 }
 
\newcommand{\ry}[2][yellow]{{%
{#2}}%
}
\newcommand{\rp}[2][pink]{{%
{#2}}%
}
\newcommand{\ro}[2][Tan]{{%
{#2}}%
}

\newcommand{\roo}[2][orange]{{%
{#2}}%
}

\newcommand{\rg}[2][green]{{%
{#2}}%
 }
\newcommand{\rblue}[2][cyan]{{%
{#2}}%
}

\newcommand\rbc[1]{\textcolor{black}{{#1}}}

\newcommand\hj[1]{{\color{black}{#1}}}
\newcommand\ha[1]{{\color{black}{#1}}}
\newcommand\hc[1]{{\color{black}{#1}}}
\newcommand\hd[1]{{\color{black}{#1}}}
\newcommand\he[1]{{\color{black}{#1}}}
\newcommand\jkt[1]{{\color{black}{#1}}}
\newcommand\jkth[1]{{\color{black}{#1}}}

\newcommand{\hym}[2][yellow]{{%
{#2}}%
}

\newcommand{\hoo}[2][orange]{{%
{#2}}%
}

\newcommand{\htann}[2][Tan]{{%
{#2}}%
}

\newcommand{\hlgg}[2][SpringGreen]{{%
{#2}}%
}

\newcommand{\hsb}[2][SkyBlue]{{%
{#2}}%
}

\newcommand{\hy}[2][yellow]{{%
{#2}}%
}
\newcommand{\hp}[2][pink]{{%
{#2}}%
}

\newcommand{\hg}[2][green]{{%
{#2}}%
}

\newcommand{\hb}[2][cyan]{{%
{#2}}%
}

\newcommand{\hgrey}[2][Gray]{{%
{#2}}%
}

\newcommand{\hly}[2][yellow]{{%
{#2}}%
}
\newcommand{\hlp}[2][pink]{{%
{#2}}%
}

\newcommand{\hlg}[2][green]{{%
{#2}}%
}

\newcommand{\tech}{BurstLink\xspace}

\newcommand*\circled[1]{\tikz[baseline=(char.base)]{
            \node[shape=circle,fill,inner sep=0.5pt] (char) {\textcolor{white}{#1}};}}
\newcommand*\circleda[1]{\tikz[baseline=(char.base)]{
            \node[shape=circle,fill,inner sep=1.2pt] (char) {\textcolor{white}{#1}};}}            
\newcommand*\circledg[1]{\tikz[baseline=(char.base)]{
            \node[shape=circle,fill=black!50,inner sep=0.5pt] (char) {\textcolor{white}{#1}};}}

\newcommand*\circledw[1]{\tikz[baseline=(char.base)]{
            \node[shape=circle,draw,fill=white,inner sep=0.5pt] (char) {\textcolor{black}{#1}};}}

\newcommand\fig[1]{Fig.~{#1}\xspace}

\newcommand\sect[1]{Section~{#1}\xspace}

\newcommand\head[1]{\noindent\textbf{#1.\xspace}}

\newcommand\pr[1]{{\color{black}{#1}}} 

\newcommand\jsrm{\bgroup\markoverwith{\textcolor{purple}{\rule[0.5ex]{2pt}{1pt}}}\ULon}

\definecolor{dollarbill}{rgb}{0.52, 0.73, 0.4}

\newcommand\jsout{\bgroup\markoverwith
{\textcolor{purple}{\rule[.5ex]{2pt}{0.4pt}}}\ULon}

\definecolor{dgreen}{rgb}{0.00, 0.75, 0.00}

\newcommand{\juang}[1]{{\color{black}#1}}

\newcommand\jh[1]{{\color{black}{#1}}}

\newcommand\js[1]{{\color{black}{#1}}}
\newcommand\haj[1]{{\color{black}{#1}}}
\newcommand\jk[1]{{\color{black}{#1}}}
\newcommand{\taha}[1]{{\color{black}#1}}

\definecolor{amber}{rgb}{1.0, 0.49, 0.0}

\newcommand{\m}[1]{{\color{black}#1}}

\begin{document}




\title{BurstLink: Techniques for Energy-Efficient Video Display \\ 
for Conventional and Virtual Reality Systems}

\settopmatter{authorsperrow=4} 

\author{Jawad Haj-Yahya}
\affiliation{
  \institution{ETH Z{\"u}rich}
  \streetaddress{}
  \city{}
  \state{}
  \country{}
  \postcode{}
}

\author{Jisung Park}
\affiliation{
  \institution{ETH Z{\"u}rich}
  \streetaddress{}
  \city{}
  \state{}
  \country{}
  \postcode{}
}

\author{Rahul Bera}
\affiliation{
  \institution{ETH Z{\"u}rich}
  \streetaddress{}
  \city{}
  \state{}
  \country{}
  \postcode{}
}

\author{Juan Gómez Luna}
\affiliation{
  \institution{ETH Z{\"u}rich}
  \streetaddress{}
  \city{}
  \state{}
  \country{}
  \postcode{}
}

\author{Taha Shahroodi}
\affiliation{
  \institution{ETH Z{\"u}rich, TU Delft}
  \streetaddress{}
  \city{}
  \state{}
  \country{}
  \postcode{}
}

\author{Jeremie S. Kim}
\affiliation{
  \institution{ETH Z{\"u}rich}
  \streetaddress{}
  \city{}
  \state{}
  \country{}
  \postcode{}
}

\author{Efraim Rotem}
\affiliation{
  \institution{Intel Corporation}
  \streetaddress{}
  \city{}
  \state{}
  \country{}
  \postcode{}
}

\author{Onur Mutlu}
\affiliation{
  \institution{ETH Z{\"u}rich}
  \streetaddress{}
  \city{}
  \state{}
  \country{}
  \postcode{}
}

\renewcommand{\shortauthors}{}

\renewcommand{\shortauthors}{Haj-Yahya, et al.}

\begin{abstract}
 
Conventional planar video streaming is the most popular application in mobile systems\hj{. The} rapid growth of 360$^{\circ}$ video content and virtual reality (VR) devices \hj{is} accelerating the adoption of VR video streaming.
Unfortunately, video streaming consumes significant system energy due to high power consumption of \hj{major} system components (e.g., DRAM, display interfaces, and display panel) involved in \hj{the video streaming} process. 
\hj{For example, in conventional planar video streaming, the video decoder (in the processor) decodes video frames and stores them in the DRAM main memory before the display controller (in the processor) transfers} 
\hj{decoded frames from DRAM to the display panel.
This system architecture causes large amount of data movement to/from DRAM as well as high DRAM bandwidth usage. As a result, DRAM by itself consumes more than $30\%$ of the video streaming energy.}

We propose \tech, a novel system-level technique that improves the energy efficiency of planar and VR video streaming.
\tech is based on two key ideas.
First, \tech \emph{directly} transfers a decoded video frame from the video decoder or \hj{the} GPU to the  display panel, \jh{completely} bypassing the host DRAM.
To this end, we extend the display panel with a \emph{double remote frame buffer (DRFB)}\haj{ instead of DRAM's double  frame buffer} so that the system can directly update the DRFB with a new frame while updating the \hj{display} panel's pixels with the current frame stored in the DRFB.
Second, \tech transfers a \emph{\haj{complete} decoded frame} to the display panel \emph{in a \haj{single} burst}, using the maximum bandwidth of modern display interfaces. 

Unlike conventional systems where the frame transfer rate is limited by the pixel-update throughput of the display panel, \tech can always take full advantage of the high bandwidth of modern display interfaces by decoupling the frame transfer from the pixel update as enabled by the DRFB.
This direct and burst frame transfer of \hj{capability} \tech significantly reduces energy consumption \hj{of} video display by 1) reducing accesses to DRAM, 2) increasing  system's residency at idle power states, and 3) enabling temporal power gating of several system components after quickly transferring each frame into the DRFB. 

{\tech} can be easily implemented in modern mobile systems with minimal changes to the video display pipeline. 
We evaluate \tech using an analytical power model that we rigorously validate on \hj{an Intel Skylake mobile system}. 
Our evaluation shows that \tech reduces system energy consumption for 4K planar and VR video streaming by 41\% and 33\%, respectively\hj{. \tech} provides an even higher \hj{energy} reduction \hj{in future video streaming systems with higher} display resolutions and/or display refresh rates.



\end{abstract}




\maketitle


 \section{Introduction}\label{sec:intro}

Conventional planar \hj{(i.e., 2-dimensional)} video streaming is {the most prevalent application in mobile devices}~\cite{youtube_2019}. Virtual reality (VR) video streaming is emerging as one of the most important applications in the entertainment market \hj{\cite{zyda2005visual}}. 
Cisco predicts that video streaming will generate more than $79\%$ of mobile data traffic by 2022~\cite{cisco_2019}, and Goldman Sachs predicts that around 79 million users will use VR video streaming by 2025~\cite{vr_2025}.
To provide users with an immersive experience, video formats and mobile display panels support increasingly high resolutions (e.g., 4K~\cite{4k_puppies, 4k_laptop}) and refresh rates (e.g., 120Hz \cite{samsung_s20}). 
These trends come at the cost of significantly higher energy consumption of video display, which negatively impacts the battery life of \hj{a} mobile device~\cite{tomsguide}. \hj{Mobile} systems need an efficient planar/VR video display {architecture} that provides high energy efficiency while enabling high video/display resolutions and refresh rates.

\begin{figure}[h]
\begin{center}
\includegraphics[trim=.7cm .8cm 0.7cm .7cm, clip=true,width=0.75\linewidth,keepaspectratio]{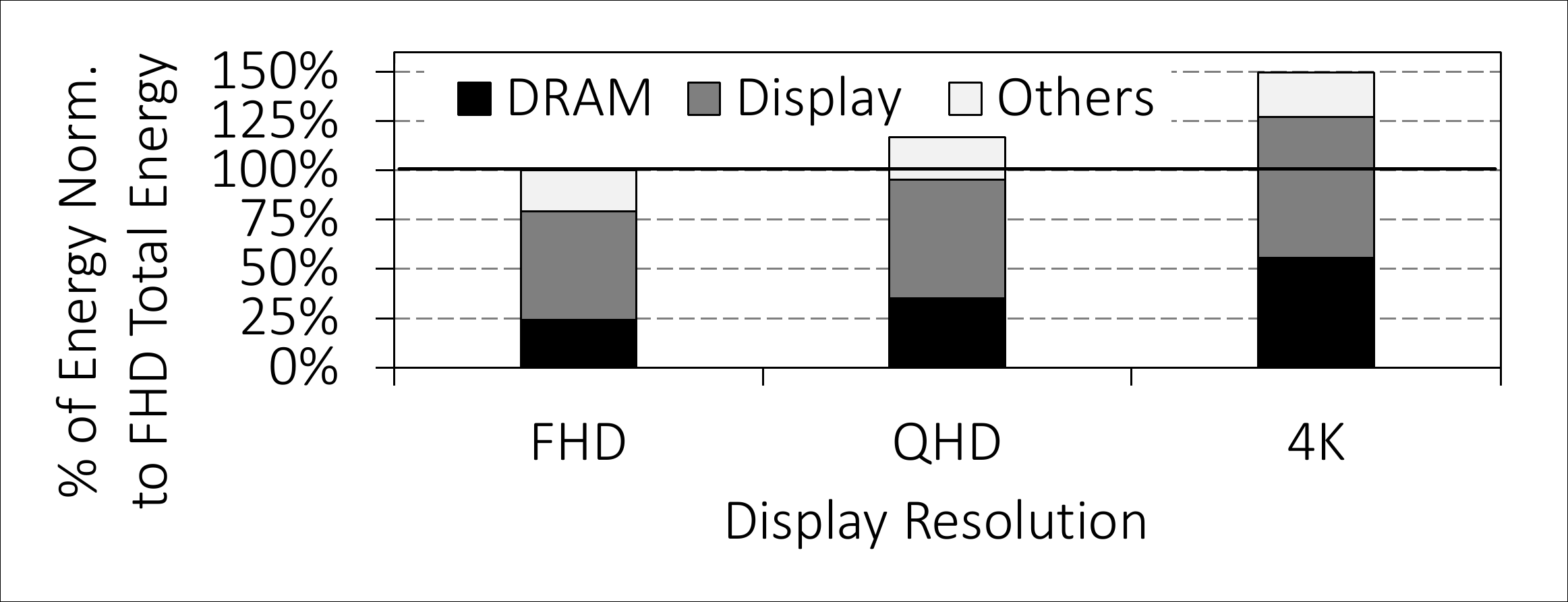}
\caption{Energy consumption of a modern mobile \hd{computing} system while streaming videos at various resolutions\hj{, normalized to FHD}.}
\label{fig:video_energy}
\end{center}
\end{figure}

\begin{sloppypar}
In conventional mobile systems, {DRAM main memory and the display panel consume} the majority of planar video streaming energy. 
\fig{\ref{fig:video_energy}} shows the \hj{measured} energy consumption \hj{breakdown of} a real Intel Skylake \hj{\cite{tam2018skylake,21_doweck2017inside}} mobile system while streaming \haj{30 frames-per-second (FPS)} videos \hj{of} full-high-definition (FHD, 1920$\times$1080), quad-high-definition (QHD, 2560$\times$1440), and 4K (3840$\times$2160) resolutions.\footnote{Recent works also show similar trends on ARM system-on-chips (SoCs)~\cite{nachiappan2015vip,yedlapalli2014short}.}
\rg{We break down the system energy consumption into three major components: {\textsf{DRAM} (main memory), \hsb{\textsf{Display} (all components in an LCD \hj{\cite{choi2002low}} display panel)}, and \textsf{Others}, which includes \hj{three main components\hj{:}} \hsb{the processor (including the video decoder and display controller)}, network (WiFi), and storage (eMMC).}}
As a single decoded frame's size becomes as large as tens of megabytes for a \hj{very} high-resolution video (e.g., 24MB for a 4K video), DRAM alone contributes more than $30\%$ of the total system energy consumption. 
State-of-the-art VR streaming schemes \cite{leng2019energy,zhao2020deja}, which significantly reduce the energy for \hj{\textsf{Others} (\hj{especially the processor})} compared to \hj{\ha{unoptimized} VR video streaming}  \cite{he2018rubiks,graf2017towards}, \hj{do not affect \textsf{DRAM} \& \textsf{Display} and } show similar trends in the energy consumption breakdown \hj{(not shown here)}.
 \end{sloppypar}

In this paper, we find two inefficiencies of conventional video display schemes that result in the underutilization of advanced architectural features widely available in modern mobile \hd{computing} systems.
\hj{First, the video decoder or GPU stores video frames in the DRAM main memory before the display controller (in the processor) transfers these frames from DRAM to the display panel. This system architecture causes large amount of data movement to/from DRAM as well as high DRAM bandwidth usage.
Modern mobile systems commonly employ a \emph{remote frame buffer (RFB) inside the display panel} to improve energy efficiency in a static-image display (see \sect{\ref{sec:display_refresh}} for more details). Unfortunately, current video streaming architectures do not utilize the RFB due to \ha{a} few limitations. For example,  when other \emph{planes} (e.g., graphical user interface (GUI) and cursor planes) must be displayed in addition to the video plane, the display controller (DC) must first merge them with the video plane to generate a single frame that the DC sends to the display panel.}
\ha{\hc{However}, when there is no interaction between the video plane and other planes, we could significantly reduce energy consumption if we send the decoded video frames directly to the display and bypass host DRAM. Unfortunately, current video streaming \hc{architectures do} not utilize this opportunity due to limitations in \hc{datapaths} of both the processor and the display panel.} 

Second, conventional video processing and display schemes underutilize the high bandwidth provided by modern display interfaces.
The state-of-the-art embedded-DisplayPort (eDP) \hj{\cite{edp1_4b,edp_psr}} interface supports a peak bandwidth of 25.92 Gbps~\cite{edp1_4b} \haj{to cope with high-quality videos with high resolutions and frame rates}.
However, even \hj{when} displaying a 4K 60FPS video, 
conventional mobile  systems send each frame at a transfer rate of about 11.3 Gbps.
\haj{This underutilization is because the display panel’s pixel-update bandwidth dictates the frame transfer rate between the host system and the display panel.}
The eDP-bandwidth underutilization negatively affects energy efficiency since a long time spent \hj{for} actively transferring frames reduces the time that the system can spend in low-power states.
This inefficiency is expected to remain for the coming years, considering \hj{the large} quality gap between displays and video contents~\cite{res_gap_1,res_gap_2,res_gap_3}\ha{:}
while modern displays increasingly support higher resolutions and refresh rates, a significant majority of video streaming content is still in high definition (HD) or standard definition (SD) resolution~\cite{res_gap_1, res_gap_2}. 
For example, a recent study~\cite{res_gap_1} reports that 4K TVs account for about 55\% of the TV market, while only less than 10\% of video content provided by major streaming platforms (e.g., Netflix, Amazon Prime Video, and YouTube) \js{is} in 4K resolution with frame rates higher than 30FPS.

Based on \hj{these two major} observations, we propose \tech, a new system-level technique that improves \hj{both}  planar and VR video \haj{streaming energy efficiency}\hj{. \tech exploits} advanced architectural features that are available yet underutilized in modern mobile systems.
\tech is based on two \textbf{key ideas}.
First, \ha{we introduce} \emph{Frame Buffer Bypassing}, a novel mechanism that enables \js{\emph{direct} transfer of} a decoded video frame from the host system \hj{directly} to the display panel\hj{,} \ha{completely} bypassing the host DRAM.
\haj{To} allow the system to transfer a new frame while the display panel updates pixels with the current frame in the remote frame buffer, we extend the display panel with a \emph{double remote frame buffer (DRFB)}. \haj{Using the DRFB, the system can directly update one of the buffers of the DRFB with a new frame while updating the panel's pixels with the current frame stored in the other buffer of the DRFB.}
Second, \ha{we introduce} \emph{Frame Bursting}, a novel mechanism that enables \haj{transferring} a \emph{whole decoded frame} to the display panel \emph{in a \haj{single} burst} using the maximum bandwidth of modern display interfaces. 
Unlike conventional systems where \haj{the pixel-update throughput of the display panel limits} the frame transfer rate, \tech can always take full advantage of the high bandwidth of modern display interfaces by decoupling the frame transfer from the pixel update using the DRFB.

The direct and bulk frame transfer enables \tech to significantly reduce energy consumption over a conventional video display scheme in two ways.
First, by directly transferring processed video frames from the video decoder or GPU to the display panel \emph{without buffering} them into the host DRAM \hj{(i.e., Frame Buffer Bypassing)}\haj{,} 
\tech saves a significant fraction of the DRAM energy consumption.
Second, by transferring an entire frame at the maximum \hj{display interface} bandwidth \hj{(i.e., Frame  Bursting)}\haj{,}
\tech reduces the \hj{usage} of the processor and the display subsystem\haj{ since they are active only during the burst period}, thereby allowing the system to enter deep low-power states more frequently by turning off unused resources (e.g., the display controller, display interface, and host DRAM). In addition to planar/VR video display, Frame Bursting is also applicable \hj{to} other important mobile workloads like casual gaming and office productivity \cite{mobilemark,19_MSFT}. 

We evaluate \tech using an analytical power model that we rigorously validate with a real modern \hj{Intel Skylake \cite{tam2018skylake}} mobile system. 
Our evaluation shows that \tech{} \hj{1)} reduces system energy consumption for 4K 60FPS planar and \haj{360$^{\circ}$} VR video streaming by 41\% and 33\%, respectively\haj{, and \hj{2)} provides an even higher reduction as
display resolution and/or display refresh rate increases}.
\tech also reduces system energy consumption for video conferencing, MobileMark~\cite{mobilemark}, and casual gaming \cite{19_MSFT} workloads by 30\%, 28\%, and 27\%, respectively\hj{, mainly by utilizing the Frame Bursting technique of \tech.}

\hj{\tech aims to improve the energy efficiency of video streaming\ha{,} one of the most important application scenarios in modern mobile \hd{computing} systems. However, \ha{\tech}  can also be used in more general frame-based applications such as  
video capture (recording), audio streaming, video chat, social networking, and interactive games.
A general takeaway from \tech is that using main memory (DRAM) as a \emph{communication hub} between system components \ha{is} energy-inefficient. Instead, \tech uses small \emph{remote} memory near the data consumer (e.g., a display panel) to significantly reduce the number of costly main memory accesses in frame-based applications.}

We make the following \textbf{key contributions} in this work:
\begin{itemize}
\item \haj{We provide the first study that identifies the main energy inefficiencies in traditional video display schemes of mobile systems and proposes novel techniques in both the processor and display panel to address the inefficiencies.}

\item We propose \tech{}, \js{a new energy-efficient video display scheme based on two key \hj{new} ideas}: 
1) \emph{Frame Buffer Bypass\js{ing}}, which transfers a decoded video frame directly to the \hj{display} panel without buffering it in the host DRAM, and 
2) \emph{Frame Bursting}, which \hj{\ha{burst-transfers} each decoded frame to the display panel as quickly as possible and \ha{thus increases} opportunities for system idleness}.
\item We evaluate \tech using a thoroughly-validated analytical power model\ha{, which we open-source online~\cite{Burstlink-github}.} 
Our evaluation shows that \tech{} reduces system energy consumption for 4K 60FPS planar and VR video streaming by 41\% and 33\%, respectively.
\tech's energy reduction increases with higher display resolutions, making \tech an even better fit for next-generation high-resolution displays. 
\end{itemize}

\section{Background}\label{sec:bg}


\subsection{Mobile SoC Microarchitecture}\label{subsec:soc_arch} 

The microarchitecture of a mobile \hj{system-on-chip (SoC)} typically consists of the following components. 

\head{\hj{Main} SoC Domains}
A high-end mobile processor (e.g., Intel Skylake~\cite{21_doweck2017inside}, AMD Kabini~\cite{psr_5}, Samsung Exynos~\cite{psr_6}) is commonly implemented as \hj{an} SoC that integrates three main domains into a single chip:
1) \emph{compute domain}, such as CPU cores and graphics engines, 
2) \emph{IO domain}, which includes several \emph{intellectual properties} (IPs) sharing the \emph{IO interconnect} (e.g., display controller (DC), image signal processing engine (ISP), video decoders (VDs), video encoders (VEs)), and 
3) \emph{memory domain}, which includes the memory controller and \hj{the} DRAM interface.

\head{\hym{IO Interconnect}}
IO interconnects, e.g., Intel On-chip System Fabric (IOSF) \mbox{\cite{lakdawala201232}} and ARM Advanced Microcontroller Bus Architecture (AMBA) \mbox{\cite{arm_amba,patil2014functional}}, are on-chip communication technologies. \haj{IO Interconnect allows}  multiple IPs to 1) perform peer-to-peer (P2P) communication \hj{\cite{ogras2010analytical}} and 2) access main memory (DRAM) using direct memory access (DMA) \hj{\cite{lee2015decoupled}}. 

\head{P2P and DMA Engines}
\haj{IO IPs are typically equipped} with DMA and P2P engines~\cite{kogel2003modular}. 
The DMA engine enables the IP to access \haj{the} main memory \haj{directly}, while the P2P engine enables direct communication between two IPs without copying  data to  main memory.
P2P reduces the data transmission delay and increases the \hj{overall} available system bandwidth.
DMA and P2P engines each have \emph{\haj{control} registers (CRs)} that the \emph{IP driver} configures.

\begin{figure}[!h]
\begin{center}
\includegraphics[trim=.5cm 0.5cm .5cm 0.5cm, clip=true,width=1\linewidth,keepaspectratio]{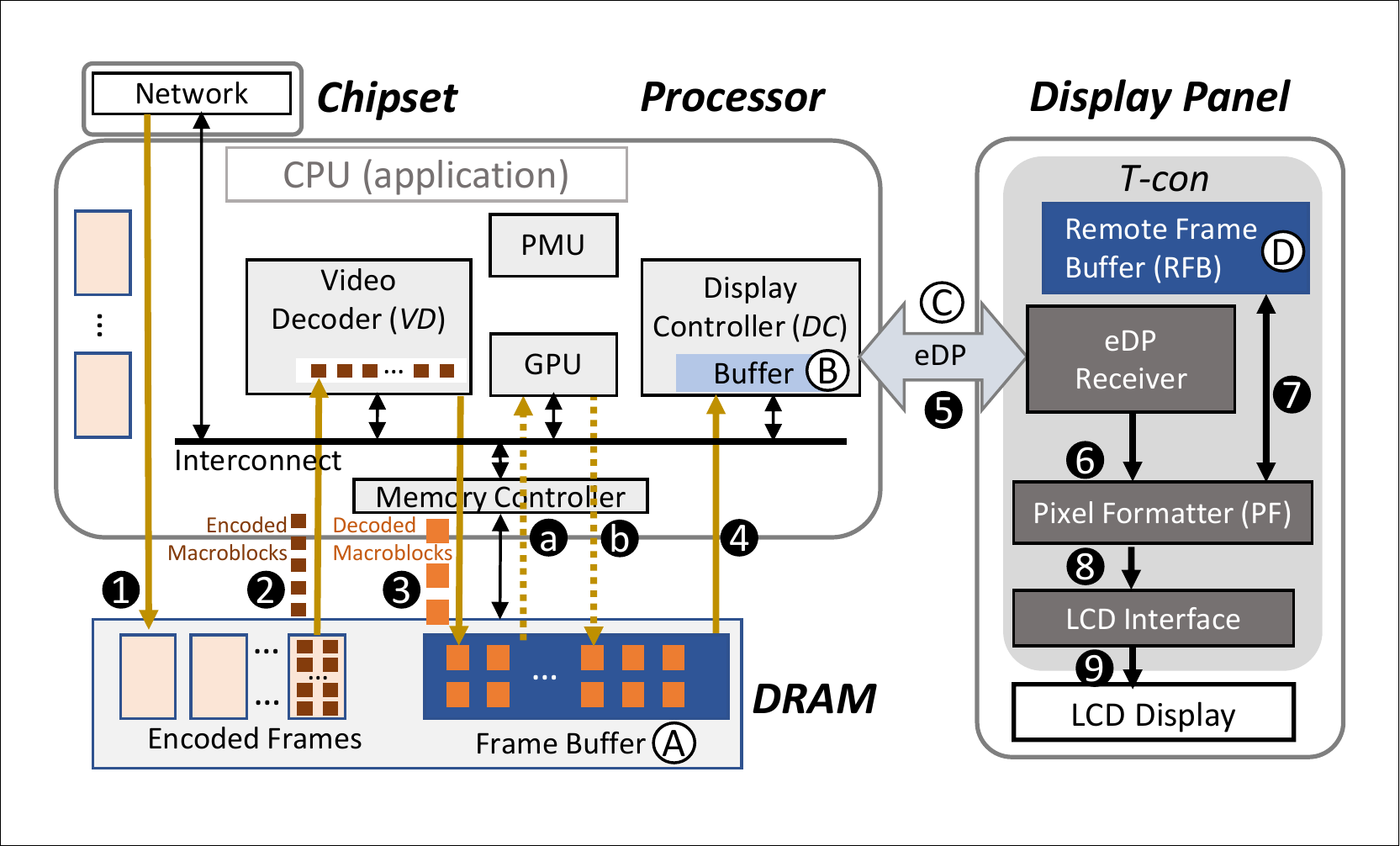}\\
 \vspace{-6pt}
\caption{\js{Overview of a conventional display subsystem.}}
\label{fig:current_pipeline}
\end{center}
 \vspace{-10pt}
\end{figure}

\head{Traditional Display Subsystem} \fig{\ref{fig:current_pipeline}} shows an overview of a conventional display subsystem, which consists of five main components:
two on the processor side (i.e., the \emph{Video Decoder (VD)} and \emph{Display Controller (DC)}) and three on the display panel side (i.e., the \emph{embedded-DisplayPort (eDP) Receiver}, \emph{Pixel Formatter (PF)}, and \emph{Remote Frame Buffer (RFB)}\hj{, all} inside the timing-controller (T-con)\footnote{A timing controller (T-con) is a circuit that processes and coordinates the coloration of the pixels in a display panel \mbox{\cite{psr_1}}.}). We explain video processing steps \hj{and power states} in more detail \hj{in Sections \ref{subsec:video_processing} and \ref{sec:syspowerstates}, respectively, using Fig. \ref{fig:current_pipeline} as a basis}.

\subsection{System Idle Power States (C-states)} \label{sec:soc_power}
The Advanced Configuration and Power Interface (ACPI \hj{\cite{acpi}})\footnote{ACPI \hj{\cite{acpi}} is an industry standard that is widely used for OS-directed configuration, power management, and thermal management of computing \haj{systems}.} defines a processor's \emph{idle power states}, commonly called \emph{C-states}~\cite{acpi}.
C-states are defined \hj{for} two \haj{primary} levels:
1) component level, such as thread ($TCi$), core ($CCi$), and graphics ($RCi$) C-states, and 2) SoC level, known as \emph{package C-states} ($PCi$ or $Ci$)~\cite{intel_skl_dev,{haj2018power}}. 

\ry{A package C-state defines an idle power state of the \haj{system} (consisting of the processor, chipset, external memory devices)\haj{. A system enters a specific package C-state} depending on each system component's idle power state (\emph{component C-state}). 
Various levels of package C-states exist to provide a range of power consumption levels with various techniques such as clock gating at the uncore level \hj{\cite{intel_skl_dev,haj2018power}} or a nearly \haj{complete} shutdown of the system \hj{\cite{haj2020techniques}}. 
The ACPI standard includes recommendations on the C-states, but manufacturers are free to define their C-states and the SoC's behavior at each C-state.
In this work, we focus on the package C-states of the Intel Skylake architecture~\cite{intel_skl_dev}, but similar idle power state definitions exist in other architectures (e.g.,  AMD~\cite{amd_specs} and ARM~\cite{qcom2018}). 
Table~\mbox{\ref{tab:p-states}} shows all package C-states of the Intel Skylake architecture and the \haj{major} conditions \hj{under} which the power management unit (PMU) places the SoC into each package C-state (a similar table exists in the Intel manual \cite{intel_skl_dev}).} 

\begin{table}[h]
\renewcommand{\arraystretch}{1.05}
\caption{Package C-states in \hj{the} Intel Skylake mobile SoC.}
\label{tab:p-states}
\resizebox{\linewidth}{!}{%
\begin{tabular}{ |c||l|l|}
\hline
\textbf{\begin{tabular}[c]{@{}l@{}}Package\\ C-state\end{tabular}} & \textbf{\haj{Major conditions to enter the package C-state}}                                                                                                                                                                                                                                                                                \\ \hline \hline
C0                                                                 &One or more cores or graphics engine \textbf{executing instructions}                                                                                                                                                                                                                                         \\ \hline
C2                                                                 & \begin{tabular}[c]{@{}l@{}}All cores in \textbf{CC3} (clocks off) \textbf{or deeper} and graphics engine  in \\ \textbf{RC6} (power-gated). DRAM is \bf{active}.\end{tabular}                                                                                                                                                                                                         \\ \hline
C3                                                                 & \begin{tabular}[c]{@{}l@{}}All cores in \textbf{CC3 or deeper} and graphics engine in \textbf{RC6}.  Last-\\Level-Cache (LLC)  may be flushed and turned off, DRAM in\\ \textbf{self-refresh (SR)}, most IO and memory  domain clocks are gated,\\ some IPs and IOs can be \textbf{active}  (e.g., \textbf{DC and Display IO}).\end{tabular}                                                         \\ \hline
C6                                                                 & \begin{tabular}[c]{@{}l@{}}All cores in \textbf{CC6} (power-gated) or \textbf{deeper} and graphics \\ engine in \textbf{RC6}. LLC may be flushed and turned off, \hj{DRAM} \\ in \textbf{self-refresh}, IO and memory domain clock generators are \\ turned off. Some IPs and IOs can be  \textbf{active (e.g., VD, DC)}.\end{tabular}                                                                                 \\ \hline
C7                                                                 & \begin{tabular}[c]{@{}l@{}}Same as Package C6 while some of the IO and memory \\ domains  are \textbf{power-gated}.\end{tabular}                                                                                                                                                                                                                                  \\ \hline
C8                                                                 & \begin{tabular}[c]{@{}l@{}}Same as Package C7 with additional \textbf{power-gating} in the IO \\ and memory domains. \textbf{Only DC and Display IO are ON}.\end{tabular}                                                                                                                                                                                                                                    \\ \hline
C9                                                                 & \begin{tabular}[c]{@{}l@{}}Same as Package C8 while all IPs must be off. Most VRs \\  voltage are reduced. \textbf{The display panel can be in PSR}.\end{tabular}                                                                                                                                                                                      \\ \hline
C10                                                                & \begin{tabular}[c]{@{}l@{}}\textbf{Same as Package C9} while \textbf{all SoC VRs} (except  always-on\\ VR) \textbf{are off}. \textbf{The display panel is off}.\end{tabular}                                                                                                                                                                                                            \\ \hline
\end{tabular}%
}
\end{table}

\subsection{Display Panel Refresh}\label{sec:display_refresh}
Current display technologies require the host SoC to refresh the display panel several tens of times every second \cite{menozzi2001crt}. For example, a display panel with a \emph{refresh rate} of 60Hz is refreshed 60 times per second. During each \haj{frame} refresh \haj{window} \haj{(i.e., 1/refresh\_rate)}, the display controller (DC) inside the host SoC transfers a full-frame to the display panel by repeatedly performing three steps: the DC 1) fetches a portion of the image data from the DRAM frame buffer (\hj{\circledw{A}} in DRAM \hj{in Fig. \ref{fig:current_pipeline}}), 
2) stores the fetched image data into the DC's local buffer \hj{(\circledw{B} in DC in Fig. \ref{fig:current_pipeline}}),  and 3) sends the buffered data to the display panel via the display interface \hj{(\circledw{C} in Fig. \ref{fig:current_pipeline}}). 


\head{Panel Self-Refresh (PSR)} 
To reduce system energy consumption when displaying \textit{static images}, VESA (Video Electronics Standards Association~\cite{vesa}) introduced the Panel-Self-Refresh (PSR) standard~\cite{psr_1,psr_2,psr_6}. 
\hj{PSR} 1) adds a local frame buffer, called a \emph{remote frame buffer (RFB,  \hj{\circledw{D} in Display in Fig. \ref{fig:current_pipeline}})}, into the panel T-con to store \emph{one} frame, and 2) defines a protocol in which the DC notifies the display panel of an unchanged image. 
\haj{These} \hj{changes} enable \hj{the} \haj{PSR mode}\hj{, where} the panel performs self-refresh using the static image stored in the RFB \emph{without} accessing DRAM main memory. 
Doing so allows many host-side components\haj{,} including DRAM, display interface, and DC to be powered down, \haj{reducing} system energy consumption.


\head{PSR Selective Updates (PSR2)} 
While a mobile system is in PSR mode, the host SoC can make selective frame updates to the RFB, also known as  PSR2~\cite{edp_psr,edp1_4b,psr_1,intel_display}. This optimization can be used, for example, to turn on/off a blinking display cursor. PSR2 is supported by the newest  embedded-Display Port (eDP) 1.4~\cite{edp1_4b}.

\subsection{Planar and VR Video Processing}
\label{subsec:video_processing}

Planar \hj{(i.e., 2-dimensional)} video processing consists of three main stages: 1) \emph{buffering} of encoded frames, 2) \emph{decoding} of the buffered frames, and 3) \emph{displaying} of the decoded frames. VR video processing requires an additional stage, called \emph{projection}, which is performed immediately before the \emph{displaying} stage. \hj{The video application and the device (e.g., GPU, display) drivers are responsible for \emph{orchestrating} the different system components \ha{(e.g., programming the DMA engines and handling
interrupts)} during these stages.}

\head{Buffering} 
\hj{For} video \emph{streaming}, the network IP receives encoded video frames. 
Similarly, \hj{for} video \emph{playback}, the application reads the frames from storage devices. 
These encoded frames, each of which is hundreds of KBytes in size, are buffered in DRAM~(\circled{1} in \fig{\ref{fig:current_pipeline}}). 
The buffering process enables the system to \hj{tolerate}  network bandwidth fluctuation~\cite{ameigeiras2012analysis} and reduce the number of storage accesses, which enables smoother and more efficient video processing. 

\head{Decoding} 
The video decoder (VD) reads an encoded frame from DRAM~(\circled{2}) and starts decoding it. 
An encoded frame consists of \haj{many } {macroblocks, each of which stores the pixel information of a small exclusive region of the frame~\cite{wiegand2003overview}.}
An encoded macroblock is the basic processing element in video decoding and typically includes 16$\times$16, 32$\times$32, or 64$\times$64 pixels \cite{mukherjee2013latest, sullivan2012overview, wiegand2003overview}. 
VD reads an encoded frame at macroblock granularity and buffers several encoded macroblocks (e.g., tens of KBytes~\cite{wiegand2003overview}). 
Each encoded macroblock first passes through a series of stages, including entropy-decoding, inverse-DCT, and inverse quantization~\cite{wiegand2003overview,rts,mukherjee2013latest, sullivan2012overview}. 
\roo{
Next, each macroblock is reconstructed in various ways depending on \haj{its} type. 
\haj{VD reconstructs an} \emph{I-Type} macroblock from its neighboring macroblocks of the \emph{same} \haj{encoded} frame\haj{. In contrast, the VD reconstructs a} \emph{P-Type} \haj{and} \emph{B-Type} macroblock from the macroblocks in the \emph{\haj{previous}} and \emph{previous/later} \haj{encoded} frames\haj{, respectively,} \hj{as} indicated by the extra information stored \hj{in macroblock} metadata (\hj{i.e.,} motion vectors).}
Finally, a decoded macroblock is written to the frame buffer in   DRAM~(\circled{3})~\cite{leng2019energy,rts} in preparation for the next stage \hj{of video processing}. 

\head{Projection} 
In planar video processing, each frame can be directly displayed once decoded. 
However, \hj{in VR video processing}\ha{, each} frame must go through a set of \emph{projective transformation} (PT) operations before being displayed.\hj{\footnote{\ha{P}rojective transformation (PT) is the process of mapping points in the 3D space that fall within the user's viewing area to pixels on a 2D plane. In this way, the VR video can be directly displayed in the same way as a conventional planar video \cite{leng2019energy,zhao2020deja}.}} 
Therefore, each decoded \hj{VR video} frame is forwarded to the GPU~(\circleda{a}), which performs PT operations and writes the processed frame back to the DRAM frame buffer~(\circled{b})~\cite{leng2019energy, zhao2020deja}.

\head{Displaying} 
The display controller (DC) reads a decoded frame from the DRAM frame buffer~(\circled{4}) at chunk granularity (e.g., 512 KB)~\cite{arm_dc, intel_display} and stores the frame data within its limited\hj{-size} internal buffer before sending it to the display panel. 
During the frame transfer \hj{from the DC internal buffer to the display panel}, host-side components\haj{,} including the CPU, network interface, and VD\haj{,} enter low-power states (e.g., DRAM \hj{is placed into} \ha{the} self-refresh \hj{(SR)} mode~\cite{intel_display}).
DC sends the frame chunks \hj{to the display panel} over the eDP interface~(\circled{5}) according to the display refresh rate. 
For instance, if the refresh rate is $60$~Hz, the DC sends the frame chunks to the display panel within a window of ${\sim}16$~ms (i.e., 1/60~sec), which we refer to as \hj{the} \emph{frame window}.
On the \hj{display} panel side, the eDP receiver forwards the chunks to the \emph{Pixel Formatter} (PF)~(\circled{6})~\cite{edp1_4b}.
PF 1) stores the decoded frame into the RFB~(\circled{7})~\cite{edp1_4b,psr_4,psr_5,psr_6}, 2) converts the decoded frame into a pixel data array, and 3) sends it to the LCD interface~(\circled{8}). 
\hj{The LCD interface uses row and column drivers to display~(\circled{9}) each frame's \ha{pixels on} the LCD display.} 

\vspace{5pt}
\subsection{System Power States in Video Processing}\label{sec:syspowerstates}

\begin{figure}[!ht]
\begin{center}
\includegraphics[trim=.7cm 1.3cm .7cm .7cm, clip=true,width=0.99\linewidth,keepaspectratio]{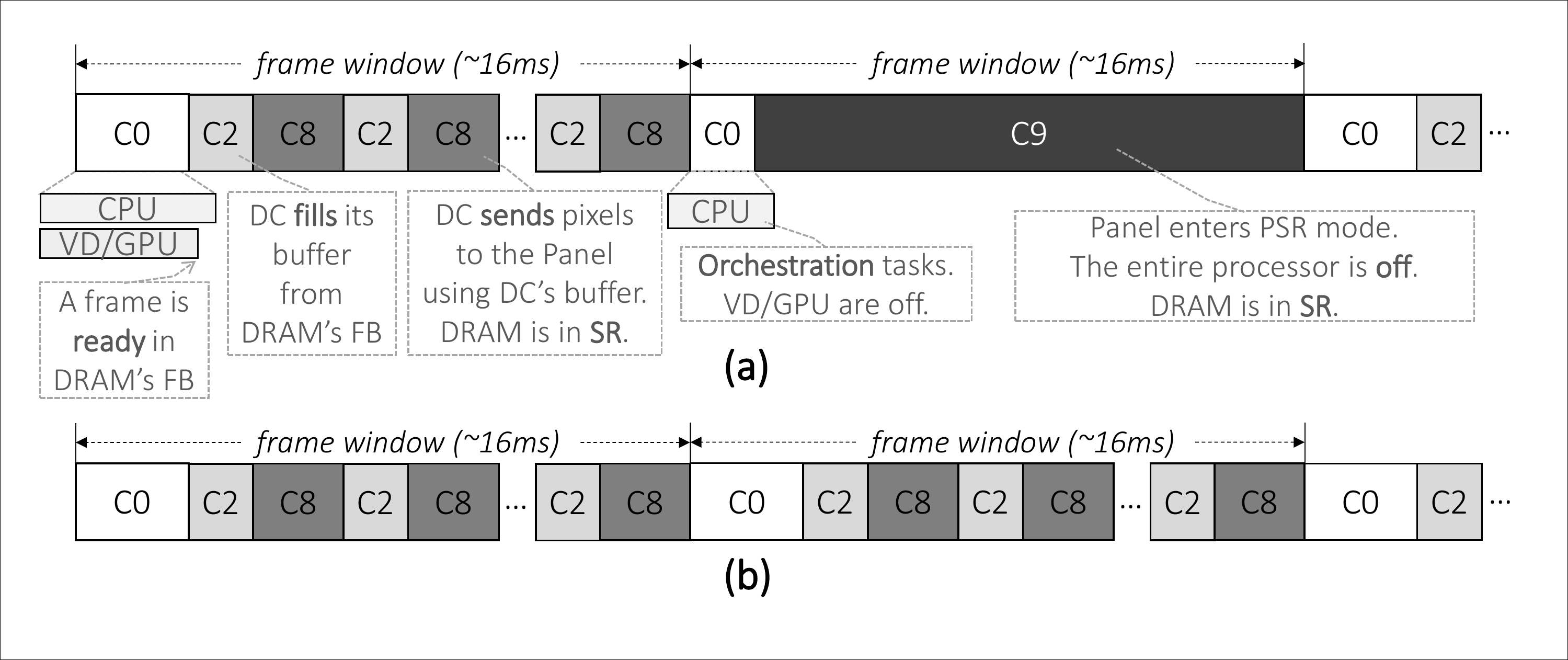}
\caption{Package C-state timeline while displaying \pr{(a)} a 30FPS video and \ry{(b) a 60FPS video on a 60Hz display panel.}}
\label{fig:power_timeline}
\end{center}
\end{figure}

During the video processing flow, the system switches between different power states. 
\fig{\ref{fig:power_timeline}} shows how the package C-state changes while an Intel Skylake mobile processor~\cite{rotem2015intel, haj2018power} renders (a) a 30FPS (frames per second) video and (b) a 60FPS video, respectively, on a 60Hz display panel (i.e., the frame window \hj{is} ${\sim}16$~ms).
At the beginning of each $16$~ms window, the \hj{system} resides in \hj{the} $C0$ power state \cite{haj2018power}, an active state where all the system components (i.e., CPU cores, GPU, video decoder (VD), display controller (DC), eDP interface, and display panel) are running. 
This $C0$ state corresponds to 
\hj{1) orchestration tasks that are run by the application and the drivers on the CPU cores,}
\hj{2)} \hlg{buffering new encoded frames} (\circled{1} in \fig{\ref{fig:current_pipeline}})\hj{, and 3)} frame decoding by the VD (\circled{2} and~\circled{3}). 
\jk{In VR video processing, the \hj{GPU} performs PT operations} while in $C0$ state (\circleda{a} and~\circled{b}).
\hoo{Once \hj{frame}  decoding state is complete, all the cores in the processor are powered off in the remaining frame window, while the DC 1) \emph{periodically} fetches a chunk of the decoded frame from DRAM to the DC buffer (}\circled{4}\hoo{) and 2) \emph{continuously} transfers the fetched data to the display panel (}\circled{5}\hoo{). 
The \hj{system} resides in \hj{the} $C2$ state while the DC fills its buffer.}
Once the DC buffer is full, the path to the host DRAM is closed\haj{,} and the system enters the $C8$ power state~\cite{haj2018power} where only the DC, eDP interface, and display panel are active.
When the DC buffer is almost empty, the DC forces the \hj{system} to return to \hj{the} $C2$ state \hj{such that it can} open the path to the host DRAM and fetch the next chunk of the decoded frame. 
\hoo{Note that the DC keeps transferring pixel data \hj{from} its buffer \hj{to display panel} at a \emph{constant} rate (which is determined by the display's pixel update rate) regardless of \hj{system} power state \hj{transitions}.}
This power state sequence repeats until the \haj{DC transfers a complete} decoded frame to the \hj{display} panel.

A display panel with a 60Hz refresh rate can support up to 60 FPS. When the video frame rate is 30 FPS, each decoded frame is updated on the panel twice (back to back), as shown in \fig{\ref{fig:power_timeline}(a)}.
During \emph{every other} frame window for a 30FPS video (\haj{e.g., second frame window in \fig{\ref{fig:power_timeline}(a)}}), the RFB provides the buffered frame data to the PF (\circled{7}) to refresh the display (PSR). During this entire frame window, the whole processor, DC, and eDP interface can be disabled, \haj{enabling} the system to enter the deep low-power state $C9$~\cite{haj2018power}. 
The PSR technology significantly reduces the energy consumption of the entire display subsystem\hc{. Therefore,} we use \ha{PSR} as \hj{the} baseline for evaluation in the rest of the paper. 
\ry{\mbox{\fig{\ref{fig:power_timeline}}}(b) shows the power C-state timeline while the same system plays a 60FPS video. 
Since the video frame rate matches the panel refresh rate, the system \ha{needs to decode and update a new frame for every} frame window, \hc{which leaves no opportunity to \hd{use} the PSR mode between frame updates}.}

\begin{figure}[!b]
\begin{center}
\vspace{-1em}
\includegraphics[trim=.7cm .7cm .7cm .7cm, clip=true,width=0.8\linewidth,keepaspectratio]{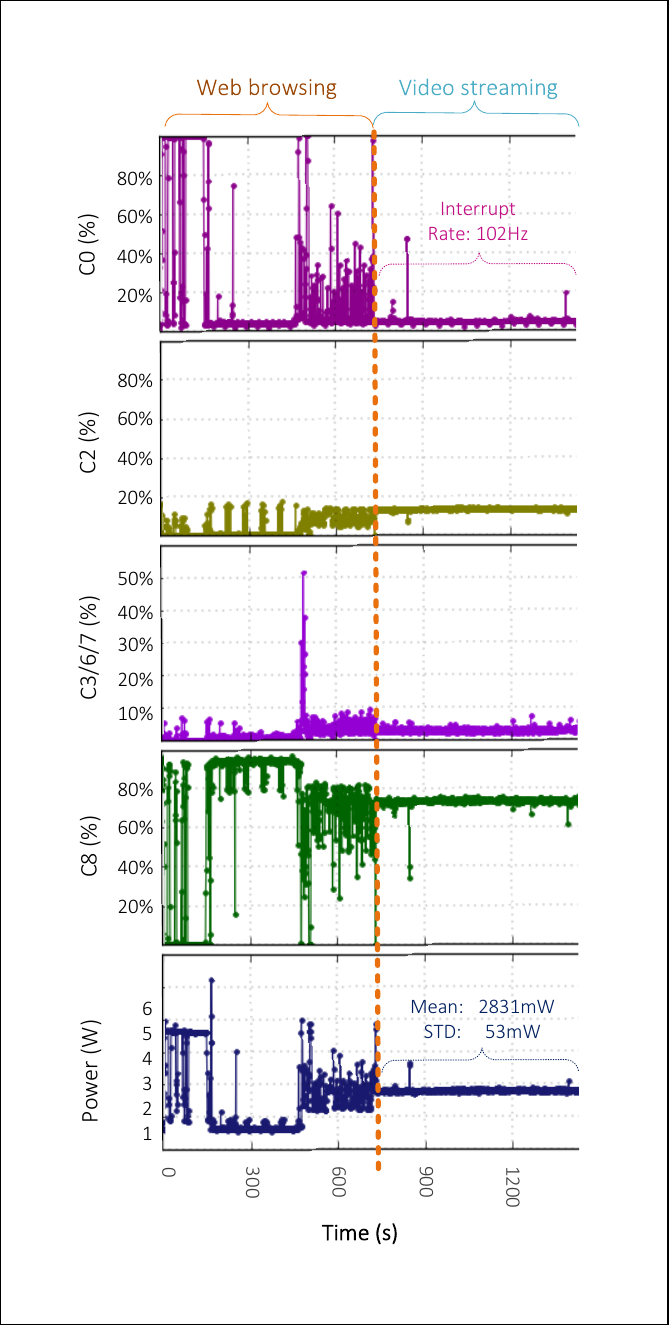}\\
\vspace{-1em}
\caption{System power consumption and \hj{p}ackage C-state residency distribution when running \hj{a} web-browsing workload followed by \hj{an} FHD 60FPS video streaming workload on a 60Hz display.}
\label{fig:pwr_resf_timeline}
\end{center}
\end{figure}

\ro{\mbox{\fig{\ref{fig:pwr_resf_timeline}}} shows the \haj{system} power consumption and package C-state residency distribution when running a web-browsing workload followed by \hj{an} FHD 60FPS video streaming workload on a 60Hz display. 
The figure shows that, while video streaming, the \haj{system primarily} resides at $C8$ (${\sim}75\%$), $C2$ (${\sim}15\%$), and $C0$ (${\sim}8\%$) states, while infrequently entering other package C-states for short times (e.g., the total residency of $C3$/$C6$/$C7$ states is less than $2\%$).}


\section{Motivation}\label{sec:motiv}

We present our key observations that motivate a new energy-efficient video display scheme in modern mobile systems. 

\head{Observation 1: Unnecessary data movement between the display subsystem and host DRAM} 
In current video processing schemes, the \hj{video decoder} (\hj{or} GPU for virtual reality (VR) videos) stores each decoded frame into the frame buffer \hj{in} the host DRAM (\circled{3} \jh{and \circled{b}} in \fig{\ref{fig:current_pipeline}}), and the \hj{display controller (DC)} fetches the decoded frame \hj{from the host DRAM} to send it to the display panel (\circled{4}). 
Doing so is necessary when there exist other planes in addition to the video plane.\footnote{A plane is a window of content to be displayed on the screen that defines an independent data stream. 
The final image is a composition (overlay) of different planes in a pre-defined order of superposition~\cite{khan2007bandwidth}.}
For example, suppose that there are four planes to display: 1) the \emph{background} plane that is typically a static image, 2) the \emph{video} plane that contains the video stream, 3) the \emph{application-graphic} plane for the graphical user interface (GUI), and 4) the \emph{cursor} plane to display the cursor.
In such a case, each plane has its frame buffer in the host DRAM. 
DC reads data chunks from each \hj{plane's frame} buffer, generates one \hj{composite} chunk out of them, and sends the \hj{composite} chunk to the display panel.
However, when the user plays a video in full-screen mode (which is \haj{typical} for planar videos and is the default for VR videos), \emph{storing the decoded frame into the host DRAM \hj{first and then reading it again from the host DRAM} is unnecessary} since there is no other plane for the DC to merge with the video plane. 
Bypassing the DRAM in \hj{such (common)} cases would significantly reduce unnecessary data movement \hj{over the power-hungry off-chip interconnects \ha{\cite{seshadri2013rowclone,david2011memory,mutlu2013memory,haj2020sysscale}}} and thus improve energy efficiency.
\hj{Our goal is to enable \emph{frame buffer bypassing} whenever it is possible, with minimal changes \hc{to} current mobile SoC microarchitectures.}

\head{Observation 2: Underutilization of the eDP interface ban\-dwidth} 
As explained in \sect{\ref{sec:syspowerstates}}, the system alternates \hj{between} power \hj{states} C2 (when reading a chunk to the DC buffer) and C8 (when the buffer is full) while the DC continuously sends a \emph{full} decoded frame (e.g., 24 MB for a 4K resolution) to the display panel, which keeps both the DC and eDP receiver active during the \hj{\emph{entire}} frame window (e.g., $\sim$16 ms in a 60Hz refresh rate).
However, the newest eDP interface~\cite{edp1_4b} supports a maximum bandwidth of $25.92$~Gbps, where it takes only $7.2$~ms to transfer an entire 4K decoded frame. 
This means that the DC and eDP receiver can potentially switch to a power-saving mode for $55$\% of the 16ms frame window after decoding and sending the entire frame in \emph{one burst}.

The root cause for this inefficiency in conventional systems is that the display controller, eDP receiver, and pixel-formatter (PF) are \emph{tightly coupled}.
\hj{For example, in a 4K display with a 60Hz refresh rate, the pixel update rate must be fixed to about \hj{11.3~Gbps} (i.e., 60 frames, each of which is 24 MB in size for 4K resolution, needs to be updated every second),
which dictates the DC's transfer rate through the eDP interface.}
The PF's pixel update rate is determined \hj{by} the \hj{display} panel's resolution and refresh rate, in order to be aligned with \hj{the update rate} of the LCD panel. 
Note that increasing the PF's pixel update rate without proper changes to the LCD panel would \hj{cause} image \emph{flickering} and \emph{distortion} \hj{\cite{psr_2}}.

\hj{Our goal is to eliminate the bottleneck in the display panel so that the system directly transfers a full decoded frame from the video-decoder (or GPU) to the display panel in a burst, exploiting the display interface’s maximum bandwidth. Doing so \ha{would} (1)  reduce the energy consumption of the host DRAM by eliminating data movement to/from the DRAM frame buffer, and (2) increase the system’s idle-power state residency by reducing the usage of the processor and the display subsystem since they are active only during the burst period. }

\section{BurstLink Design}\label{sec:technique}

To address the two inefficiencies discussed in \sect{\ref{sec:motiv}}, 
we propose \tech, \hj{a novel system-level technique that improves the energy efficiency of planar and VR video streaming.
\tech is based on two key mechanisms: \emph{Frame Buffer Bypass} and \emph{Frame Bursting}.
\emph{Frame Buffer Bypass} \emph{directly} transfers a decoded video frame from the video decoder or \hj{the} GPU to the  display panel, {completely} bypassing the host DRAM.
\emph{Frame Bursting} transfers a \emph{complete decoded frame} to the display panel \emph{in a single burst}, using the maximum bandwidth of modern display interfaces.} 
This section describes \js{how the two mechanisms} work to make video display and other \ha{frame-based} mobile workloads more energy-efficient \js{in} modern mobile systems.

\begin{figure}[ht]
\begin{center}
\includegraphics[trim=.6cm .75cm .6cm .75cm, clip=true,width=1\linewidth,keepaspectratio]{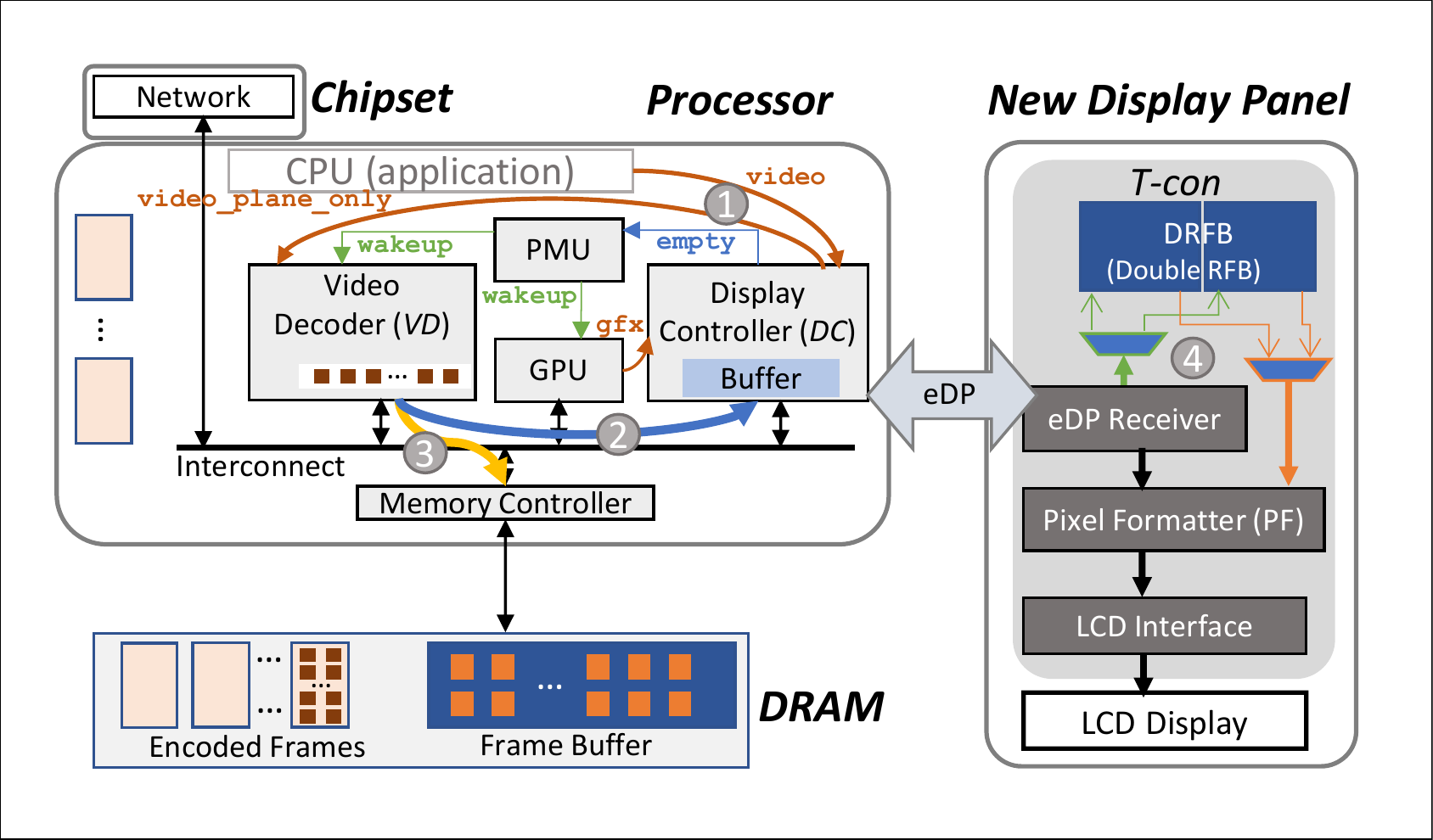}
\caption{\tech video processing and display.}\label{fig:full_screen_video}
\end{center}
\end{figure}

\subsection{Frame Buffer Bypass} \label{sec:frame_bypass}
\begin{figure}[!t]
\begin{center}
\includegraphics[trim=0.7cm 1.2cm 0.7cm 0.7cm, clip=true,width=0.99\linewidth,keepaspectratio]{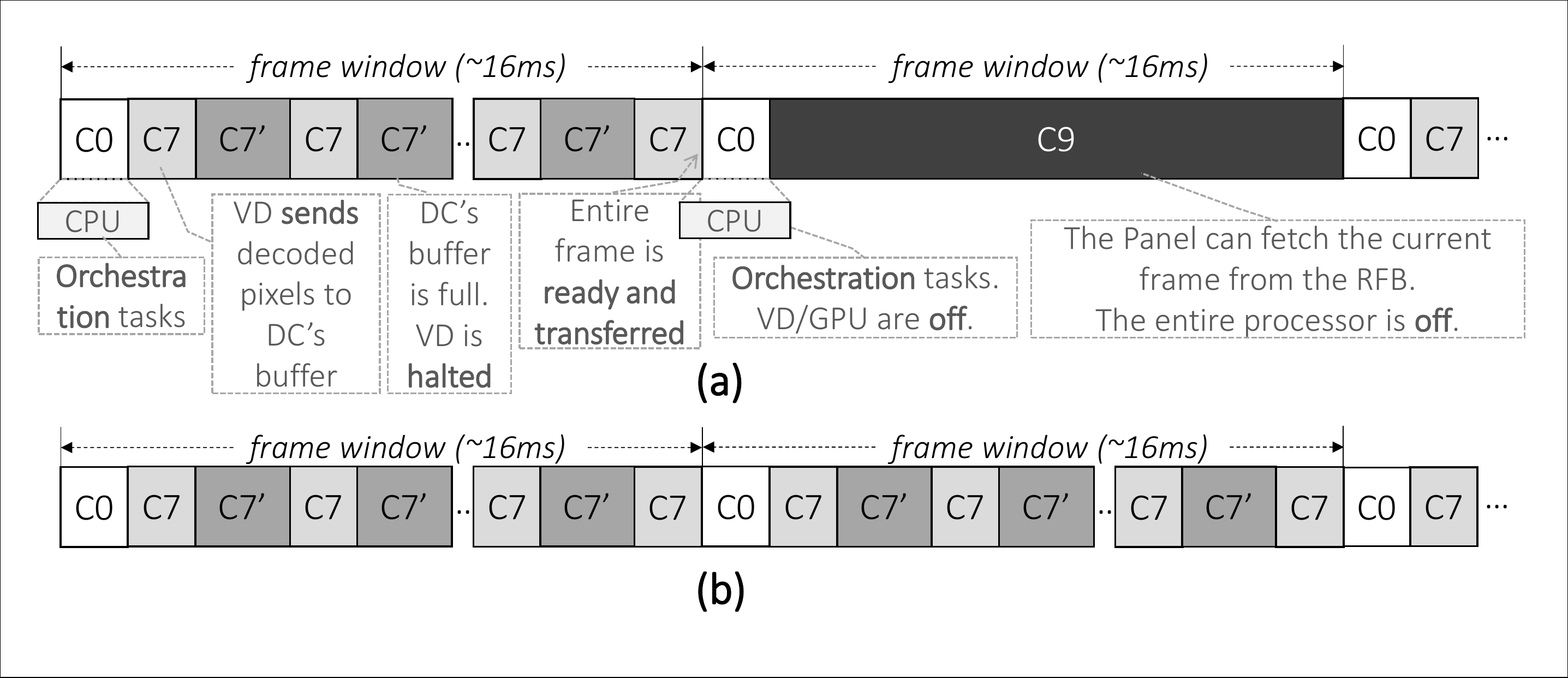}\\
\caption{\hg{Package C-state} timeline of a mobile processor using the Frame Buffer Bypass technique while displaying (a) a $30$FPS video and (b) a $60$FPS video on a $60$Hz display panel.}
\label{fig:power_timeline_bypassing}
\end{center}
\end{figure}

The \js{Frame Buffer Bypass technique} redirects the processed frame from the \m{video} decoder (VD) (or the GPU) to the display controller (DC) \js{via} the on-chip interconnect \js{(\circledg{2} in Fig.~\ref{fig:full_screen_video})}
\hd{if two conditions are satisfied\he{:}}
\hc{1) \hd{an asserted} signal from the DC {(\texttt{video\_plane\_only} \circledg{1} in Fig.~\ref{fig:full_screen_video})}} \hc{\hd{indicating that} only \hd{the} video plane needs to be displayed and thus the frame should not be merged with any other frames and 2) \hd{a set flag in the VD \he{(\texttt{single\_video})} indicating that} \he{only} a single video application is running.}
\js{Fig.~\ref{fig:power_timeline_bypassing} depicts the package C-state timeline throughout the Frame Buffer Bypass procedure. 
\haj{Once the CPU completes the orchestration tasks, the VD sends the processed frame directly to the DC buffer\footnote{The DC buffer is implemented as a \js{\emph{double-buffer}}~\cite{intel_display}, which allows the DC to send \m{one} frame\m{'s} data to the display panel while it is simultaneously receiving data for another frame.} instead of first buffering it in DRAM.}
Bypassing DRAM allows \haj{the processor to perform} this process while the system is in the low-power state $C7$ (described in Table~\ref{tab:p-states}) instead of the \ha{higher-power} $C0$ power state required by conventional systems.}
When the DC buffer is full, \hym{the VD is halted until the DC transmits} the data to the DRFB in the display panel over the eDP interface. The system power state reduces even further to $C7'$, i.e., $C7$ \m{with} VD clock-gated. 
Once the buffer is almost empty, the DC notifies the VD via the \hj{power management unit} (PMU) \texttt{empty} and \texttt{wakeup} signals \hj{(}in \mbox{\fig{\ref{fig:full_screen_video}}}) so that the VD can continue transferring the frame to the DC buffer.
The display panel receives the data over the eDP \m{interface} and stores it \jh{directly} into the \jh{DRFB} inside the display. \hly{The pixel formatter (PF) pulls the  data from the \jh{DRFB} at its own \js{IO} rate and renders the pixels on the panel.}

\js{The above process of \tech provides \haj{two} key advantages over conventional systems.
First, \tech effectively} bypasses DRAM in most of the video processing stages. \m{Doing so  reduces both} DRAM bandwidth and energy consumption.
\js{Second, \tech} \hgg{reduces the orchestration} overhead \m{\ha{that} requires} application and driver \m{involvement} (e.g., programming the DMA engines and handling interrupts).  \m{Doing so} decreases active power state residency (i.e., $C0$) and increases \haj{deep} idle power state residency \js{(e.g., $C7$ or $C7'$)}\haj{, thereby reducing system energy consumption.}

\haj{Effectively,} the Frame Buffer Bypass technique  interleaves the decoding (projection) and display stages \js{in} \m{both} planar \js{and} VR video processing. 
Unlike conventional systems that decode \js{(}or project\js{)} the \js{entire} frame at the beginning of the frame window in power state $C0$, our technique \js{distributes} the frame decoding process \js{across} the frame window while \js{keeping} intermediate data inside the DC buffer and the DRFB.
\js{Doing so allows the decoding process to be performed in the $C7$ low-power state without any performance penalty compared to conventional systems.}

\noindent \textbf{Windowed Video \js{Support}.} In addition to \pr{full-screen} \js{planar and VR videos}, \tech~\js{uses} the Frame Buffer Bypassing technique for \emph{windowed planar video\m{,}\footnote{\hj{We assume windowed video only for planar video since} VR video is \js{typically} streamed in \pr{full-screen} mode with a Head-Mounted
Display (HMD) \ha{\cite{leng2019energy}}.}} such as a video clip in a window inside the browser (e.g., YouTube \cite{youtube_2019}). 
This \haj{feature} is enabled by the selective update capability of PSR (i.e., PSR2 described in Section \ref{sec:display_refresh}) that is supported by the eDP 1.4 protocol~\cite{edp1_4b}.


\js{\tech performs w}indowed video streaming \js{in two} stages.
In the first stage (which is the same \m{as} in conventional systems), the system prepares \pr{an} initial frame with the traditional components. 
\js{Suppose that a user watches a streaming video played on a web browser.
In this case, the GPU prepares the graphical parts of the browser, and
the VD decodes the video frames downloaded from the content server (e.g., YouTube) via \haj{the} network.
The frames from the GPU and the VD are stored separately in different DRAM frame buffers.
Then the DC reads, scales (i.e., resizes the video frame to fit the browser window), and overlays the frames to generate the final integrated frame \m{that is} sent to the panel and stored in the DRFB (or the regular RFB in conventional systems).}

\js{The second stage starts once the host processor detects that the graphical frames from the GPU are not changing (i.e., only the video window is being updated).
Then the host processor informs the DC and VD so that the display subsystem operates in PSR2 mode.
In this stage, the VD continues decoding the downloaded frames and sends the decoded frames directly to the DC. The DC then directly sends only the decoded video frames (after scaling them) to the eDP receiver with the offsets of the video frames \m{that \hj{define} the} updated region\m{s in the video frames}. 
The eDP receiver selectively updates the video only at the corresponding offsets in the DRFB, and the PF renders the entire frame into the LCD \ha{display panel}. 
}



\noindent \textbf{\m{Falling \hc{Back} to \hc{the} \hc{Conventional} Display Mode}.} \m{For} all cases that \haj{\tech does} \emph{not} support, the system falls back to the \m{conventional} display mode. 
\haj{For example,}
whenever multiple planes are required, the system \js{operates using the conventional \m{video display} scheme}
\m{where} \haj{the processor transfers} all the \js{decoded frames through the DRAM frame buffer.
To \m{enable this},} \tech \emph{dynamically} selects the destination of the VD/GPU output (i.e., the decoded frame) using the \emph{destination selector} inside the VD/GPU.
As shown in \fig{\ref{fig:full_screen_video}}, the destination selector directs the decoded frame to the DC~(\circledg{2}) when \hj{1)} displaying only the video plane \hj{(\texttt{video\_plane\_only} signal from DC) and 2) a single video application is running (\texttt{single\_video} flag in the VD)}.
Otherwise, it stores the decoded frame \js{into the DRAM frame buffer~(\circledg{3}) as in conventional schemes}.

\haj{Examples of} cases that \haj{\tech does} not support include  
\haj{1}) \js{when} there is a graphics interrupt (to the DC \cite{intel_display}) indicating that a graphics plane is \hj{available} \hy{(e.g., \ry{when the application's GUI appears)}},  
\haj{2}) \js{when} the system \js{\m{exits} the} PSR2 \js{mode} (\ha{e.g.,} in case of \haj{a} windowed video \js{display}) due to a user-input interrupt (e.g., from the touch screen or keyboard), and 3) when using multiple display panels.
\hp{Note that additional content like a \emph{closed-caption} (CC) \mbox{\cite{kellicker2014closed,cc_wiki}} \js{arrives} with the video stream and \js{does} \emph{not} require multi-plane support as \haj{VD handles} \js{such content}.}   


\vspace{-3pt}
\subsection{Frame Bursting}
In \js{conventional} display subsystems, the \haj{system sets the} transfer rate between the display controller (DC) and the display panel (i.e., the eDP \m{interface} transfer rate) \js{depending on the} display resolution \js{(i.e., the number of pixels per frame)}, \m{panel} refresh rate \js{(i.e., the number of displayed frames per second)}, and color depth \js{(i.e., the number of bits per pixel (bpp))}~\cite{intel_display}. 
\js{Since the PF's throughput is dictated by the pixel-update speed of the LCD panel, conventional systems align the eDP transfer rate with the PF frequency, which leads to a far lower eDP transfer rate than the maximum bandwidth of the eDP interface (e.g., up to $25.92$Gbps in eDP 1.4~\cite{edp1_4b})}. 
\js{This \m{far-from-optimal} transfer rate bottlenecks {\tech}'s \ha{entire} video processing pipeline\haj{, limiting the system power states to $C7$ and $C7'$} \haj{(while decoding a frame and transferring its data to the display panel}, as shown in \fig{\ref{fig:power_timeline_bypassing}}).} 

To leverage the maximum eDP interface bandwidth, we propose \emph{Frame Bursting}, a technique to burst transfer the decoded frame from the processor to the display panel. 
With the Frame Bursting technique, the display panel receives \js{a full-}frame over the eDP interface and stores it \emph{directly} into the \jh{DRFB} (\circledg{4} in \fig{\ref{fig:full_screen_video}}) before transferring the frame to the PF, \js{which removes the slow update process between the PF and the LCD panel from} the critical path. 
The PF can fetch the frame data from the DRFB at the rate required \js{by a given configuration (i.e., the display resolution, refresh rate, and color depth) to generate pixels and send them into the LCD}.
This \haj{technique} reduces the utilization of the processor and the display subsystem, \js{enabling} the system to enter deep low-power states \haj{(e.g., $C9$)} after \m{quickly} transferring the decoded frame to the \jh{DRFB}. \haj{The processor turns off} all unused resources (e.g., \js{the} DC, eDP interface, \js{and} DRAM) in \m{such deep low-power} states. 

\vspace{-3pt}
\subsection{System Power States in \tech}\label{sec:techpstates}

\begin{figure}[ht]
\begin{center}
\includegraphics[trim=0.7cm 0.73cm 0.7cm 0.75cm, clip=true,width=1\linewidth,keepaspectratio]{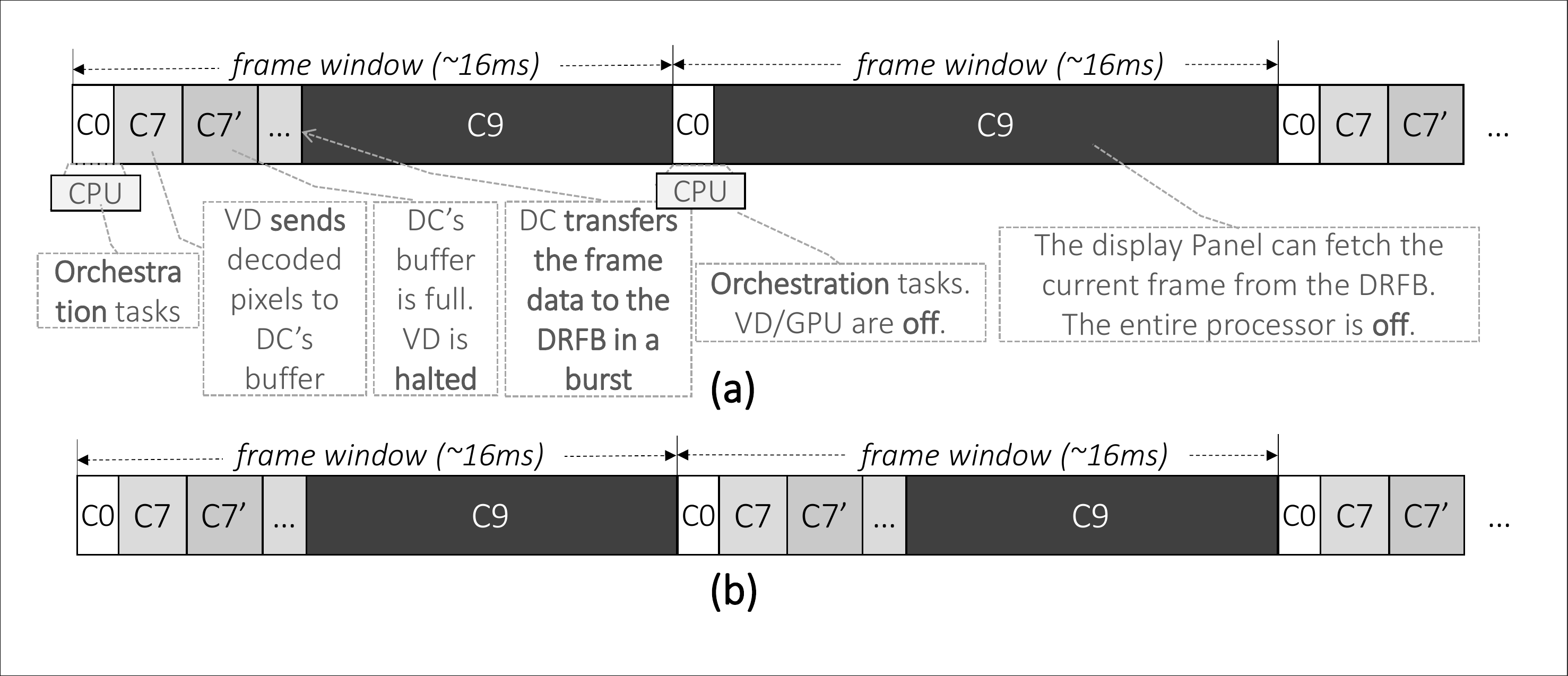}
\caption{\hj{Package C-state} timeline of a mobile processor with complete \tech \hym{for (a) 30FPS and (b) 60FPS videos \hj{on a 60Hz display panel}.} 
}\label{fig:burstlink_power}
\end{center}
\end{figure}

\fig{\ref{fig:burstlink_power}} shows the \hj{power state (i.e., package C-state)} timeline of a system that supports \js{\tech (i.e., both Frame Buffer Bypass and Frame Bursting)} while rendering a 30FPS (frames per second)  \hym{and 60FPS} \jh{planar} video 
on a $60$Hz display panel. 
The system resides in  \hj{power state} $C0$ at the beginning of a frame window, where the display driver sends the encoded video frame to the VD \hj{and then remains idle for the rest of the frame}.
\hj{Once the driver is idle, the} system then \hj{\ha{alternate} between $C7$ and $C7'$ power states, where the} VD decodes the video frame in \js{multiple} chunks 
and sends them to the DC, which transfers \js{the chunks} to the \jh{DRFB} over the eDP interface at maximum bandwidth.
Once the entire decoded frame is sent to the \jh{DRFB}, the system \js{enters} the deep low-power state $C9$, in which most components, including \js{the} VD, DC\js{,} and eDP interfaces (both on processor and panel side\js{s}), are power-gated. \hym{For every other frame of a 30FPS  video, the system enters \m{directly} into $C9$ \hj{at the beginning of the frame (after \ha{a} short driver orchestration task in $C0$)} state since the entire \m{frame} already exists in the DRFB.}

\js{Table~\ref{tab:power_comparison} compares \tech to conventional video processing (Baseline) in terms of the average power consumption in each power state $Ci$ and the \emph{power-state residency (Residency)}, i.e., the percentage of a frame window that the system resides in state $Ci$.} \haj{The last column shows the average power (AvgP) of each scheme.}
\js{As shown in Table~\ref{tab:power_comparison}, \tech significantly reduces the average power consumption over conventional video processing by more than 40\% by 1) enabling the system to reside in lower power states (\hj{e.g.,} $C9$) and 2) reducing the time that the system resides in higher power states \haj{(e.g., $C0$)}.
We discuss \tech's power/energy savings in more detail in  Sections~\ref{subsec:setting}~and~\ref{sec:eval}.
}

\begin{table}[ht]
\renewcommand{\arraystretch}{1.1}
\centering
\caption{Power consumption comparison of \tech and conventional video processing on \m{an} Intel Skylake SoC when playing a\m{n} FHD video at 30FPS on a 60Hz display.}
\label{tab:power_comparison}
\resizebox{\linewidth}{!}{%
\begin{tabular}{llccccc|c}
\hline
                          & \textbf{Package  C-\m{s}tates} & $C0$   & $C2$   & $C7$   & $C8$   & $C9$   &\textbf{\rp{AvgP (mW)}}         \\ \hline \hline
\multirow{2}{*}{\textbf{Baseline} (measured)}  & \textbf{ Power (mW)}       & $5940$ & $5445$ & $1385$ & $1285$ & $1090$ & \multirow{2}{*}{\textbf{2162} } \\
                          & \textbf{Residency (\%)}      & $9\%$  & $11\%$ & -      & $80\%$ & -      &                         \\ \hline
\multirow{2}{*}{\textbf{\tech} (estimated)}     & \textbf{ Power (mW)}       & $6090$ & $5740$ & $1530$ & $1435$ & $1090$ & \multirow{2}{*}{\textbf{1274}} \\
                          & \textbf{Residency (\%)}      & $2\%$  & -      & $19\%$ & -      & $79\%$ &                         \\ \hline
\end{tabular}%
}
\end{table}

\begin{sloppypar}
\subsection{Implementation and Hardware Cost}
\label{subsec:design}
Frame Buffer Bypass and Frame Bursting can be implemented independently \ha{on top of} a conventional display subsystem\ha{.} \tech \hlp{incorporates both mechanisms to \js{maximize}  energy \js{efficiency}. 
\tech requires \m{mainly} three changes to a conventional system.} 
\end{sloppypar}

\noindent\textbf{\js{DRFB.}} 
\rg{While using the \textit{double} remote frame buffer (DRFB) introduces additional cost, power, and area overheads\m{, it} would not be \haj{a severe obstacle to} wide adoption of \mbox{\tech} due to \haj{two} reasons.
First, according to the bill-of-materials (\rpp{BOM}\footnote{\rpp{Given a product, a BOM is a list of product's immediate components with which it is built and the components' relationships.}}\cite{jiao2000generic}) cost estimation of Microsoft Surface Pro~\mbox{\cite{msft_srfc_pro_bom}} (our evaluated system \m{in Section~\ref{sec:eval}}), \mbox{DRAM’s} BOM cost is \hym{\$13.9/GB} while the full HD display panel costs \hym{\$100.4}~\mbox{\cite{msft_srfc_pro_bom}}.
Based on this estimation, doubling the existing \mbox{Tcon’s} RFB from $24$MB to $48$MB will increase its BOM cost by only $32.5$ cents (i.e., $13.9 \times \m{\frac{24}{1024}}$), which is \m{only} a $0.3\%$ increase in the total BOM cost of the display panel (\m{corresponding to} $0.05\%$ of the mobile device BOM cost).
Second, according to our estimation based on Samsung's recent proposal for a cost-effective RFB implementation~\mbox{\cite{hwang201716}}, the additional power overhead from doubling the \haj{size of the} RFB is $58$~mW (6\% of the display panel power), which is significantly lower than \mbox{\tech's} \ha{overall} power savings.
Note that the DRFB does not increase the panel size since the DRAM of the DRFB is mounted on a flexible printed circuit board (FPCB) rather the panel \mbox{\cite{hwang201716}}.
}


\noindent \textbf{Destination Selector.} 
\js{As shown in \fig{\ref{fig:full_screen_video}}, \tech dynamically selects the destination of the video decoder's (VD) output depending on the 1) \texttt{single\_video} flag in the VD and 2) \texttt{video\_plane\_only} \ha{(\circledg{1} in Fig.~\ref{fig:full_screen_video})} signal from the display controller (DC) to the destination selector.}
The flag and signal can be determined using two data elements that are already stored in configuration registers in the VD and DC.
First, since each video application injects its requests into the VD using the driver API~\mbox{\cite{rts,nachiappan2015vip,yedlapalli2014short,grafika}}, the VD already keeps track of \emph{the number of concurrently running video applications} (and their requirements) in its control and status registers (CSRs). 
Second, \m{since} each application also sends its requirements to the DC~\cite{intel_display,msft_display_driver}, the number of used planes and each plane's type (e.g., video, graphics, or cursor) are \m{already} available in the DC CSRs (e.g., SR02 and GRX registers in Intel DC~\mbox{\cite{intel_display}}).
\hj{Therefore, implementing \ha{the} destination selector is straightforward.}

\noindent \textbf{PMU Firmware Changes.} \tech requires changes to the \js{PMU firmware} to 1) enable \js{the processor} to enter power state $C9$ when Frame Buffer Bypassing is \js{enabled}, 2) \js{wake} up the VD \js{(i.e., switch to power state $C7$)} to resume frame data decoding once the DC buffer is empty, and 3) \js{enable the DC to transfer the decoded frame data using the maximum eDP bandwidth} when Frame Bursting is activated. 
We estimate \js{that} these changes \js{increase the power-management firmware code (e.g., Pcode~\cite{gough2015cpu} in the Intel Skylake SoC)} by only a few \hlg{tens of lines,\footnote{This estimation is based \js{on} the \js{number} of lines \m{of code} for Pcode in \js{Intel} Skylake systems to fill the DC buffer \hj{\cite{gough2015cpu}}.}} 
\js{which leads to only a 0.004\% increase in the processor's die area for the Intel Skylake SoC~\cite{21_doweck2017inside}}. 

\vspace{20pt}
\subsection{Generalization of {\tech} Techniques}
\js{{\tech} aims to improve the energy efficiency of video streaming (which is one of the most important application scenarios in modern mobile systems). However, the proposed techniques can also be used in more general frame-based applications} such as 
video \hj{capture} \js{(recording)}, audio streaming, video chat, social networking, and interactive games.
\ry{A general takeaway from \mbox{\tech} is that using main memory (DRAM) as a \emph{communication hub} between system components \ha{is} energy-inefficient. Instead, \mbox{\tech} uses small \emph{remote} memory (e.g., 48MB DRFB, which is only $0.3\%$ of 16GB DRAM) near the data consumer (e.g., a display panel) or the data producer (e.g., a camera sensor) to significantly reduce the number of costly main memory accesses in frame-based applications.} 
\section{Experimental Methodology}
\label{subsec:setting}

\jkt{We outline} our methodology for evaluating \tech. First, we describe our workloads. Second, we introduce our \jkt{new industry-grade} analytical power model for evaluating the baseline \ha{system} and \tech. \ha{We open-source our model online~\cite{Burstlink-github}.} Third, we discuss our process for validating our model against \emph{power measurements} \jkt{from} a real modern mobile device \jkt{that is based on the Intel Skylake system}.  



 
\subsection{Workloads \label{sec:workloads}}

We evaluate \tech with \jh{planar and VR} video-streaming workloads~\jh{\cite{19_MSFT,corbillon2017360}}, which are used in standard industrial benchmarks for battery-life~\cite{19_MSFT,bench1,bench2,bench3,bench4} and academic evaluations of  video-streaming optimizations~\hj{\cite{yedlapalli2014short,rts,leng2019energy,nachiappan2015vip,zhao2020deja,boroumand2018google,nachiappan2015domain,rao2017application,gaudette2016improving,duanmu2017view}}.
\haj{These workloads typically assume that}
 only a single application (e.g., video streaming in our evaluation) \haj{runs} on the system.
\ry{In smartphones and tablets, the currently used application typically runs in full-screen mode, while the other opened applications are normally sent to \mbox{\taha{the}} background and moved to \mbox{\taha{a}} suspended state. This \haj{assumption} is also true of our evaluated system, the Microsoft Surface \mbox{\cite{msft_surface}} when running in Tablet-Mode \mbox{\cite{tidrow2015windows}}. }

\subsection{Analytical Power Model}\label{sec:power_model}

\noindent \textbf{Modeling the Baseline System.} \jkt{We develop a new} analytical power model \jkt{that} estimates the average system \jkt{power, $Power_{avg}$,} within a frame window, as follows: 




\vspace{10pt}


\noindent    \resizebox{0.99\hsize}{!}{$Power_{avg} = \sum_{i=0}^{10} P_{C_i} \cdot R_{C_i} +  P\_en_{C_i} \cdot Lat_\_en_{C_i} + P\_ex_{C_i} \cdot Lat\_ex_{C_i}$}


\vspace{10pt}
\noindent
$P_{C_i}$ denotes the \jkt{average} system power consumption in power state $Ci$ \hj{(e.g., package C-states in Table \ref{tab:p-states})}. 
\hg{$P\_en_{C_i}$ and $P\_ex_{C_i}$ denote the average power consumption  while entering and \pr{exiting} state $Ci$, respectively.
$R_{C_i}$ denotes the residency at power state $Ci$, i.e., 
the percentage of the total time the system spends at power state $Ci$. $Lat\_en_{C_i}$ ($Lat\_ex_{C_i}$) denotes the latency for entering (exiting) power state $Ci$.} 
\hlgg{We use a synthetic benchmark to 1) place the system in different power states, and 2) measure entry and exit latencies~\mbox{\cite{schone2015wake}}. We obtain power state residency using processor's residency reporting counters~\mbox{\cite{intel_skl_dev}}}.
\jkt{Our power model inherently accounts for major system parameters such as 1)} DC buffer size, 2) DRAM capacity, 3) DRAM bandwidth, and 4) eDP bandwidth. 
\haj{These parameters can directly affect each state's residency and power consumption, and the frequency \jkt{with which} the system switches between power states.}

\rblue{
\hj{For power states in which the DRAM is active (i.e., not in self-refresh, such as in package $C0$ and $C2$ \ha{states} as shown in Table \ref{tab:p-states}), the power state's average system power consumption (i.e., $P_{C_i}$) depends also on DRAM power, which is correlated to DRAM bandwidth.} \haj{We model} \jkt{DRAM} power consumption in two parts: 1) \emph{background power}, which is consumed regardless of memory access characteristics and only depends on DRAM power states (i.e., self-refresh, CKE-High (active), CKE-Low (fast power-down)~\mbox{\cite{david2011memory}}), and 2) \emph{operating power}, which highly depends on DRAM read/write bandwidth~\cite{david2011memory, chang2018voltron, ghose2018your}.  
First, to model background power, we record the time spent in each DRAM power state.
We weight the power consumption in each state by the measured time values 
to obtain the average background power.
Note that the DRAM power states in our processor are correlated to the package C-states. 
For example, 
DRAM is in \haj{the} active (CKE-High) \haj{state} only in $C0$ and $C2$ states 
while it \jkt{is in} \haj{the} self-refresh state in all other package C-states.   
Second, \haj{we model the} \jkt{operating} power by multiplying the average power per unit read/write bandwidth (e.g., 1 GB/s) by the actual read/write bandwidth consumed. 
To determine the average power per unit read (write) bandwidth, 
 we 1) run a memory benchmark that generates reads (writes) \jkt{at different bandwidth values (similar to~\cite{subramanian2013mise})},
 2) measure DRAM power consumption, and
 3) extrapolate the power consumption per 1GB/s reads (writes). 
}

\noindent \textbf{Modeling the {\tech} System.} We model the power consumption 
of the system enhanced with the two techniques of \tech using  measured data \jkt{from our baseline power} model. The two techniques of \tech affect both the residency and power level at each package C-state. For example, $C9$ residency increases as the system finishes transferring each decoded frame more quickly. 

We carefully model the \emph{\jkt{estimated}} average power consumption (i.e., $P_{C_i}$) and the  residency (i.e., $R_{C_i}$) at each power state $Ci$ \haj{with \tech}, taking into account two essential factors. 1) \emph{Inactive} system components (e.g., power-gated DC or DRAM in \haj{the} self-refresh state) in each power state. 2) Changes in each SoC component's operating frequency (e.g., the DC and eDP interface consume more power when using the maximum eDP bandwidth for Frame Bursting). 
We \jkt{plug in} the new values in our analytical model to \jkt{estimate the average system} power consumption when applying each \taha{of} \tech's technique\taha{s}.



\subsection{Measurements and Power Model Validation}\label{sec:model_valid}

\noindent \textbf{Baseline System.}  
We use an \haj{\emph{Intel reference design}} for high-end tablet devices \cite{zahir2013medfield}, such as the Microsoft Surface Pro~\cite{msft_surface}. \haj{Our baseline system} is equipped with an Intel Skylake~\cite{21_doweck2017inside,fayneh20164} processor (whose specifications are summarized in Table~\ref{tbl:sys_setup}), and multiple debug/configuration capabilities. 


\begin{table}[h]
\renewcommand{\arraystretch}{1.0}
\centering
\caption{Baseline system.}
\label{tbl:sys_setup}
\resizebox{0.9\linewidth}{!}{
\begin{tabular}{ll}
\hline
\textbf{Processor} & \begin{tabular}[c]{@{}l@{}} 
Intel i5-6300U Skylake \hj{\cite{tam2018skylake}}, 14 nm, TDP: 15 W\\ Frequencies: 800-2400 MHz, L3: 3 MB \end{tabular} \\ \hline
\textbf{Memory}    & \begin{tabular}[c]{@{}l@{}}LPDDR3-1866MHz \hj{\cite{no2013jesd79}}, 8 GB, dual-channel  \end{tabular} \\ \hline
\end{tabular}
}
\end{table}

To validate our power \jkt{model} for the baseline \js{and \tech} system\js{s}, 
we carry out the following \haj{steps}: 
1) \haj{we measure} the average power and residency at each power state in the baseline system, 
2) \jkt{we break down} the measured power \ha{into system components},  
and 3) \jkt{we measure} the effect of frequency/bandwidth changes on the average power and residency at each power-state. \jkt{We compare all our power measurements to the estimations provided by our power model.}

\begin{sloppypar}
\noindent \textbf{Measurement Setup.}
For the \haj{system} power measurements, we use a Keysight N6705B DC power analyzer \cite{keysight_N6705B} equipped with an N6781A source measurement unit (SMU) \cite{keysight_acc}. 
The N6705B is normally used for \jkt{high-accuracy (around $99.975\%$~\cite{keysight_acc})} power measurement of low-power devices (e.g., smartphones and tablets). 
The power analyzer measures and logs the instantaneous power consumption of different device components. Control/analysis software (14585A \hj{\cite{keysight_N6705B}}) for data visualization and management runs on a separate laptop connected to the power analyzer.

\begin{figure}[!h]
\begin{center}
\includegraphics[trim=.75cm .68cm .75cm .75cm,clip=true,width=1\linewidth,keepaspectratio]{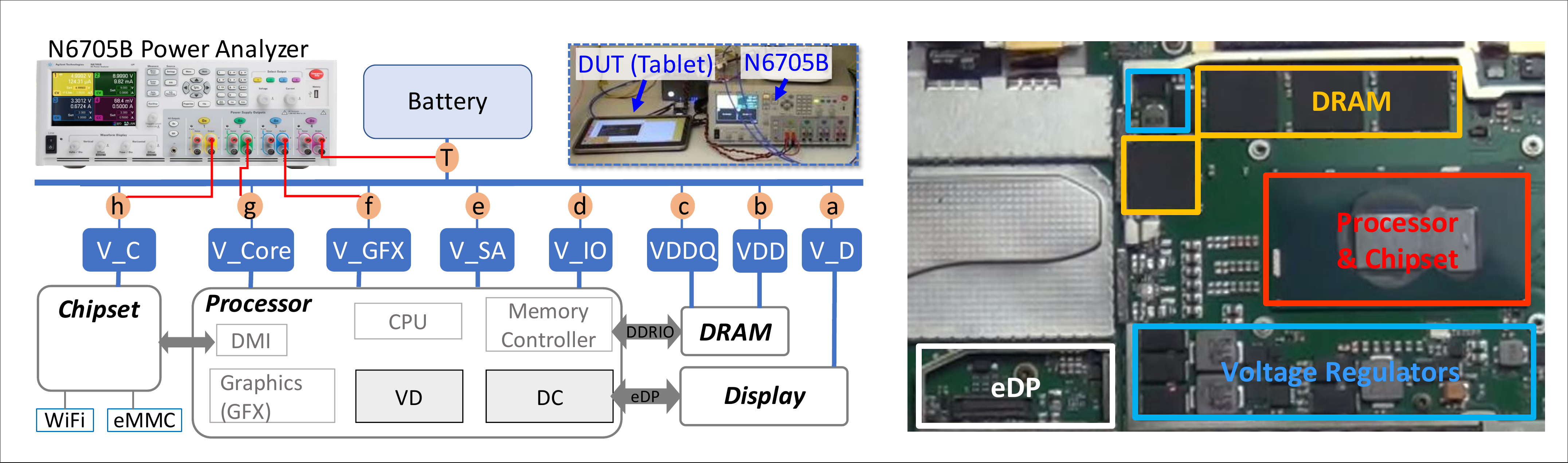}
\caption{Power measurement \jkt{of the Intel Skylake mobile system} using the Keysight N6705B DC power analyzer \cite{keysight_N6705B} \ha{(left)}. \hj{Illustration of Microsoft Surface Pro tablet's system components~\cite{msft_surface} \ha{(right)}.}  
}\label{fig:measurements}
\end{center}
\end{figure}

\fig{\ref{fig:measurements}} shows the \jkt{power} measurement setup of \jkt{the Intel} Skylake mobile system \cite{21_doweck2017inside,fayneh20164,haj2019comprehensive,yasin2019metric} \jkt{under study}. The system has a battery and multiple power supplies (i.e., voltage regulators \ha{\cite{haj2021ichannels,haj2020flexwatts}}) for the mobile system components. We refer the reader to \jkt{the Keysight} manual \cite{keysight_N6705B} for more \jkt{detail} on the actual connections of measurement wires to \haj{the} N6705B power analyzer, the design under test (DUT), DUT's battery, and control/analysis software.
The power analyzer can measure the power consumption of the different power states ($C0$, $C2$, and $C7$--$9$) in a single experiment. 
We measure the residency of each C-state 
using the Intel VTune profiler~\cite{vtune} \jkt{on our evaluated workloads}.

\noindent \textbf{Baseline Power Measurements}. We carry out multiple measurements for different \haj{system} components, including the processor, DRAM, chipset, and display. As illustrated in \fig{\ref{fig:measurements}}, \haj{we connect} the power analyzer's four channels 
to measurement points \textsf{a}-\textsf{f} for individual power domains\taha{,} and measurement point \textsf{T} for the total system power drained out of the battery. For the processor power, we measure four voltage domains: 1) V\_Core, the voltage supply for cores and last-level-cache (LLC), 2) V\_GFX, the voltage supply for the graphics engine and the VD, 3) V\_IO, the voltage supply for the IOs including the eDP \jkt{DRAM}, DDRIO (digital part), and the interface to \taha{the} chipset, and 4) V\_SA, the voltage supply for the \emph{system agent}\footnote{\hj{SA stands for System Agent which houses the traditional Northbridge chip \ha{\cite{radhakrishnan2007blackford,conway2007amd}}. {SA} contains several functionalities, such as the memory controller and the IO controllers/engines~\ha{\cite{sunrise_point_skl,haj2020sysscale,haj2020flexwatts,haj2020techniques}}.}} that contains several controllers, including the memory controller and DC. Each measurement uses four analog channels with a 50-$\mu$s sampling interval. 
\end{sloppypar}

\noindent \textbf{Power Breakdown \jkt{\ha{into} System Components}.}
\haj{We} further break down the measured processor power consumption into \hj{processor's sub-}components \hj{(e.g., VD, DC, eDP, memory-controller)} using power estimation techniques~\cite{gunther2001managing}. \pr{Using} the design characteristics of these components (such as capacitance, leakage, operational frequency\taha{,} and voltage), we estimate their relative power consumption. \pr{Next}, using \jkt{the measured power consumption of the system}, we estimate the power of each component. Note that other power estimation techniques can \jkt{also be used} to determine this \jkt{component-level} power breakdown~\cite{bertran2013systematic,brooks2003new,landman1996high}. 

\noindent \textbf{Power Model Accuracy.}
\hym{To validate our model, we run four representative battery life workloads \mbox{\cite{19_MSFT}}}: \jkt{web browsing, light gaming, video conferencing, and video playback} with multiple display resolution setups.   
We measure average \jkt{system} power (as explained above) and collect package C-states residencies along with each run. We \jkt{use} our analytical power model to estimate the average power consumption of these workloads. Then\taha{,} we compare the measured vs. estimated average power consumption. We \jkt{find} that the accuracy of our analytical power model is approximately $96\%$ \hb{for the \jkt{evaluated} workloads. The accuracy \jkt{for} each of the four \jkt{mainly-used power states in our battery-life} workloads, $C0$, $C2$, $C7$, and $C8$, is $97.4\%$, $96.2\%$, $95.1\%$, and $94.7\%$, respectively.}    

\section{Evaluation}
\label{sec:eval}



We evaluate \tech against the baseline \jkt{video display system} (described in \fig{\ref{fig:current_pipeline}}) with \haj{\emph{five}} studies. 
1) \juang{We study} energy reduction for four different display resolutions \jh{for planar video streaming}.
\jh{2) \juang{We show} energy reduction of different workloads and resolutions for VR video streaming.} 
3) We show the effect of frame rate on \tech energy reduction.
\haj{4}) We compare \tech to state-of-the-art techniques that reduce \haj{the} energy \haj{consumption} of video processing. 
\haj{5}) We evaluate the benefits of \tech for other mobile workloads \jkt{than video} \jkth{display}.  


\subsection{Planar Video Streaming Energy Reduction}
\begin{figure}[!h]
\begin{center}
\includegraphics[trim=.7cm .7cm .7cm .7cm, clip=true,width=1\linewidth,keepaspectratio]{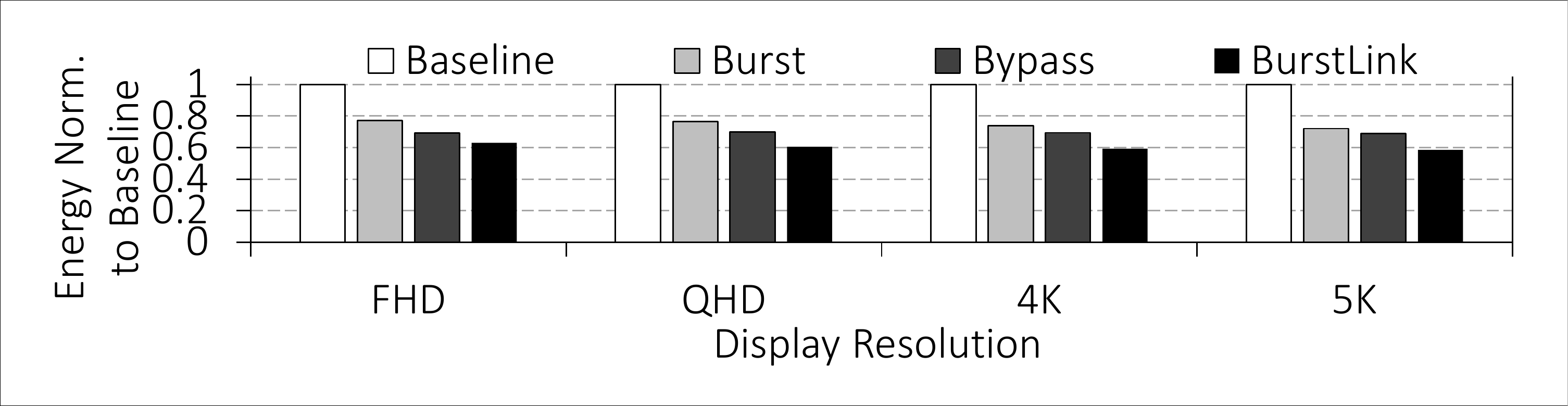}
\caption{\haj{Total system} energy reduction of Frame Bursting, Frame \hc{Buffer} Bypassing, and \tech for $30$~FPS HD \jkt{video}.
}\label{fig:results_30_60}
\end{center}
\end{figure}

\fig{\ref{fig:results_30_60}} shows the \pr{energy} \jkth{consumption} of each  technique of \tech (i.e., Frame Bursting \hd{and} Frame \hc{Buffer} Bypassing) and \jkt{the full \tech\hc{,} normalized} to the baseline \jkt{system,} averaged across frame windows \haj{of} $30$~FPS \jkt{videos} displaying on a $60$~Hz panel. 

We make two major observations. First, \tech reduces the overall \haj{system} energy consumption by $37\%$ for \jkth{an} $FHD$ display. Frame Bursting and Frame Buffer Bypassing \jkth{reduce} overall energy by $23\%$ and $31\%$ compared to the baseline, respectively. Second, \tech's energy reduction \jkt{increases as display resolution increases}. For a \haj{$5K$} display, \tech~\jkth{reduces} the overall \jkt{system} energy by \haj{${\sim}42$\%}.

\begin{figure}[!h]
\begin{center}
\includegraphics[trim=.7cm .9cm .7cm .7cm, clip=true,width=1\linewidth,keepaspectratio]{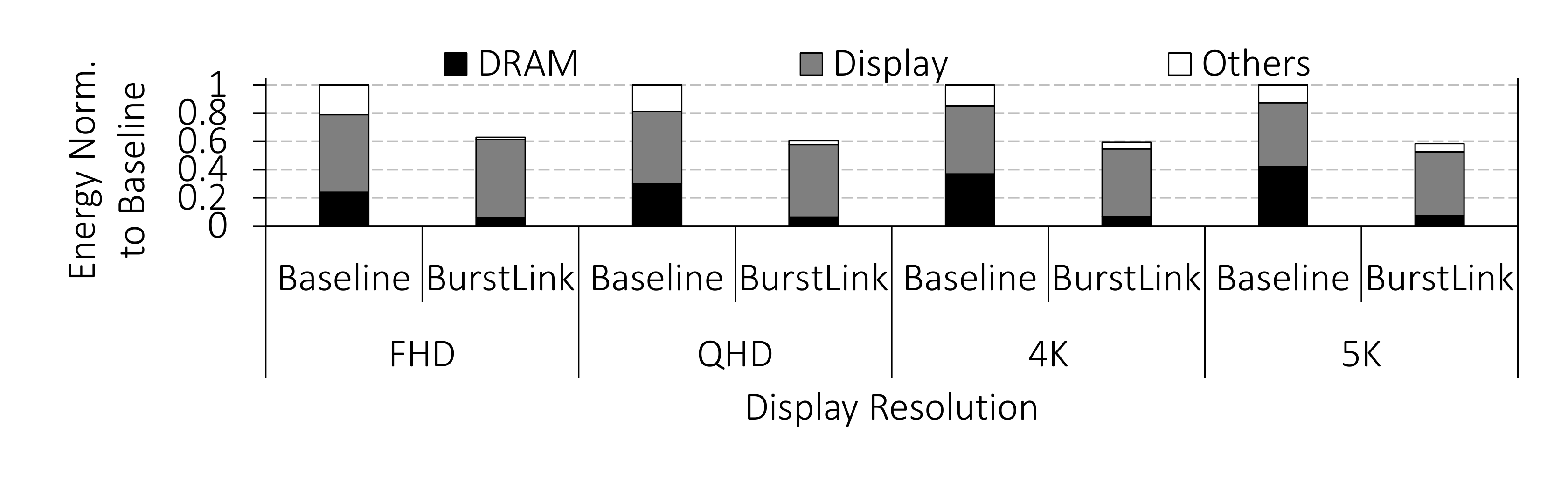}
\caption{\jkth{Energy breakdown} into \hc{system} components.
}
\label{fig:video_energy_result}
\end{center}
\end{figure}

\fig{\ref{fig:video_energy_result}} compares \tech's \haj{system} energy consumption breakdown across \pr{three} \jkt{major system} components\pr{, i.e., DRAM, display, and others (\juang{which includes} \haj{the} processor, 
\juang{WiFi network card, and eMMC storage})} with that of the baseline \haj{system}. \tech reduces the total dissipated energy \jkt{of} DRAM by $3.8\times$ and $5.7\times$ for $FHD$ and $5K$ 
\juang{resolutions}, respectively. \hj{The higher the video resolution, the higher the DRAM\hc{'s} relative energy consumption out of the entire system energy consumption, \jkth{and therefore, the higher the energy reduction of \tech}.} \hj{\tech reduces the total dissipated energy of \jkth{\emph{others}} by $13.1\times$ and $2.1\times$ for $FHD$ and $5K$ 
resolutions, respectively. The higher the video resolution, the lower the processor's relative energy consumption out of the entire system energy consumption, \jkth{and therefore, the lower the energy reduction of \tech}.}

\subsection{VR Video Streaming Energy Reduction}

\begin{figure}[!h]
\begin{center}
\includegraphics[trim=1.3cm .75cm 1.cm .7cm, clip=true,width=1\linewidth,keepaspectratio]{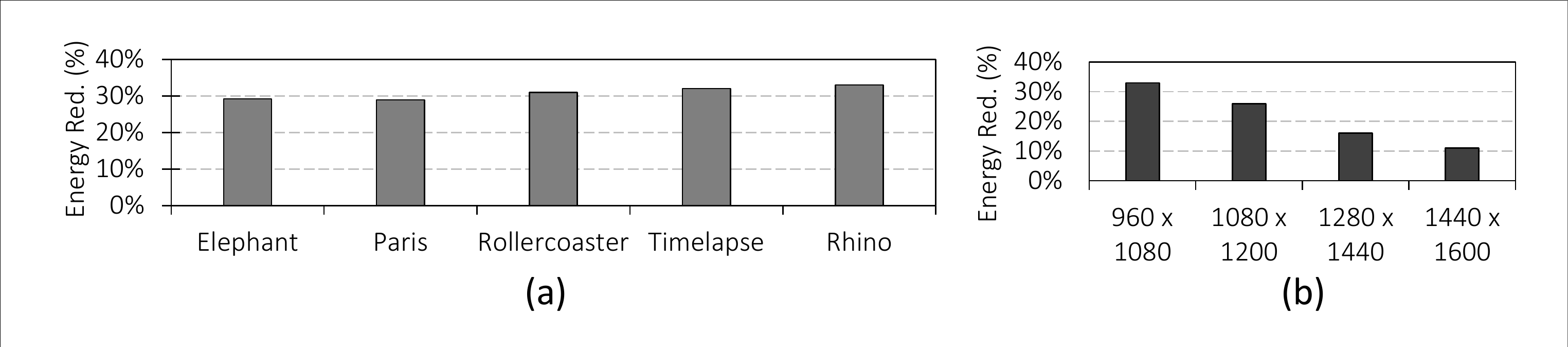}
\caption{\jh{(a) VR video streaming energy reduction for five VR workloads. (b) Energy reduction for different VR display resolutions of the Rhino workload (other workloads are similar).} 
}
\label{fig:vr_workloads}
\end{center}
\end{figure}

\jh{Fig. \ref{fig:vr_workloads} shows the energy reduction of five 360$^{\circ}$ VR streaming workloads \cite{corbillon2017360} when running with \tech. We assume an optimized state-of-the-art VR streaming scheme \cite{leng2019energy,zhao2020deja} \jkt{in the baseline}, which significantly reduces the compute energy compared to traditional schemes.  
}

\jh{We make two major observations. 
First, \tech~reduces the overall \haj{system} energy consumption by up to $33\%$. \jkt{Compute-energy-dominant} (mainly GPU) workloads have lower benefits compared to \jkt{memory-energy-dominant} workloads \jkt{(as would be expected), since \tech~ \jkth{greatly} reduces memory energy}. Second, \tech's benefits decrease as VR display resolution increases. This \juang{is} mainly because compute energy becomes more dominant in VR workloads as \jkt{display} resolution increases \cite{leng2019energy,zhao2020deja}, which leaves \jkt{less relative potential} for \tech~ \jkt{to save in} memory energy.    
}

\subsection{Effect of Video Frame Rate}\label{sec:frame_rate}

\begin{figure}[!h]
\vspace{-5pt}
\begin{center}
\includegraphics[trim=.7cm 0.8cm .7cm 0.7cm, clip=true,width=1\linewidth,keepaspectratio]{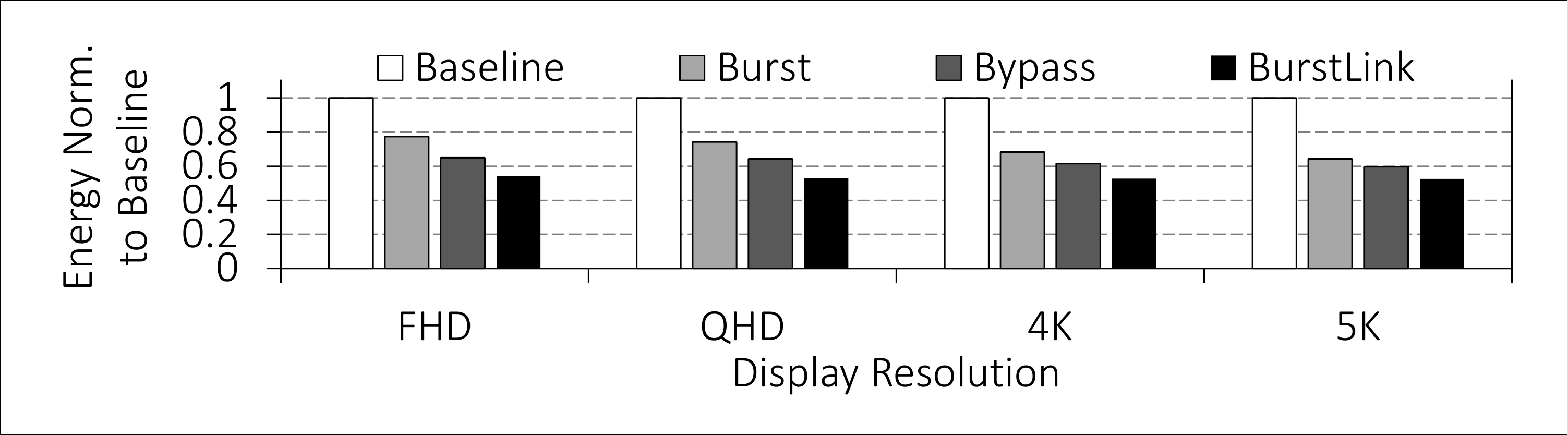}
\caption{Total system energy reduction of Frame Bursting, Frame \hc{Buffer}  Bypassing, and \tech for $60$~FPS HD videos. 
}\label{fig:results_60_60}
\end{center}
\end{figure}

We evaluate \tech~with $60$~FPS HD videos. \tech's energy reduction \jkt{increases as} video frame rate \jkt{increases \hj{from $30$~FPS (in Fig. \ref{fig:results_30_60})} to  $60$~FPS}. As shown in \fig{\ref{fig:results_60_60}}, \tech~reduces overall energy consumption by $46\%$ and $47\%$ for 
$FHD$ and $5K$ display \juang{resolutions}, respectively.

Based on these results, we make \hj{two} key observations. 
First, workloads with 60 FPS 
\juang{obtain} higher energy benefits \haj{compared to \jkth{those with} 30 FPS}. This \haj{result} is expected 
\hj{since, workloads with 30 FPS \hj{have reduced baseline bandwidth and DRAM energy consumption} as they utilize the optimization described in  Fig. \ref{fig:power_timeline}. Therefore, the relative DRAM energy saving of \tech~is lower than \hc{that for} \jkth{workloads with} 60 FPS}.
\hgrey{
Second, DRAM and eDP bandwidth consumption increase \jkt{as display resolution increases} and/or \jkt{as refresh rate increases}. As a result, \hsb{\tech's energy reduction also increases with increased} display resolution and/or refresh rate.
This \haj{observation} makes \mbox{\tech} even more \haj{critical} for future display technology, given the trends of increasing resolutions and refresh rates.}
\subsection{Benefits over Existing Techniques} \label{sec:exsiting_tech}
We 
study the benefits of using \tech with state-of-the-art techniques such as 1) \hj{video} compression schemes~\cite{disp_comp_nvidia,disp_comp_amd,disp_comp_arm, TE_arm, dyke2013method, shim2004compressed, rasmusson2011frame, intel_display}, 
2) a recent technique by Zhang \textit{et al.}~\cite{rts} that 
\juang{combines} race-to-sleep \hj{\cite{das2016slowdown,efraim2012energy}}, content caching \hj{\cite{mukherjee2013latest,sullivan2012overview}}, and display caching \hj{\cite{yang2002performance}} techniques,
\haj{and 3) Virtualizing IP \hj{chains} (VIP) \mbox{\cite{nachiappan2015vip}} technique.}

\noindent \textbf{Effect of Frame Buffer Compression (FBC).}
The evaluated Intel Skylake system supports FBC \cite{intel_display}.
FBC compresses a decoded video frame before storing it in the frame buffer region in host DRAM to reduce the data movement overhead and DRAM bandwidth consumption. 
Modern FBC techniques can compress video frames \juang{with} up to $50\%$ \juang{compression} rate.
\haj{FBC} leads to high computational overheads \cite{disp_comp_arm, shim2004compressed} and significant storage overhead. 
\ha{For example, Intel's FBC uses a region of memory reserved for graphics data to store the compressed frame buffer \cite{intel_display}.}

Moreover, FBC \hj{techniques are error-prone} due to \juang{the} design complexity of compression blocks and algorithms \cite{enable_fbc}. Therefore, several systems allow the display driver to enable or disable this feature~\cite{intel_display,disp_comp_amd,disp_comp_arm}.

\begin{figure}[h]
\begin{center}
\includegraphics[trim=.8cm .8cm .8cm .8cm, clip=true,width=1\linewidth,keepaspectratio]{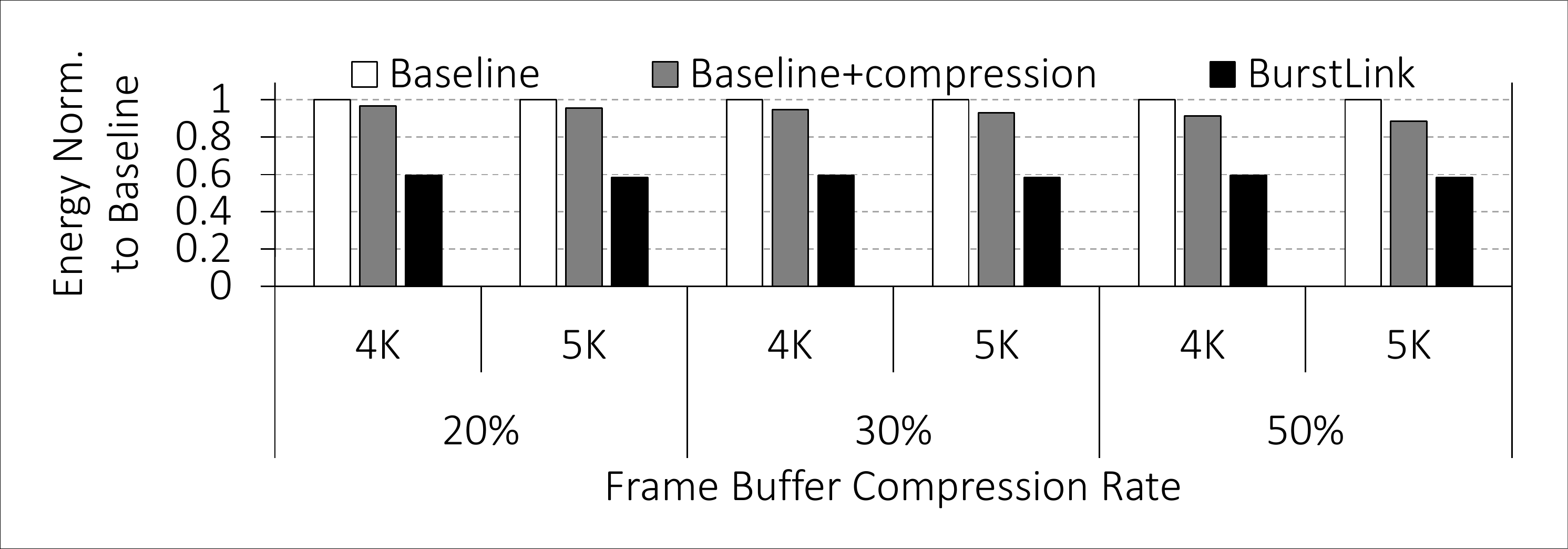}
\caption{Energy reduction of \tech~ 
over \jkt{a baseline} with frame buffer compression (refresh rate: $60$Hz).
}\label{fig:compression}
\end{center}
\end{figure}

\fig{\ref{fig:compression}} compares the energy consumption of \tech to the baseline system with FBC enabled. As the effect of FBC is more prominent in \hoo{higher resolution videos}, we only show the energy consumption for $4K$ and $5K$ resolution displays with a $60$Hz refresh rate.  As shown in \fig{\ref{fig:compression}}, \jkt{FBC} with 
$50\%$ compression 
\juang{rate} can reduce overall \haj{system} energy consumption by $9\%$ for a $4K$ resolution display. 

On the other hand, \tech eliminates the DRAM storage overhead by bypassing DRAM and \juang{directly} transferring the decoded frame 
to PSR \jkth{in burst}. Besides DRAM bandwidth reduction, \tech~ \juang{enables} the system to spend more time in deep idle power \jkth{states where several components are power-gated}  (e.g., DC and eDP interface in the processor and the display panel). As shown in \fig{\ref{fig:compression}}, \tech 
reduces overall energy \jkt{consumption} by $40.6\%$ for $4K$ displays.

\begin{sloppypar}
\noindent \textbf{Race-to-Sleep, Content Caching, and Display Caching.}  \jkt{Zhang} \textit{et al.} \cite{rts} propose a mechanism to save video processing energy consumption by applying three techniques: 1) batching several \hj{encoded} frames and boosting the VD frequency to decode multiple frames \hj{at once to increase system idle periods, allowing the system to enter deep idle power states}, 
2) caching decoded macroblocks in VD to reduce the data written to DRAM, 
and 3) 
\juang{using} a cache in DC to reduce the amount of data that \haj{the system} reads from DRAM. 
\hlp{This mechanism is an extension 
\juang{of} a \jkt{prior} mechanism, \emph{short-circuiting} \mbox{\cite{yedlapalli2014short}}, 
\juang{which} proposes techniques 2) and 3)}.

\jkt{Zhang \textit{et al.} \cite{rts}'s} mechanism has significant design complexity because 
\juang{it requires} \haj{implementing} two special \jkt{caching schemes, frame batching and concurrent decoding of several frames}. The average DRAM bandwidth 
\juang{reduction} of this mechanism is $34\%$ \cite{rts} (for the three techniques combined), \haj{reducing} the overall \haj{system} energy consumption \haj{of} \hoo{$4K$ video streaming} by $6\%$. On the other hand, \tech eliminates the DRAM overhead \jkt{completely} by 
\juang{transferring} the decoded frame directly to PSR and reduces the overall \haj{system} energy consumption by $40.6\%$ for $4K$ displays. 
\hlp{We conclude that {\tech} has higher energy reduction than \haj{\mbox{\cite{rts}}'s} three techniques combined (including short-circuiting \mbox{\cite{yedlapalli2014short}})}.
\end{sloppypar}

\noindent \hly{\textbf{Virtualizing IP \hj{Chains (VIP)}.}}
VIP \cite{nachiappan2015vip} proposes two main mechanisms. 
First, it reduces the CPU core orchestration overhead of invoking several IPs when running frame-based applications (e.g., video playback) by 
1)  \hj{chaining of IPs (i.e., the output of an IP in the chain is an input to \jkth{the} next IP in the chain) to avoid data transfer through the DRAM} 
and 2) initiating the handling of multiple frames at the same time.
{\tech} \htann{also reduces the CPU core orchestration overhead} (from approximately $10\%$ to less than $5\%$ of the frame time) \htann{by offloading part of the orchestration task to the PMU firmware} (as we discuss in Sec \mbox{\ref{sec:frame_bypass}}). \htann{One of the main} \juang{advantages} of {\tech} over VIP is that unlike VIP, which requires substantial changes to the software stack, \tech is transparent to the application and requires only a few changes to the drivers (e.g., video decoder driver).

\hly{Second, VIP reduces the traffic to DRAM by \jkt{enabling} direct communication between IPs (IP-to-IP chaining) instead of using DRAM. Compared to {\tech}, 
\htann{VIP is limited due to two main reasons}. 
\htann{First,} the traffic of frame-based applications is not always a chain. 
One IP waits for traffic from two (or more) 
\juang{other} IPs/sources to complete its task \haj{in many cases}. 
\juang{For example,} in \jkt{windowed} video playback (discussed in Section~\ref{sec:frame_bypass}), the DC needs the data of \emph{multiple} frame buffers to generate the final image. 
\htann{Second,} VIP does not solve the key bottleneck in the display data flow, 
\juang{where} the decoding and displaying \juang{processes occupy} the entire frame window (e.g., within $16ms$) because the display panel consumes the frame data within the entire frame time. 
\juang{This \haj{bottleneck} forces} the VD, DC, and \jkt{the} eDP interface (at both SoC and display panel) \juang{to remain} active across the entire frame.} 
\juang{\tech avoids this bottleneck with the Frame Bursting technique.}

\htann{In addition to the above advantages of {\tech}} over VIP, our results (when modeling VIP using our power model) show that\htann{ {\tech} has higher energy reduction over VIP} for 4K video streaming workloads due to the ability of {\tech} to turn off the VD, DC, and \jkt{the} eDP interface during the majority of the frame window.  

\subsection{Benefits on Other Mobile Workloads}
\label{sec:otherworkloads}
\begin{figure}[b]
\begin{center}
\includegraphics[trim=.8cm 1.2cm .8cm .8cm, clip=true,width=1\linewidth,keepaspectratio]{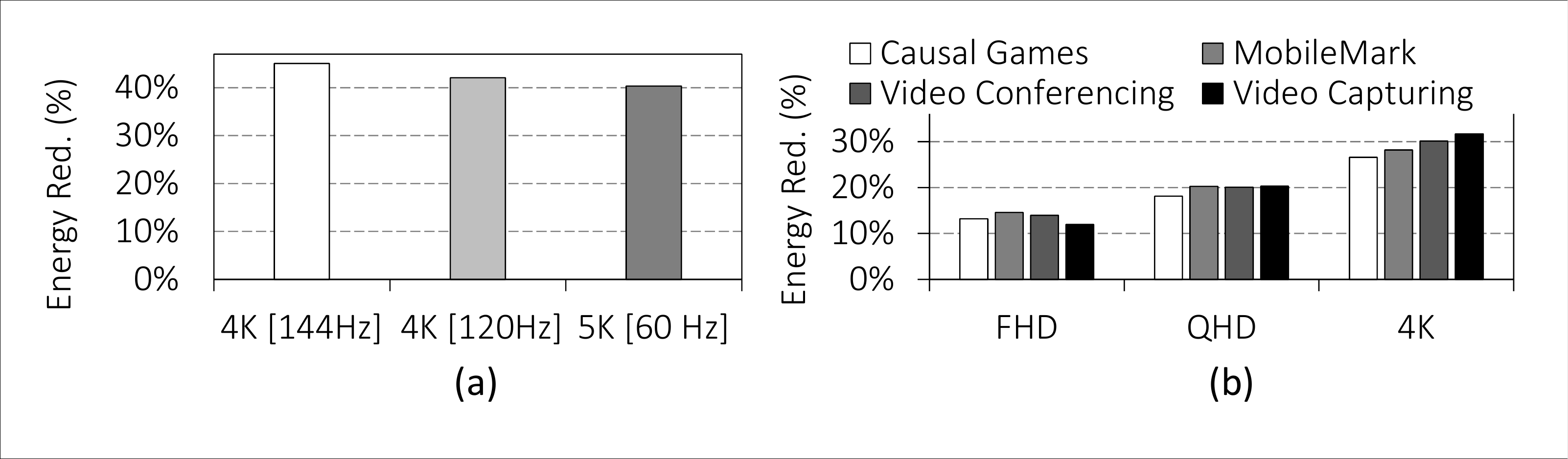}
\vspace{-.5em}
\caption{\hy{(a) \hgg{Energy reduction of \hc{Frame Buffer} Bypassing  for three high video and display resolutions.}} (b) Energy reduction of \haj{\hc{Frame} Bursting}  for four mobile \hd{computing} workloads.
}\label{fig:other_wls}
\end{center}
\end{figure}

Besides video streaming, other popular mobile \hd{computing} workloads can 
benefit from {\tech}'s \hy{two techniques \jkt{(i.e., Frame Buffer Bypassing and Frame Bursting)}. 
We \jkth{demonstrate} this with the following two examples.

First, \hgg{high resolution local (i.e., not streaming) video-playback (e.g., 4K with 120/144Hz or 5K with 60Hz)} \jkt{can exhibit \jkth{lower} DRAM energy consumption when using the DRAM \hj{Frame Buffer Bypassing} technique.}
\mbox{\fig{\ref{fig:other_wls}(a)}} \jkt{shows the large (more than 40\%)} energy reduction of \juang{our} Frame Buffer Bypassing technique when playing three local video files with different resolutions and/or frame rates.}

Second, \ha{we demonstrate \tech's benefits under four} 
mobile workloads\ha{:} video capturing, video conferencing, casual gaming~\cite{19_MSFT}, and MobileMark~\hc{\cite{mobilemark}.}\footnote{MobileMark is an application-based benchmark that reflects usage patterns of business mobile users in the areas of office productivity, creativity and web browsing.} 
In a mobile device (e.g., \juang{a} tablet), these workloads render images on display using one plane (usually the graphics plane). 
In these applications, the DC transfers the frame data from \juang{the} DRAM frame buffer into the display panel. \hlg{When \juang{the} DC detects that a single plane \juang{exists,}\footnote{\hlg{\juang{The} number of active planes is known to the DC. \juang{The} DC uses this information in multi-plane cases, \juang{where} 
it reads from multiple frame buffers to overlay the planes \mbox{\cite{intel_display}}.}} it activates the \juang{Frame} Bursting technique of {\tech}, which transfers the frame buffer from DRAM into the \pr{DRFB} in a burst}. 
Frame Bursting reduces energy consumption by 1) increasing \juang{the} idle time of these workloads, as the workloads spend less time transferring the frame buffer and more time in low-power states, and 2) power-gating the DC and \jkt{the} eDP interface at both the processor and the display panel.  \fig{\ref{fig:other_wls}(b)} shows the \jkt{large} energy reduction of Frame Bursting for four workloads 
\juang{on a high-end tablet.}

\section{Related Work}
\label{sec:related}


To our knowledge, this is the first work to leverage \hj{Panel-Self-Refresh} memory to improve the energy efficiency of video processing in modern mobile \hd{computing} systems.
\tech significantly reduces costly data movement between DRAM and display subsystem components and enables frame transfer with the maximum \hj{display} bandwidth, which allows the system to reside in lower power states.
We briefly discuss related prior work that we classify into five categories. 

\noindent \textbf{Buffer Compression.}
Many prior works~\cite{huang1998method, yabe1998compression, shim2004compressed, rasmusson2011frame, dyke2013method, disp_comp_arm, TE_arm} propose various frame buffer compression techniques, such as run-length encoding (RLE)\hj{~\cite{shim2004compressed}} and differential pulse code modulation (DPCM)\hj{~\cite{shim2004compressed}}, which increase the effective DRAM bandwidth (by \rbc{reducing} the amount of transferred frame data) and thus improve display processing performance.
We already \jkth{show} that \tech outperforms frame buffer \jkth{compression (Section \ref{sec:exsiting_tech})}.

\noindent \textbf{Batching of Decoded Frames.}
Zhang \textit{et al.}~\cite{rts} \js{propose} 
three \rbc{optimizations} to increase the idle time (race to halt) and reduce bandwidth from VD to DRAM and DRAM to DC in video processing. Their proposals lie across batch processing, content and display caching and frequency/voltage optimization. 
\jkth{We already demonstrate the clear advantages of \tech~ over these optimizations \hc{(}Section \ref{sec:exsiting_tech}\hc{)}.}

Other prior works~\cite{chen2004analysis, jin2007efficient, tu2010batch} propose different batching and pipelining techniques at a macro-block level. As {\tech} offers to achieve energy efficiency through leveraging PSR memory directly, similar techniques for batch processing can be \hj{easily combined \jkth{with} our proposal} to increase energy improvement.



\noindent \textbf{Optimizing Frequency/Voltage.}
Several prior works \jkth{improve energy efficiency by reducing the voltage and frequency of \ha{1)} the logic used for decoding/encoding}~\cite{nachiappan2015domain, raffin2016low,rao2017application} \ha{or 2) multiple system \hc{domains} \cite{haj2020sysscale}}.
{\tech}, on the other hand, does not depend on frequency/voltage scaling for its energy improvement.

\noindent \textbf{Other PSR-based proposals.}
Prior works~\cite{psr_1,psr_2, edp_psr,psr_3,psr_4,psr_5,psr_6} use the PSR frame buffer mainly for \emph{static images} rather than bypassing DRAM. \tech uses the PSR for static and dynamic images to skip costly data movement and significantly reduce DRAM energy consumption. 

\section{Conclusion
}
\label{sec:conclusion}
We introduce \tech, a \hj{new} energy-efficient \jh{planar and VR} video display scheme that \hj{uses} display panel local memory \hj{to eliminate buffering frames in main memory}. \tech \emph{directly transfers} a full decoded frame from the video-decoder (or GPU) to the \hj{display panel} \emph{in a burst}, exploiting  \rbc{the display interface's maximum bandwidth}. \hj{Doing so} (1) reduces the energy consumption of the host DRAM by \hj{eliminating} \jkth{data movement} to\hj{/from} the DRAM frame buffer, and (2) increases the system's idle-power state residency \hj{by reducing the usage of the processor and the display subsystem since they are active only during the burst period}. 
\tech reduces system energy consumption for 4K planar and VR video streaming by 41\% and 33\%, respectively, and its benefits increase \jkth{as display resolution and/or refresh rate increases}.
As video consumption in mobile devices continues to increase sharply, along with an increase in display resolution to meet user satisfaction,
\tech~\hj{is poised to} provide high energy efficiency in current and next-generation \haj{mobile} processors.
\hj{\tech's techniques can also be used in more general frame-based applications such as  
video capture (recording), audio streaming, video chat, social networking, and interactive games.

A general takeaway from \tech is that using main memory (DRAM) as a \emph{communication hub} between system components \jkth{is} energy-inefficient. Instead, \tech uses small \emph{remote} memory near the data consumer (e.g., a display panel) to significantly reduce the number of costly main memory accesses in frame-based applications.}


\begin{acks}
We thank the anonymous reviewers of ASPLOS 2021, ISCA 2021, and MICRO 2021 for feedback. We thank the
SAFARI Research Group members for valuable feedback and the stimulating intellectual environment they provide. We acknowledge the generous gifts provided by our industrial partners: Google, Huawei, Intel, Microsoft, and VMware. \ha{An earlier version of this paper was placed on arxiv.org on 11 April 2021 \cite{haj2021burstlink}}.
\end{acks}

\bibliographystyle{ACM-Reference-Format}
\bibliography{refs}


\begin{thebibliography}{117}


\ifx \showCODEN    \undefined \def \showCODEN     #1{\unskip}     \fi
\ifx \showDOI      \undefined \def \showDOI       #1{#1}\fi
\ifx \showISBNx    \undefined \def \showISBNx     #1{\unskip}     \fi
\ifx \showISBNxiii \undefined \def \showISBNxiii  #1{\unskip}     \fi
\ifx \showISSN     \undefined \def \showISSN      #1{\unskip}     \fi
\ifx \showLCCN     \undefined \def \showLCCN      #1{\unskip}     \fi
\ifx \shownote     \undefined \def \shownote      #1{#1}          \fi
\ifx \showarticletitle \undefined \def \showarticletitle #1{#1}   \fi
\ifx \showURL      \undefined \def \showURL       {\relax}        \fi
\providecommand\bibfield[2]{#2}
\providecommand\bibinfo[2]{#2}
\providecommand\natexlab[1]{#1}
\providecommand\showeprint[2][]{arXiv:#2}

\bibitem[\protect\citeauthoryear{??}{Bur}{2021}]%
        {Burstlink-github}
\bibinfo{title}{{BurstLink Source Code}}.
\newblock
\newblock
\newblock
\shownote{https://github.com/CMU-SAFARI/BurstLink.}
 \bibinfo{year}{2021}\natexlab{}.


\bibitem[\protect\citeauthoryear{{9to5mac}}{{9to5mac}}{2020}]%
        {bench4}
\bibfield{author}{\bibinfo{person}{{9to5mac}}.}
\newblock \bibinfo{title}{{{iPhone 11 and iPhone 11 Pro Improves on iPhone XR
  Stellar Battery Life}}}.
\newblock
\newblock
\newblock
\shownote{https://bit.ly/34Rlx5A.}
 \bibinfo{year}{2020}\natexlab{}.


\bibitem[\protect\citeauthoryear{{AMD}}{{AMD}}{2016}]%
        {disp_comp_amd}
\bibfield{author}{\bibinfo{person}{{AMD}}.}
\newblock \bibinfo{title}{{{RADEON: Dissecting the Polaris Architecture}}}.
\newblock
\newblock
\newblock
\shownote{{AMD Whitepaper}.}
 \bibinfo{year}{2016}\natexlab{}.


\bibitem[\protect\citeauthoryear{{AMD}}{{AMD}}{2020}]%
        {amd_specs}
\bibfield{author}{\bibinfo{person}{{AMD}}.}
\newblock \bibinfo{title}{{AMD Processor Specifications}}.
\newblock
\newblock
\newblock
\shownote{https://bit.ly/3zQV46u.}
 \bibinfo{year}{2020}\natexlab{}.


\bibitem[\protect\citeauthoryear{Ameigeiras, Ramos-Munoz, Navarro-Ortiz, and
  Lopez-Soler}{Ameigeiras et~al\mbox{.}}{2012}]%
        {ameigeiras2012analysis}
\bibfield{author}{\bibinfo{person}{Pablo Ameigeiras}, \bibinfo{person}{Juan~J
  Ramos-Munoz}, \bibinfo{person}{Jorge Navarro-Ortiz}, {and}
  \bibinfo{person}{Juan~M Lopez-Soler}.}
\newblock \showarticletitle{{Analysis and Modeling of YouTube Traffic}}.
\newblock \bibinfo{journal}{\emph{{ETT}}}.
 \bibinfo{year}{2012}\natexlab{}.
\newblock


\bibitem[\protect\citeauthoryear{Anandtech}{Anandtech}{2020}]%
        {4k_laptop}
\bibfield{author}{\bibinfo{person}{Anandtech}.}
\newblock \bibinfo{title}{{Lenovo Launches Legion Y740S Ultra-Thin 4K 15.6-Inch
  Gaming Notebook.}}
\newblock
\newblock
\newblock
\shownote{https://bit.ly/39kxz7J.}
 \bibinfo{year}{2020}\natexlab{}.


\bibitem[\protect\citeauthoryear{{Anandtech}}{{Anandtech}}{2020}]%
        {19_MSFT}
\bibfield{author}{\bibinfo{person}{{Anandtech}}.}
\newblock \bibinfo{title}{{The Microsoft Surface Pro (2017) Review:
  Evaluation}}.
\newblock
\newblock
\newblock
\shownote{https://bit.ly/2WCB3yZ.}
 \bibinfo{year}{2020}\natexlab{}.


\bibitem[\protect\citeauthoryear{{ARM}}{{ARM}}{2016a}]%
        {disp_comp_arm}
\bibfield{author}{\bibinfo{person}{{ARM}}.}
\newblock \bibinfo{title}{{{ARM Frame Buffer Compression}}}.
\newblock
\newblock
\newblock
\shownote{https://bit.ly/2WsvG5X.}
 \bibinfo{year}{2016}\natexlab{a}.


\bibitem[\protect\citeauthoryear{{ARM}}{{ARM}}{2016b}]%
        {TE_arm}
\bibfield{author}{\bibinfo{person}{{ARM}}.}
\newblock \bibinfo{title}{{{Transaction Elimination}}}.
\newblock
\newblock
\newblock
\shownote{https://bit.ly/2qI3rmy.}
 \bibinfo{year}{2016}\natexlab{b}.


\bibitem[\protect\citeauthoryear{{ARM}}{{ARM}}{2020}]%
        {arm_amba}
\bibfield{author}{\bibinfo{person}{{ARM}}.}
\newblock \bibinfo{title}{{{AMBA: The Standard for On-Chip Communication}}}.
\newblock \bibinfo{howpublished}{\url{https://bit.ly/2UEzQ8l}}.
\newblock
 \bibinfo{year}{2020}\natexlab{}.


\bibitem[\protect\citeauthoryear{{ARM}}{{ARM}}{2021}]%
        {arm_dc}
\bibfield{author}{\bibinfo{person}{{ARM}}.}
\newblock \bibinfo{title}{{{ARM Mali Display Processors.}}}
\newblock
\newblock
\newblock
\shownote{https://bit.ly/33b01Yk.}
 \bibinfo{year}{2021}\natexlab{}.


\bibitem[\protect\citeauthoryear{{BAPCo}}{{BAPCo}}{2019}]%
        {mobilemark}
\bibfield{author}{\bibinfo{person}{{BAPCo}}.}
\newblock \bibinfo{title}{{{MobileMark 2014}}}.
\newblock
\newblock
\newblock
\shownote{https://bapco.com/products/mobilemark-2018.}
 \bibinfo{year}{2019}\natexlab{}.


\bibitem[\protect\citeauthoryear{Barkay, Engineer, Buchholz, Codorean,
  DeVetter, and Stevens}{Barkay et~al\mbox{.}}{2013}]%
        {psr_4}
\bibfield{author}{\bibinfo{person}{Omri Barkay}, \bibinfo{person}{WLAN
  Engineer}, \bibinfo{person}{IT~Dave Buchholz}, \bibinfo{person}{IT~Dan
  Codorean}, \bibinfo{person}{IT~Doug DeVetter}, {and} \bibinfo{person}{Baz
  Stevens}.}
\newblock \showarticletitle{{Evaluating Intel{\textregistered} Pro Wireless
  Display for Enterprise Use}}.
\newblock \bibinfo{journal}{\emph{Intel White Paper}}.
 \bibinfo{year}{2013}\natexlab{}.
\newblock


\bibitem[\protect\citeauthoryear{Bertran, Gonzalez, Martorell, Navarro, and
  Ayguade}{Bertran et~al\mbox{.}}{2013}]%
        {bertran2013systematic}
\bibfield{author}{\bibinfo{person}{Ramon Bertran}, \bibinfo{person}{Marc
  Gonzalez}, \bibinfo{person}{Xavier Martorell}, \bibinfo{person}{Nacho
  Navarro}, {and} \bibinfo{person}{Eduard Ayguade}.}
\newblock \showarticletitle{{A Systematic Methodology to Generate Decomposable
  and Responsive Power Models for CMPs}}.
\newblock \bibinfo{journal}{\emph{TC}}.
 \bibinfo{year}{2013}\natexlab{}.
\newblock


\bibitem[\protect\citeauthoryear{Boroumand, Ghose, Kim, Ausavarungnirun, Shiu,
  Thakur, et~al\mbox{.}}{Boroumand et~al\mbox{.}}{2018}]%
        {boroumand2018google}
\bibfield{author}{\bibinfo{person}{Amirali Boroumand}, \bibinfo{person}{Saugata
  Ghose}, \bibinfo{person}{Youngsok Kim}, \bibinfo{person}{Rachata
  Ausavarungnirun}, \bibinfo{person}{Eric Shiu}, \bibinfo{person}{Rahul
  Thakur}, {et~al\mbox{.}}}
\newblock \showarticletitle{{Google Workloads for Consumer Devices: Mitigating
  Data Movement Bottlenecks}}. In \bibinfo{booktitle}{\emph{ASPLOS}}.
 \bibinfo{year}{2018}\natexlab{}.
\newblock


\bibitem[\protect\citeauthoryear{Bouvier, Cohen, Fry, Godey, and
  Mantor}{Bouvier et~al\mbox{.}}{2013}]%
        {psr_5}
\bibfield{author}{\bibinfo{person}{Dan Bouvier}, \bibinfo{person}{Brad Cohen},
  \bibinfo{person}{Walter Fry}, \bibinfo{person}{Sreekanth Godey}, {and}
  \bibinfo{person}{Michael Mantor}.}
\newblock \showarticletitle{{AMD Kabini APU SoC}}. In
  \bibinfo{booktitle}{\emph{Hot Chips}}.
 \bibinfo{year}{2013}\natexlab{}.
\newblock


\bibitem[\protect\citeauthoryear{Brooks, Bose, Srinivasan, Gschwind, Emma, and
  Rosenfield}{Brooks et~al\mbox{.}}{2003}]%
        {brooks2003new}
\bibfield{author}{\bibinfo{person}{David Brooks}, \bibinfo{person}{Pradip
  Bose}, \bibinfo{person}{Viji Srinivasan}, \bibinfo{person}{Michael~K
  Gschwind}, \bibinfo{person}{Philip~G Emma}, {and} \bibinfo{person}{Michael~G
  Rosenfield}.}
\newblock \showarticletitle{{New Methodology for Early-stage,
  Microarchitecture-level Power-Performance Analysis of Microprocessors}}.
\newblock \bibinfo{journal}{\emph{IBM J. Res. Dev}}.
 \bibinfo{year}{2003}\natexlab{}.
\newblock


\bibitem[\protect\citeauthoryear{Burgess and Green}{Burgess and Green}{2018}]%
        {youtube_2019}
\bibfield{author}{\bibinfo{person}{Jean Burgess} {and} \bibinfo{person}{Joshua
  Green}.}
\newblock \bibinfo{booktitle}{\emph{{YouTube: Online Video and Participatory
  Culture}}}.
\newblock \bibinfo{publisher}{John Wiley \& Sons}.
 \bibinfo{year}{2018}\natexlab{}.
\newblock


\bibitem[\protect\citeauthoryear{Chang, Yagl{\i}k{\c{c}}{\i}, Ghose, Agrawal,
  Chatterjee, Kashyap, et~al\mbox{.}}{Chang et~al\mbox{.}}{2017}]%
        {chang2018voltron}
\bibfield{author}{\bibinfo{person}{Kevin~K Chang},
  \bibinfo{person}{Abdullah~Giray Yagl{\i}k{\c{c}}{\i}},
  \bibinfo{person}{Saugata Ghose}, \bibinfo{person}{Aditya Agrawal},
  \bibinfo{person}{Niladrish Chatterjee}, \bibinfo{person}{Abhijith Kashyap},
  {et~al\mbox{.}}}
\newblock \showarticletitle{{Voltron: Understanding and Exploiting the
  Voltage-Latency-Reliability Trade-Offs in Modern DRAM Chips to Improve Energy
  Efficiency}}.
\newblock \bibinfo{journal}{\emph{SIGMETRICS}}.
 \bibinfo{year}{2017}\natexlab{}.
\newblock


\bibitem[\protect\citeauthoryear{Chen, Huang, and Chen}{Chen
  et~al\mbox{.}}{2004}]%
        {chen2004analysis}
\bibfield{author}{\bibinfo{person}{Tung-Chien Chen}, \bibinfo{person}{Yu-Wen
  Huang}, {and} \bibinfo{person}{Liang-Gee Chen}.}
\newblock \showarticletitle{{Analysis and Design of Macroblock Pipelining for
  H.264/AVC VLSI Architecture}}. In \bibinfo{booktitle}{\emph{ISCAS}}.
 \bibinfo{year}{2004}\natexlab{}.
\newblock


\bibitem[\protect\citeauthoryear{Choi, Shim, and Chang}{Choi
  et~al\mbox{.}}{2002}]%
        {choi2002low}
\bibfield{author}{\bibinfo{person}{Inseok Choi}, \bibinfo{person}{Hojun Shim},
  {and} \bibinfo{person}{Naehyuck Chang}.}
\newblock \showarticletitle{{Low-power Color TFT LCD Display for Hand-held
  Embedded Systems}}. In \bibinfo{booktitle}{\emph{ISLPED}}.
 \bibinfo{year}{2002}\natexlab{}.
\newblock


\bibitem[\protect\citeauthoryear{{Cisco}}{{Cisco}}{2019}]%
        {cisco_2019}
\bibfield{author}{\bibinfo{person}{{Cisco}}.}
\newblock \bibinfo{title}{{{Cisco Visual Networking Index: Global Mobile Data
  Traffic Forecast Update}}}.
\newblock \bibinfo{howpublished}{White Paper}.
\newblock
\newblock
\shownote{https://bit.ly/2Y2wqhb.}
 \bibinfo{year}{2019}\natexlab{}.


\bibitem[\protect\citeauthoryear{Conway and Hughes}{Conway and Hughes}{2007}]%
        {conway2007amd}
\bibfield{author}{\bibinfo{person}{Pat Conway} {and} \bibinfo{person}{Bill
  Hughes}.}
\newblock \showarticletitle{{The AMD Opteron northbridge architecture}}.
\newblock \bibinfo{journal}{\emph{IEEE Micro}}.
 \bibinfo{year}{2007}\natexlab{}.
\newblock


\bibitem[\protect\citeauthoryear{Corbillon, De~Simone, and Simon}{Corbillon
  et~al\mbox{.}}{2017}]%
        {corbillon2017360}
\bibfield{author}{\bibinfo{person}{Xavier Corbillon},
  \bibinfo{person}{Francesca De~Simone}, {and} \bibinfo{person}{Gwendal
  Simon}.}
\newblock \showarticletitle{{360-Degree Video Head Movement Dataset}}. In
  \bibinfo{booktitle}{\emph{MMSys}}.
 \bibinfo{year}{2017}\natexlab{}.
\newblock


\bibitem[\protect\citeauthoryear{Das, Merrett, and Al-Hashimi}{Das
  et~al\mbox{.}}{2016}]%
        {das2016slowdown}
\bibfield{author}{\bibinfo{person}{Anup Das}, \bibinfo{person}{Geoff~V
  Merrett}, {and} \bibinfo{person}{Bashir~M Al-Hashimi}.}
\newblock \showarticletitle{{The slowdown or race-to-idle question:
  Workload-aware energy optimization of SMT multicore platforms under process
  variation}}. In \bibinfo{booktitle}{\emph{DATE}}.
 \bibinfo{year}{2016}\natexlab{}.
\newblock


\bibitem[\protect\citeauthoryear{David, Fallin, Gorbatov, Hanebutte, and
  Mutlu}{David et~al\mbox{.}}{2011}]%
        {david2011memory}
\bibfield{author}{\bibinfo{person}{Howard David}, \bibinfo{person}{Chris
  Fallin}, \bibinfo{person}{Eugene Gorbatov}, \bibinfo{person}{Ulf~R
  Hanebutte}, {and} \bibinfo{person}{Onur Mutlu}.}
\newblock \showarticletitle{{Memory Power Management via Dynamic
  Voltage/Frequency Scaling}}. In \bibinfo{booktitle}{\emph{ICAC}}.
 \bibinfo{year}{2011}\natexlab{}.
\newblock


\bibitem[\protect\citeauthoryear{Doweck, Kao, Lu, Mandelblat, Rahatekar,
  Rappoport, et~al\mbox{.}}{Doweck et~al\mbox{.}}{2017}]%
        {21_doweck2017inside}
\bibfield{author}{\bibinfo{person}{Jack Doweck}, \bibinfo{person}{Wen-Fu Kao},
  \bibinfo{person}{Allen Kuan-yu Lu}, \bibinfo{person}{Julius Mandelblat},
  \bibinfo{person}{Anirudha Rahatekar}, \bibinfo{person}{Lihu Rappoport},
  {et~al\mbox{.}}}
\newblock \showarticletitle{{Inside 6th-Generation Intel Core: New
  Microarchitecture Code-Named Skylake}}.
\newblock \bibinfo{journal}{\emph{MICRO}}.
 \bibinfo{year}{2017}\natexlab{}.
\newblock


\bibitem[\protect\citeauthoryear{Duanmu, Kurdoglu, Liu, and Wang}{Duanmu
  et~al\mbox{.}}{2017}]%
        {duanmu2017view}
\bibfield{author}{\bibinfo{person}{Fanyi Duanmu}, \bibinfo{person}{Eymen
  Kurdoglu}, \bibinfo{person}{Yong Liu}, {and} \bibinfo{person}{Yao Wang}.}
\newblock \showarticletitle{{View Direction and Bandwidth Adaptive 360 Degree
  Video Streaming Using a Two-tier System}}. In
  \bibinfo{booktitle}{\emph{ISCAS}}.
 \bibinfo{year}{2017}\natexlab{}.
\newblock


\bibitem[\protect\citeauthoryear{Dyke}{Dyke}{2013}]%
        {dyke2013method}
\bibfield{author}{\bibinfo{person}{Kenneth~C Dyke}.}
\newblock \bibinfo{title}{{Method for Reducing Frame Buffer Memory Accesses}}.
\newblock
\newblock
\newblock
\shownote{{US Patent 8,358,314}.}
 \bibinfo{year}{2013}\natexlab{}.


\bibitem[\protect\citeauthoryear{Efraim, Ginosar, Weiser, and Mendelson}{Efraim
  et~al\mbox{.}}{2012}]%
        {efraim2012energy}
\bibfield{author}{\bibinfo{person}{Rotem Efraim}, \bibinfo{person}{Ran
  Ginosar}, \bibinfo{person}{C Weiser}, {and} \bibinfo{person}{Avi Mendelson}.}
\newblock \showarticletitle{{Energy Aware Race to Halt: A Down to EARtH
  Approach for Platform energy management}}.
\newblock \bibinfo{journal}{\emph{CAL}}.
 \bibinfo{year}{2012}\natexlab{}.
\newblock


\bibitem[\protect\citeauthoryear{Fayneh, Yuffe, Knoll, Zelikson, Abozaed,
  Talker, et~al\mbox{.}}{Fayneh et~al\mbox{.}}{2016}]%
        {fayneh20164}
\bibfield{author}{\bibinfo{person}{Eyal Fayneh}, \bibinfo{person}{Marcelo
  Yuffe}, \bibinfo{person}{Ernest Knoll}, \bibinfo{person}{Michael Zelikson},
  \bibinfo{person}{Muhammad Abozaed}, \bibinfo{person}{Yair Talker},
  {et~al\mbox{.}}}
\newblock \showarticletitle{{4.1 14nm 6th-Generation Core Processor SoC with
  Low Power Consumption and Improved Performance}}. In
  \bibinfo{booktitle}{\emph{ISSCC}}. IEEE.
 \bibinfo{year}{2016}\natexlab{}.
\newblock


\bibitem[\protect\citeauthoryear{FlatpanelsHD}{FlatpanelsHD}{2020}]%
        {res_gap_1}
\bibfield{author}{\bibinfo{person}{FlatpanelsHD}.}
\newblock \bibinfo{title}{The Resolution Gap}.
\newblock
\newblock
\newblock
\shownote{https://bit.ly/3djTHm1.}
 \bibinfo{year}{2020}\natexlab{}.


\bibitem[\protect\citeauthoryear{Gaudette, Wu, and Vrudhula}{Gaudette
  et~al\mbox{.}}{2016}]%
        {gaudette2016improving}
\bibfield{author}{\bibinfo{person}{Benjamin Gaudette},
  \bibinfo{person}{Carole-Jean Wu}, {and} \bibinfo{person}{Sarma Vrudhula}.}
\newblock \showarticletitle{{Improving Smartphone User Experience by Balancing
  Performance and Energy with Probabilistic QoS Guarantee}}. In
  \bibinfo{booktitle}{\emph{HPCA}}.
 \bibinfo{year}{2016}\natexlab{}.
\newblock


\bibitem[\protect\citeauthoryear{Ghose, Yaglik{\c{c}}i, Gupta, Lee, Kudrolli,
  Liu, et~al\mbox{.}}{Ghose et~al\mbox{.}}{2018}]%
        {ghose2018your}
\bibfield{author}{\bibinfo{person}{Saugata Ghose},
  \bibinfo{person}{Abdullah~Giray Yaglik{\c{c}}i}, \bibinfo{person}{Raghav
  Gupta}, \bibinfo{person}{Donghyuk Lee}, \bibinfo{person}{Kais Kudrolli},
  \bibinfo{person}{William~X Liu}, {et~al\mbox{.}}}
\newblock \showarticletitle{{What Your DRAM Power Models are not Telling You:
  Lessons from a Detailed Experimental Study}}.
\newblock \bibinfo{journal}{\emph{SIGMETRICS}}.
 \bibinfo{year}{2018}\natexlab{}.
\newblock


\bibitem[\protect\citeauthoryear{Google}{Google}{2020}]%
        {grafika}
\bibfield{author}{\bibinfo{person}{Google}.}
\newblock \bibinfo{title}{{Grafika Video Application}}.
\newblock
\newblock
\newblock
\shownote{https://github.com/google/grafika.}
 \bibinfo{year}{2020}\natexlab{}.


\bibitem[\protect\citeauthoryear{Gough, Steiner, and Saunders}{Gough
  et~al\mbox{.}}{2015}]%
        {gough2015cpu}
\bibfield{author}{\bibinfo{person}{Corey Gough}, \bibinfo{person}{Ian Steiner},
  {and} \bibinfo{person}{Winston Saunders}.}
\newblock \showarticletitle{CPU Power Management}.
\newblock In \bibinfo{booktitle}{\emph{Energy Efficient Servers: Blueprints for
  Data Center Optimization}}.
 \bibinfo{year}{2015}\natexlab{}.
\newblock


\bibitem[\protect\citeauthoryear{Graf, Timmerer, and Mueller}{Graf
  et~al\mbox{.}}{2017}]%
        {graf2017towards}
\bibfield{author}{\bibinfo{person}{Mario Graf}, \bibinfo{person}{Christian
  Timmerer}, {and} \bibinfo{person}{Christopher Mueller}.}
\newblock \showarticletitle{{Towards Bandwidth Efficient Adaptive Streaming of
  Omnidirectional Video over http: Design, Implementation, and Evaluation}}. In
  \bibinfo{booktitle}{\emph{MMSys}}.
 \bibinfo{year}{2017}\natexlab{}.
\newblock


\bibitem[\protect\citeauthoryear{Gunther}{Gunther}{2001}]%
        {gunther2001managing}
\bibfield{author}{\bibinfo{person}{Stephen~H Gunther}.}
\newblock \showarticletitle{{Managing the Impact of Increasing Microprocessor
  Power Consumption}}.
\newblock \bibinfo{journal}{\emph{Intel Technology Journal}}.
 \bibinfo{year}{2001}\natexlab{}.
\newblock


\bibitem[\protect\citeauthoryear{Haj-Yahya, Alser, Kim,
  Ya{\u{g}}l{\i}k{\c{c}}{\i}, Vijaykumar, Rotem, et~al\mbox{.}}{Haj-Yahya
  et~al\mbox{.}}{2020b}]%
        {haj2020sysscale}
\bibfield{author}{\bibinfo{person}{Jawad Haj-Yahya}, \bibinfo{person}{Mohammed
  Alser}, \bibinfo{person}{Jeremie Kim}, \bibinfo{person}{A~Giray
  Ya{\u{g}}l{\i}k{\c{c}}{\i}}, \bibinfo{person}{Nandita Vijaykumar},
  \bibinfo{person}{Efraim Rotem}, {et~al\mbox{.}}}
\newblock \showarticletitle{{SysScale: Exploiting Multi-domain Dynamic Voltage
  and Frequency Scaling for Energy Efficient Mobile Processors}}. ISCA.
 \bibinfo{year}{2020}\natexlab{b}.
\newblock


\bibitem[\protect\citeauthoryear{Haj-Yahya, Alser, Kim, Orosa, Rotem,
  Mendelson, et~al\mbox{.}}{Haj-Yahya et~al\mbox{.}}{2020a}]%
        {haj2020flexwatts}
\bibfield{author}{\bibinfo{person}{Jawad Haj-Yahya}, \bibinfo{person}{Mohammed
  Alser}, \bibinfo{person}{Jeremie~S Kim}, \bibinfo{person}{Lois Orosa},
  \bibinfo{person}{Efraim Rotem}, \bibinfo{person}{Avi Mendelson},
  {et~al\mbox{.}}}
\newblock \showarticletitle{{FlexWatts: A Power-and Workload-Aware Hybrid Power
  Delivery Network for Energy-Efficient Microprocessors}}. In
  \bibinfo{booktitle}{\emph{MICRO}}.
 \bibinfo{year}{2020}\natexlab{a}.
\newblock


\bibitem[\protect\citeauthoryear{Haj-Yahya, Kim, Yaglikci, Puddu, Orosa, Luna,
  et~al\mbox{.}}{Haj-Yahya et~al\mbox{.}}{2021a}]%
        {haj2021ichannels}
\bibfield{author}{\bibinfo{person}{Jawad Haj-Yahya}, \bibinfo{person}{Jeremie~S
  Kim}, \bibinfo{person}{A~Giray Yaglikci}, \bibinfo{person}{Ivan Puddu},
  \bibinfo{person}{Lois Orosa}, \bibinfo{person}{Juan~G{\'o}mez Luna},
  {et~al\mbox{.}}}
\newblock \showarticletitle{{IChannels: Exploiting Current Management
  Mechanisms to Create Covert Channels in Modern Processors}}.
\newblock \bibinfo{journal}{\emph{ISCA}}.
 \bibinfo{year}{2021}\natexlab{a}.
\newblock


\bibitem[\protect\citeauthoryear{Haj-Yahya, Mendelson, Asher, and
  Chattopadhyay}{Haj-Yahya et~al\mbox{.}}{2018}]%
        {haj2018power}
\bibfield{author}{\bibinfo{person}{Jawad Haj-Yahya}, \bibinfo{person}{Avi
  Mendelson}, \bibinfo{person}{Yosi~Ben Asher}, {and} \bibinfo{person}{Anupam
  Chattopadhyay}.}
\newblock \showarticletitle{{Power Management of Modern Processors}}.
\newblock In \bibinfo{booktitle}{\emph{EEHPC}}.
 \bibinfo{year}{2018}\natexlab{}.
\newblock


\bibitem[\protect\citeauthoryear{Haj-Yahya, Park, Bera, Luna, Rotem, Shahroodi,
  et~al\mbox{.}}{Haj-Yahya et~al\mbox{.}}{2021b}]%
        {haj2021burstlink}
\bibfield{author}{\bibinfo{person}{Jawad Haj-Yahya}, \bibinfo{person}{Jisung
  Park}, \bibinfo{person}{Rahul Bera}, \bibinfo{person}{Juan~G{\'o}mez Luna},
  \bibinfo{person}{Efraim Rotem}, \bibinfo{person}{Taha Shahroodi},
  {et~al\mbox{.}}}
\newblock \showarticletitle{{BurstLink: Techniques for Energy-Efficient
  Conventional and Virtual Reality Video Display}}.
\newblock \bibinfo{journal}{\emph{arXiv preprint arXiv:2104.05119}}.
 \bibinfo{year}{2021}\natexlab{b}.
\newblock


\bibitem[\protect\citeauthoryear{Haj-Yahya, Rotem, Mendelson, and
  Chattopadhyay}{Haj-Yahya et~al\mbox{.}}{2019}]%
        {haj2019comprehensive}
\bibfield{author}{\bibinfo{person}{Jawad Haj-Yahya}, \bibinfo{person}{Efraim
  Rotem}, \bibinfo{person}{Avi Mendelson}, {and} \bibinfo{person}{Anupam
  Chattopadhyay}.}
\newblock \showarticletitle{{A Comprehensive Evaluation of Power Delivery
  Schemes for Modern Microprocessors}}. In \bibinfo{booktitle}{\emph{ISQED}}.
 \bibinfo{year}{2019}\natexlab{}.
\newblock


\bibitem[\protect\citeauthoryear{Haj-Yahya, Sazeides, Alser, Rotem, and
  Mutlu}{Haj-Yahya et~al\mbox{.}}{2020c}]%
        {haj2020techniques}
\bibfield{author}{\bibinfo{person}{Jawad Haj-Yahya}, \bibinfo{person}{Yanos
  Sazeides}, \bibinfo{person}{Mohammed Alser}, \bibinfo{person}{Efraim Rotem},
  {and} \bibinfo{person}{Onur Mutlu}.}
\newblock \showarticletitle{Techniques for Reducing the Connected-Standby
  Energy Consumption of Mobile Devices}. In \bibinfo{booktitle}{\emph{HPCA}}.
 \bibinfo{year}{2020}\natexlab{c}.
\newblock


\bibitem[\protect\citeauthoryear{Hammarlund, Martinez, Bajwa, Hill, Hallnor,
  Jiang, et~al\mbox{.}}{Hammarlund et~al\mbox{.}}{2013}]%
        {psr_3}
\bibfield{author}{\bibinfo{person}{Per Hammarlund}, \bibinfo{person}{Alberto~J
  Martinez}, \bibinfo{person}{Atiq Bajwa}, \bibinfo{person}{David~L Hill},
  \bibinfo{person}{Erik Hallnor}, \bibinfo{person}{Hong Jiang},
  {et~al\mbox{.}}}
\newblock \showarticletitle{{4th Generation Intel Core Processor, Codenamed
  Haswell}}. In \bibinfo{booktitle}{\emph{Hot chips}}.
 \bibinfo{year}{2013}\natexlab{}.
\newblock


\bibitem[\protect\citeauthoryear{Han, Yun, Hwang, Kim, and Kim}{Han
  et~al\mbox{.}}{2013}]%
        {psr_6}
\bibfield{author}{\bibinfo{person}{Sodam Han}, \bibinfo{person}{Yonghee Yun},
  \bibinfo{person}{Eunju Hwang}, \bibinfo{person}{Wook Kim}, {and}
  \bibinfo{person}{Young~Hwan Kim}.}
\newblock \showarticletitle{{Samsung Exynos 5410 Processor-Experience the
  Ultimate Performance and Versatility}}.
\newblock \bibinfo{journal}{\emph{White Paper}}.
 \bibinfo{year}{2013}\natexlab{}.
\newblock


\bibitem[\protect\citeauthoryear{He, Qureshi, Qiu, Li, Li, and Han}{He
  et~al\mbox{.}}{2018}]%
        {he2018rubiks}
\bibfield{author}{\bibinfo{person}{Jian He}, \bibinfo{person}{Mubashir~Adnan
  Qureshi}, \bibinfo{person}{Lili Qiu}, \bibinfo{person}{Jin Li},
  \bibinfo{person}{Feng Li}, {and} \bibinfo{person}{Lei Han}.}
\newblock \showarticletitle{Rubiks: Practical 360-Degree Streaming for
  Smartphones}. In \bibinfo{booktitle}{\emph{MobiSys}}.
 \bibinfo{year}{2018}\natexlab{}.
\newblock


\bibitem[\protect\citeauthoryear{Heather, Bellini, amd M.~Sugiyama, Shin, Alam,
  , et~al\mbox{.}}{Heather et~al\mbox{.}}{2020}]%
        {vr_2025}
\bibfield{author}{\bibinfo{person}{C. Heather}, \bibinfo{person}{Bellini},
  \bibinfo{person}{W.~Chen amd M.~Sugiyama}, \bibinfo{person}{M. Shin},
  \bibinfo{person}{S. Alam}, \bibinfo{person}{}, {et~al\mbox{.}}}
\newblock \bibinfo{title}{{Virtual and Augmented Reality.}}
\newblock
\newblock
\newblock
\shownote{https://bit.ly/3nULNE0.}
 \bibinfo{year}{2020}\natexlab{}.


\bibitem[\protect\citeauthoryear{HowtoGeek}{HowtoGeek}{2020}]%
        {res_gap_3}
\bibfield{author}{\bibinfo{person}{HowtoGeek}.}
\newblock \bibinfo{title}{The Resolution Gap}.
\newblock
\newblock
\newblock
\shownote{https://bit.ly/3dhCGJc.}
 \bibinfo{year}{2020}\natexlab{}.


\bibitem[\protect\citeauthoryear{Huang, Yao, and Chen}{Huang
  et~al\mbox{.}}{1998}]%
        {huang1998method}
\bibfield{author}{\bibinfo{person}{Hung-ju Huang}, \bibinfo{person}{Jo-tan
  Yao}, {and} \bibinfo{person}{Chung-heng Chen}.}
\newblock \bibinfo{title}{{Method and System for Segment Encoded Graphic Data
  Compression}}.
\newblock
\newblock
\newblock
\shownote{{US} Patent 5,748,904.}
 \bibinfo{year}{1998}\natexlab{}.


\bibitem[\protect\citeauthoryear{Hwang, Choe, Kim, Park, Bae, Choe,
  et~al\mbox{.}}{Hwang et~al\mbox{.}}{2017}]%
        {hwang201716}
\bibfield{author}{\bibinfo{person}{Moon-Sang Hwang}, \bibinfo{person}{Deok-Jun
  Choe}, \bibinfo{person}{Do-Wan Kim}, \bibinfo{person}{Joon-Bae Park},
  \bibinfo{person}{Jun-Woo Bae}, \bibinfo{person}{Won-jun Choe},
  {et~al\mbox{.}}}
\newblock \showarticletitle{{16-2: Cost-effective Driver IC Architecture using
  Low-power Memory Interface for Mobile Display Application}}. In
  \bibinfo{booktitle}{\emph{SID Symposium Digest of Technical Papers}}. Wiley
  Online Library.
 \bibinfo{year}{2017}\natexlab{}.
\newblock


\bibitem[\protect\citeauthoryear{Intel}{Intel}{[n. d.]}]%
        {vtune}
\bibfield{author}{\bibinfo{person}{Intel}.}
\newblock \showarticletitle{{Intel VTune Profiler User Guide}}.
  \bibinfo{howpublished}{online, accessed March 2020}.

\newblock
\newblock
\shownote{https://software.intel.com/en-us/vtune-help-window-cpu-c-p-states-platform-power-analysis.}


\bibitem[\protect\citeauthoryear{{Intel}}{{Intel}}{2016}]%
        {intel_display}
\bibfield{author}{\bibinfo{person}{{Intel}}.}
\newblock \bibinfo{title}{{{Open Source Graphics}}}.
\newblock \bibinfo{howpublished}{\url{https://bit.ly/2qQgHp7}}.
\newblock
 \bibinfo{year}{2016}\natexlab{}.


\bibitem[\protect\citeauthoryear{{Intel}}{{Intel}}{2020}]%
        {intel_skl_dev}
\bibfield{author}{\bibinfo{person}{{Intel}}.}
\newblock \bibinfo{title}{{6th Generation Intel® Processor for U/Y-Platforms
  Datasheet}}.
\newblock
\newblock
\newblock
\shownote{https://intel.ly/37rtnU7.}
 \bibinfo{year}{2020}\natexlab{}.


\bibitem[\protect\citeauthoryear{JEDEC}{JEDEC}{2013}]%
        {no2013jesd79}
\bibfield{author}{\bibinfo{person}{JEDEC}.}
\newblock \showarticletitle{{JESD79-3-1.35 V DDR3L-800, DDR3L-1066, DDR3L-1333,
  DDR3L-1600, and DDR3L-1866}}.
\newblock \bibinfo{journal}{\emph{JEDEC Std., JESD79-3-1A}}
  \bibinfo{volume}{1}.
 \bibinfo{year}{2013}\natexlab{}.
\newblock


\bibitem[\protect\citeauthoryear{Jiao, Tseng, Ma, and Zou}{Jiao
  et~al\mbox{.}}{2000}]%
        {jiao2000generic}
\bibfield{author}{\bibinfo{person}{Jianxin Jiao}, \bibinfo{person}{Mitchell~M
  Tseng}, \bibinfo{person}{Qinhai Ma}, {and} \bibinfo{person}{Yi Zou}.}
\newblock \showarticletitle{{Generic Bill-of-Materials-and-Operations for
  High-Variety Production Management}}.
\newblock \bibinfo{journal}{\emph{Concurrent Engineering}}.
 \bibinfo{year}{2000}\natexlab{}.
\newblock


\bibitem[\protect\citeauthoryear{Jin, Jung, and Lee}{Jin et~al\mbox{.}}{2007}]%
        {jin2007efficient}
\bibfield{author}{\bibinfo{person}{Genhua Jin}, \bibinfo{person}{Jin-Su Jung},
  {and} \bibinfo{person}{Hyuk-Jae Lee}.}
\newblock \showarticletitle{{An Efficient Pipelined Architecture for H.264/AVC
  Intra Frame Processing}}. In \bibinfo{booktitle}{\emph{ISCAS}}.
 \bibinfo{year}{2007}\natexlab{}.
\newblock


\bibitem[\protect\citeauthoryear{Kellicker}{Kellicker}{2014}]%
        {kellicker2014closed}
\bibfield{author}{\bibinfo{person}{Scott Kellicker}.}
\newblock \bibinfo{title}{Closed Captions for Live Streams}.
\newblock
\newblock
\newblock
\shownote{US Patent 8,782,721.}
 \bibinfo{year}{2014}\natexlab{}.


\bibitem[\protect\citeauthoryear{Keysight}{Keysight}{2019a}]%
        {keysight_N6705B}
\bibfield{author}{\bibinfo{person}{Keysight}.}
\newblock \bibinfo{title}{{IntKeysight N6705B DC Power Analyzer}}.
\newblock
\newblock
\newblock
\shownote{https://bit.ly/2MZ9Hhv.}
 \bibinfo{year}{2019}\natexlab{a}.


\bibitem[\protect\citeauthoryear{Keysight}{Keysight}{2019b}]%
        {keysight_acc}
\bibfield{author}{\bibinfo{person}{Keysight}.}
\newblock \bibinfo{title}{{IntKeysight Source Measure Units Power Modules}}.
\newblock
\newblock
\newblock
\shownote{https://bit.ly/38kkStt.}
 \bibinfo{year}{2019}\natexlab{b}.


\bibitem[\protect\citeauthoryear{Khan and Rengarajan}{Khan and
  Rengarajan}{2007}]%
        {khan2007bandwidth}
\bibfield{author}{\bibinfo{person}{Moinul~H Khan} {and}
  \bibinfo{person}{Srikanth Rengarajan}.}
\newblock \showarticletitle{{Bandwidth-efficient Display Controller for Low
  Power Devices in Presence of Occlusion}}. In
  \bibinfo{booktitle}{\emph{International Conference on Consumer Electronics}}.
 \bibinfo{year}{2007}\natexlab{}.
\newblock


\bibitem[\protect\citeauthoryear{Kogel, Doerper, Wieferink, Leupers, Ascheid,
  Meyr, et~al\mbox{.}}{Kogel et~al\mbox{.}}{2003}]%
        {kogel2003modular}
\bibfield{author}{\bibinfo{person}{Tim Kogel}, \bibinfo{person}{Malte Doerper},
  \bibinfo{person}{Andreas Wieferink}, \bibinfo{person}{Rainer Leupers},
  \bibinfo{person}{Gerd Ascheid}, \bibinfo{person}{Heinrich Meyr},
  {et~al\mbox{.}}}
\newblock \showarticletitle{{A Modular Simulation Framework for Architectural
  Exploration of On-chip Interconnection Networks}}. In
  \bibinfo{booktitle}{\emph{CODES/ISSS}}.
 \bibinfo{year}{2003}\natexlab{}.
\newblock


\bibitem[\protect\citeauthoryear{Kwa, Hayek, Shah, and Bhowmik}{Kwa
  et~al\mbox{.}}{2012}]%
        {psr_2}
\bibfield{author}{\bibinfo{person}{Seh Kwa}, \bibinfo{person}{George~R Hayek},
  \bibinfo{person}{Kamal~R Shah}, {and} \bibinfo{person}{Achintya~K Bhowmik}.}
\newblock \showarticletitle{{Panel Self-Refresh Technology: Decoupling Image
  Update from LCD Panel Refresh in Mobile Computing Systems}}. In
  \bibinfo{booktitle}{\emph{SID Symposium Digest of Technical Papers}}.
 \bibinfo{year}{2012}\natexlab{}.
\newblock


\bibitem[\protect\citeauthoryear{Lakdawala, Schaecher, Fu, Limaye, Duster, Tan,
  et~al\mbox{.}}{Lakdawala et~al\mbox{.}}{2012}]%
        {lakdawala201232}
\bibfield{author}{\bibinfo{person}{Hasnain Lakdawala}, \bibinfo{person}{Mark
  Schaecher}, \bibinfo{person}{Chang-Tsung Fu}, \bibinfo{person}{Rahul Limaye},
  \bibinfo{person}{Jon Duster}, \bibinfo{person}{Yulin Tan}, {et~al\mbox{.}}}
\newblock \showarticletitle{{A 32 nm SoC with Dual Core ATOM Processor and RF
  WiFi Transceiver}}.
\newblock \bibinfo{journal}{\emph{JSSC}}.
 \bibinfo{year}{2012}\natexlab{}.
\newblock


\bibitem[\protect\citeauthoryear{Landman}{Landman}{1996}]%
        {landman1996high}
\bibfield{author}{\bibinfo{person}{Paul Landman}.}
\newblock \showarticletitle{{High-level Power Estimation}}. In
  \bibinfo{booktitle}{\emph{ISLPED}}.
 \bibinfo{year}{1996}\natexlab{}.
\newblock


\bibitem[\protect\citeauthoryear{Lee, Subramanian, Ausavarungnirun, Choi, and
  Mutlu}{Lee et~al\mbox{.}}{2015}]%
        {lee2015decoupled}
\bibfield{author}{\bibinfo{person}{Donghyuk Lee}, \bibinfo{person}{Lavanya
  Subramanian}, \bibinfo{person}{Rachata Ausavarungnirun},
  \bibinfo{person}{Jongmoo Choi}, {and} \bibinfo{person}{Onur Mutlu}.}
\newblock \showarticletitle{{Decoupled Direct Memory Access: Isolating CPU and
  IO Traffic by Leveraging a Dual-Data-Port DRAM}}. In
  \bibinfo{booktitle}{\emph{PACT}}.
 \bibinfo{year}{2015}\natexlab{}.
\newblock


\bibitem[\protect\citeauthoryear{Leng, Chen, Sun, Huang, and Zhu}{Leng
  et~al\mbox{.}}{2019}]%
        {leng2019energy}
\bibfield{author}{\bibinfo{person}{Yue Leng}, \bibinfo{person}{Chi-Chun Chen},
  \bibinfo{person}{Qiuyue Sun}, \bibinfo{person}{Jian Huang}, {and}
  \bibinfo{person}{Yuhao Zhu}.}
\newblock \showarticletitle{{Energy-Efficient Video Processing for Virtual
  Reality}}. In \bibinfo{booktitle}{\emph{ISCA}}.
 \bibinfo{year}{2019}\natexlab{}.
\newblock


\bibitem[\protect\citeauthoryear{Menozzi, Lang, Naepflin, Zeller, and
  Krueger}{Menozzi et~al\mbox{.}}{2001}]%
        {menozzi2001crt}
\bibfield{author}{\bibinfo{person}{M Menozzi}, \bibinfo{person}{F Lang},
  \bibinfo{person}{U Naepflin}, \bibinfo{person}{C Zeller}, {and}
  \bibinfo{person}{H Krueger}.}
\newblock \showarticletitle{CRT Versus LCD: Effects of Refresh Rate, Display
  Technology and Background Luminance in Visual Performance}.
\newblock \bibinfo{journal}{\emph{Displays}}.
 \bibinfo{year}{2001}\natexlab{}.
\newblock


\bibitem[\protect\citeauthoryear{Microsoft}{Microsoft}{2020}]%
        {msft_display_driver}
\bibfield{author}{\bibinfo{person}{Microsoft}.}
\newblock \bibinfo{title}{{Programming Reference for Windows Driver Kit -
  Display}}.
\newblock
\newblock
\newblock
\shownote{https://bit.ly/33T8D7L.}
 \bibinfo{year}{2020}\natexlab{}.


\bibitem[\protect\citeauthoryear{Mukherjee, Bankoski, Grange, Han, Koleszar,
  Wilkins, et~al\mbox{.}}{Mukherjee et~al\mbox{.}}{2013}]%
        {mukherjee2013latest}
\bibfield{author}{\bibinfo{person}{Debargha Mukherjee}, \bibinfo{person}{Jim
  Bankoski}, \bibinfo{person}{Adrian Grange}, \bibinfo{person}{Jingning Han},
  \bibinfo{person}{John Koleszar}, \bibinfo{person}{Paul Wilkins},
  {et~al\mbox{.}}}
\newblock \showarticletitle{{The Latest Open-source Video Codec VP9-an Overview
  and Preliminary Results}}. In \bibinfo{booktitle}{\emph{PCS}}.
 \bibinfo{year}{2013}\natexlab{}.
\newblock


\bibitem[\protect\citeauthoryear{Mutlu}{Mutlu}{2013}]%
        {mutlu2013memory}
\bibfield{author}{\bibinfo{person}{Onur Mutlu}.}
\newblock \showarticletitle{{Memory Scaling: A Systems Architecture
  Perspective}}. In \bibinfo{booktitle}{\emph{IMW}}.
 \bibinfo{year}{2013}\natexlab{}.
\newblock


\bibitem[\protect\citeauthoryear{Nachiappan, Yedlapalli, Soundararajan,
  Sivasubramaniam, Kandemir, Iyer, et~al\mbox{.}}{Nachiappan
  et~al\mbox{.}}{2015a}]%
        {nachiappan2015domain}
\bibfield{author}{\bibinfo{person}{Nachiappan~Chidambaram Nachiappan},
  \bibinfo{person}{Praveen Yedlapalli}, \bibinfo{person}{Niranjan
  Soundararajan}, \bibinfo{person}{Anand Sivasubramaniam},
  \bibinfo{person}{Mahmut~T Kandemir}, \bibinfo{person}{Ravi Iyer},
  {et~al\mbox{.}}}
\newblock \showarticletitle{{Domain Knowledge Based Energy Management in
  Handhelds}}. In \bibinfo{booktitle}{\emph{HPCA}}.
 \bibinfo{year}{2015}\natexlab{a}.
\newblock


\bibitem[\protect\citeauthoryear{Nachiappan, Zhang, Ryoo, Soundararajan,
  Sivasubramaniam, Kandemir, et~al\mbox{.}}{Nachiappan et~al\mbox{.}}{2015b}]%
        {nachiappan2015vip}
\bibfield{author}{\bibinfo{person}{Nachiappan~Chidambaram Nachiappan},
  \bibinfo{person}{Haibo Zhang}, \bibinfo{person}{Jihyun Ryoo},
  \bibinfo{person}{Niranjan Soundararajan}, \bibinfo{person}{Anand
  Sivasubramaniam}, \bibinfo{person}{Mahmut~T Kandemir}, {et~al\mbox{.}}}
\newblock \showarticletitle{{VIP: Virtualizing IP chains on Handheld
  Platforms}}. In \bibinfo{booktitle}{\emph{ISCA}}.
 \bibinfo{year}{2015}\natexlab{b}.
\newblock


\bibitem[\protect\citeauthoryear{{Nvidia}}{{Nvidia}}{2016}]%
        {disp_comp_nvidia}
\bibfield{author}{\bibinfo{person}{{Nvidia}}.}
\newblock \bibinfo{title}{{{NVIDIA GeDorce GTX 1080}}}.
\newblock
\newblock
\newblock
\shownote{Nvidia Whitepaper.}
 \bibinfo{year}{2016}\natexlab{}.


\bibitem[\protect\citeauthoryear{Ogras, Bogdan, and Marculescu}{Ogras
  et~al\mbox{.}}{2010}]%
        {ogras2010analytical}
\bibfield{author}{\bibinfo{person}{Umit~Y Ogras}, \bibinfo{person}{Paul
  Bogdan}, {and} \bibinfo{person}{Radu Marculescu}.}
\newblock \showarticletitle{{An Analytical Approach for Network-on-Chip
  Performance Analysis}}.
\newblock \bibinfo{journal}{\emph{TCAD}}.
 \bibinfo{year}{2010}\natexlab{}.
\newblock


\bibitem[\protect\citeauthoryear{Patil, Jas, Lisherness, and Carrieri}{Patil
  et~al\mbox{.}}{2014}]%
        {patil2014functional}
\bibfield{author}{\bibinfo{person}{Srinivas Patil}, \bibinfo{person}{Abhijit
  Jas}, \bibinfo{person}{Peter Lisherness}, {and} \bibinfo{person}{Enrico
  Carrieri}.}
\newblock \bibinfo{title}{Functional Fabric-based Test Controller for
  Functional and Structural Test and Debug}.
\newblock
\newblock
\newblock
\shownote{US Patent 8,793,095.}
 \bibinfo{year}{2014}\natexlab{}.


\bibitem[\protect\citeauthoryear{{Pcmag}}{{Pcmag}}{2020a}]%
        {bench1}
\bibfield{author}{\bibinfo{person}{{Pcmag}}.}
\newblock \bibinfo{title}{{{How We Test Laptops}}}.
\newblock
\newblock
\newblock
\shownote{https://bit.ly/3lJZZ1C.}
 \bibinfo{year}{2020}\natexlab{a}.


\bibitem[\protect\citeauthoryear{{Pcmag}}{{Pcmag}}{2020b}]%
        {bench2}
\bibfield{author}{\bibinfo{person}{{Pcmag}}.}
\newblock \bibinfo{title}{{{Samsung Galaxy S20 Review}}}.
\newblock
\newblock
\newblock
\shownote{https://bit.ly/3dsF14q.}
 \bibinfo{year}{2020}\natexlab{b}.


\bibitem[\protect\citeauthoryear{{Phonearena}}{{Phonearena}}{2020}]%
        {bench3}
\bibfield{author}{\bibinfo{person}{{Phonearena}}.}
\newblock \bibinfo{title}{{{PhoneArena Battery Test Results}}}.
\newblock
\newblock
\newblock
\shownote{https://bit.ly/36Z3Ljr.}
 \bibinfo{year}{2020}\natexlab{}.


\bibitem[\protect\citeauthoryear{{Phoronix}}{{Phoronix}}{2019}]%
        {enable_fbc}
\bibfield{author}{\bibinfo{person}{{Phoronix}}.}
\newblock \bibinfo{title}{{{Linux 4.11 To Enable Frame-Buffer Compression By
  Default For Skylake+}}}.
\newblock \bibinfo{howpublished}{online, accessed March 2020}.
\newblock
\newblock
\shownote{https://bit.ly/3a59yCP.}
 \bibinfo{year}{2019}\natexlab{}.


\bibitem[\protect\citeauthoryear{{Qualcomm Technologies}}{{Qualcomm
  Technologies}}{2018}]%
        {qcom2018}
\bibfield{author}{\bibinfo{person}{{Qualcomm Technologies}}.}
\newblock \bibinfo{title}{{{Qualcomm Snapdragon 410E (APQ8016E) Processor
  Device Specification}}}.
\newblock \bibinfo{howpublished}{online}.
\newblock
\newblock
\shownote{https://developer.qualcomm.com/qfile/28813/lm80-p0436-7\_f\_410e\_proc\_apq8016e\_device\_spec.pdf.}
 \bibinfo{year}{2018}\natexlab{}.


\bibitem[\protect\citeauthoryear{Radhakrishnan, Chinthamani, and
  Cheng}{Radhakrishnan et~al\mbox{.}}{2007}]%
        {radhakrishnan2007blackford}
\bibfield{author}{\bibinfo{person}{Sivakumar Radhakrishnan},
  \bibinfo{person}{Sundaram Chinthamani}, {and} \bibinfo{person}{Kai Cheng}.}
\newblock \showarticletitle{{The Blackford Northbridge Chipset for the Intel
  5000}}.
\newblock \bibinfo{journal}{\emph{IEEE Micro}}.
 \bibinfo{year}{2007}\natexlab{}.
\newblock


\bibitem[\protect\citeauthoryear{Raffin, Nogues, Hamidouche, Tomperi, Pelcat,
  and Menard}{Raffin et~al\mbox{.}}{2016}]%
        {raffin2016low}
\bibfield{author}{\bibinfo{person}{Erwan Raffin}, \bibinfo{person}{Erwan
  Nogues}, \bibinfo{person}{Wassim Hamidouche}, \bibinfo{person}{Seppo
  Tomperi}, \bibinfo{person}{Maxime Pelcat}, {and} \bibinfo{person}{Daniel
  Menard}.}
\newblock \showarticletitle{{Low Power HEVC Software Decoder for Mobile
  Devices}}.
\newblock \bibinfo{journal}{\emph{JRTIP}}.
 \bibinfo{year}{2016}\natexlab{}.
\newblock


\bibitem[\protect\citeauthoryear{Rao, Wang, Yalamanchili, Wardi, and
  Handong}{Rao et~al\mbox{.}}{2017}]%
        {rao2017application}
\bibfield{author}{\bibinfo{person}{Karthik Rao}, \bibinfo{person}{Jun Wang},
  \bibinfo{person}{Sudhakar Yalamanchili}, \bibinfo{person}{Yorai Wardi}, {and}
  \bibinfo{person}{Ye Handong}.}
\newblock \showarticletitle{{Application-specific Performance-aware Energy
  Optimization on Android Mobile Devices}}. In
  \bibinfo{booktitle}{\emph{HPCA}}.
 \bibinfo{year}{2017}\natexlab{}.
\newblock


\bibitem[\protect\citeauthoryear{Rasmusson, Akenine-M{\"o}ller, Hasselgren, and
  Munkberg}{Rasmusson et~al\mbox{.}}{2011}]%
        {rasmusson2011frame}
\bibfield{author}{\bibinfo{person}{Jim Rasmusson}, \bibinfo{person}{Tomas
  Akenine-M{\"o}ller}, \bibinfo{person}{Jon Hasselgren}, {and}
  \bibinfo{person}{Jacob Munkberg}.}
\newblock \bibinfo{title}{{Frame Buffer Compression and Decompression Method
  for Graphics Rendering}}.
\newblock
\newblock
\newblock
\shownote{US Patent 8,031,937.}
 \bibinfo{year}{2011}\natexlab{}.


\bibitem[\protect\citeauthoryear{Rotem}{Rotem}{2015}]%
        {rotem2015intel}
\bibfield{author}{\bibinfo{person}{Efraim Rotem}.}
\newblock \showarticletitle{{Intel Architecture, Code Name Skylake Deep Dive: A
  New Architecture to Manage Power Performance and Energy Efficiency}}. In
  \bibinfo{booktitle}{\emph{IDF}}.
 \bibinfo{year}{2015}\natexlab{}.
\newblock


\bibitem[\protect\citeauthoryear{SAMPLES}{SAMPLES}{2012}]%
        {4k_puppies}
\bibfield{author}{\bibinfo{person}{4K SAMPLES}.}
\newblock \bibinfo{title}{{Puppies Bath in 4K.}}
\newblock \bibinfo{howpublished}{online, accessed Jan 2020}.
\newblock
\newblock
\shownote{https://bit.ly/39jHIS2.}
 \bibinfo{year}{2012}\natexlab{}.


\bibitem[\protect\citeauthoryear{{Samsung}}{{Samsung}}{2020}]%
        {samsung_s20}
\bibfield{author}{\bibinfo{person}{{Samsung}}.}
\newblock \bibinfo{title}{{{Samsung Galaxy S20 Specification }}}.
\newblock
\newblock
\newblock
\shownote{https://www.samsung.com/global/galaxy/galaxy-s20/specs.}
 \bibinfo{year}{2020}\natexlab{}.


\bibitem[\protect\citeauthoryear{Sch{\"o}ne, Molka, and Werner}{Sch{\"o}ne
  et~al\mbox{.}}{2015}]%
        {schone2015wake}
\bibfield{author}{\bibinfo{person}{Robert Sch{\"o}ne}, \bibinfo{person}{Daniel
  Molka}, {and} \bibinfo{person}{Michael Werner}.}
\newblock \showarticletitle{{Wake-up Latencies for Processor Idle States on
  Current x86 Processors}}.
\newblock \bibinfo{journal}{\emph{Computer Science-Research and Development}}.
 \bibinfo{year}{2015}\natexlab{}.
\newblock


\bibitem[\protect\citeauthoryear{Seshadri, Kim, Fallin, Lee, Ausavarungnirun,
  Pekhimenko, et~al\mbox{.}}{Seshadri et~al\mbox{.}}{2013}]%
        {seshadri2013rowclone}
\bibfield{author}{\bibinfo{person}{Vivek Seshadri}, \bibinfo{person}{Yoongu
  Kim}, \bibinfo{person}{Chris Fallin}, \bibinfo{person}{Donghyuk Lee},
  \bibinfo{person}{Rachata Ausavarungnirun}, \bibinfo{person}{Gennady
  Pekhimenko}, {et~al\mbox{.}}}
\newblock \showarticletitle{{RowClone: Fast and Energy-efficient in-DRAM Bulk
  Data Copy and Initialization}}. In \bibinfo{booktitle}{\emph{MICRO}}.
 \bibinfo{year}{2013}\natexlab{}.
\newblock


\bibitem[\protect\citeauthoryear{Shah, Kwa, and Hayek}{Shah
  et~al\mbox{.}}{2013}]%
        {psr_1}
\bibfield{author}{\bibinfo{person}{Kamal Shah}, \bibinfo{person}{Seh Kwa},
  {and} \bibinfo{person}{George Hayek}.}
\newblock \showarticletitle{{Extending Battery Life of Ultrabook™ Through use
  of Panel Self Refresh Technology (PSR)}}. In \bibinfo{booktitle}{\emph{SID
  Symposium Digest of Technical Papers}}.
 \bibinfo{year}{2013}\natexlab{}.
\newblock


\bibitem[\protect\citeauthoryear{Shim, Chang, and Pedram}{Shim
  et~al\mbox{.}}{2004}]%
        {shim2004compressed}
\bibfield{author}{\bibinfo{person}{Hojun Shim}, \bibinfo{person}{Nachyuck
  Chang}, {and} \bibinfo{person}{Massoud Pedram}.}
\newblock \showarticletitle{{A Compressed Frame Buffer to Reduce Display Power
  Consumption in Mobile Systems}}. In \bibinfo{booktitle}{\emph{ASPDAC}}.
 \bibinfo{year}{2004}\natexlab{}.
\newblock


\bibitem[\protect\citeauthoryear{Subramanian, Seshadri, Kim, Jaiyen, and
  Mutlu}{Subramanian et~al\mbox{.}}{2013}]%
        {subramanian2013mise}
\bibfield{author}{\bibinfo{person}{Lavanya Subramanian}, \bibinfo{person}{Vivek
  Seshadri}, \bibinfo{person}{Yoongu Kim}, \bibinfo{person}{Ben Jaiyen}, {and}
  \bibinfo{person}{Onur Mutlu}.}
\newblock \showarticletitle{{MISE: Providing Performance Predictability and
  Improving Fairness in Shared Main Memory Systems}}. In
  \bibinfo{booktitle}{\emph{HPCA}}.
 \bibinfo{year}{2013}\natexlab{}.
\newblock


\bibitem[\protect\citeauthoryear{Sullivan, Ohm, Han, and Wiegand}{Sullivan
  et~al\mbox{.}}{2012}]%
        {sullivan2012overview}
\bibfield{author}{\bibinfo{person}{Gary~J Sullivan},
  \bibinfo{person}{Jens-Rainer Ohm}, \bibinfo{person}{Woo-Jin Han}, {and}
  \bibinfo{person}{Thomas Wiegand}.}
\newblock \showarticletitle{{Overview of the High Efficiency Video Coding
  (HEVC) Standard}}.
\newblock \bibinfo{journal}{\emph{TCSVT}}.
 \bibinfo{year}{2012}\natexlab{}.
\newblock


\bibitem[\protect\citeauthoryear{Tam, Muljono, Huang, Iyer, Royneogi, Satti,
  et~al\mbox{.}}{Tam et~al\mbox{.}}{2018}]%
        {tam2018skylake}
\bibfield{author}{\bibinfo{person}{Simon~M Tam}, \bibinfo{person}{Harry
  Muljono}, \bibinfo{person}{Min Huang}, \bibinfo{person}{Sitaraman Iyer},
  \bibinfo{person}{Kalapi Royneogi}, \bibinfo{person}{Nagmohan Satti},
  {et~al\mbox{.}}}
\newblock \showarticletitle{{SkyLake-SP: A 14nm 28-Core Xeon{\textregistered}
  Processor}}. In \bibinfo{booktitle}{\emph{ISSCC}}.
 \bibinfo{year}{2018}\natexlab{}.
\newblock


\bibitem[\protect\citeauthoryear{{Tech Thoughts}}{{Tech Thoughts}}{2021}]%
        {msft_srfc_pro_bom}
\bibfield{author}{\bibinfo{person}{{Tech Thoughts}}.}
\newblock \bibinfo{title}{{Microsoft Surface RT \& Pro: BOM \& Price
  Estimate}}.
\newblock
\newblock
\newblock
\shownote{https://bit.ly/3rspuY3.}
 \bibinfo{year}{2021}\natexlab{}.


\bibitem[\protect\citeauthoryear{TheVerge}{TheVerge}{2020}]%
        {res_gap_2}
\bibfield{author}{\bibinfo{person}{TheVerge}.}
\newblock \bibinfo{title}{YouTube Joins Netflix in Reducing Video Quality in
  Europe}.
\newblock
\newblock
\newblock
\shownote{https://bit.ly/3jQEewe.}
 \bibinfo{year}{2020}\natexlab{}.


\bibitem[\protect\citeauthoryear{Tidrow, Boyce, and Shapiro}{Tidrow
  et~al\mbox{.}}{2015}]%
        {tidrow2015windows}
\bibfield{author}{\bibinfo{person}{Rob Tidrow}, \bibinfo{person}{Jim Boyce},
  {and} \bibinfo{person}{Jeffrey~R Shapiro}.}
\newblock \bibinfo{booktitle}{\emph{{Windows 10 Bible}}}.
\newblock \bibinfo{publisher}{John Wiley \& Sons}.
 \bibinfo{year}{2015}\natexlab{}.
\newblock


\bibitem[\protect\citeauthoryear{{Toms Guide}}{{Toms Guide}}{2020}]%
        {tomsguide}
\bibfield{author}{\bibinfo{person}{{Toms Guide}}.}
  \bibinfo{year}{2020}\natexlab{}.
\newblock \bibinfo{title}{{Samsung Galaxy S20 Ultra Battery Life: 120Hz Display
  Takes 3 Hours Off}}.
\newblock
\urldef\tempurl%
\url{https://www.tomsguide.com/news/samsung-galaxy-s20-ultra-battery-life-120hz-display-takes-3-hours-off}
\showURL{%
\tempurl}


\bibitem[\protect\citeauthoryear{Tu and Hsueh}{Tu and Hsueh}{2010}]%
        {tu2010batch}
\bibfield{author}{\bibinfo{person}{Tang-Hsun Tu} {and}
  \bibinfo{person}{Chih-Wen Hsueh}.}
\newblock \showarticletitle{{Batch-Pipelining for H. 264 Decoding on Multicore
  Systems}}. In \bibinfo{booktitle}{\emph{DCC}}.
 \bibinfo{year}{2010}\natexlab{}.
\newblock


\bibitem[\protect\citeauthoryear{{UEFI.org}}{{UEFI.org}}{2021}]%
        {acpi}
\bibfield{author}{\bibinfo{person}{{UEFI.org}}.}
\newblock \bibinfo{title}{{{Advanced Configuration and Power Interface (ACPI)
  specification }}}.
\newblock
\newblock
\newblock
\shownote{https://bit.ly/2ToagrG.}
 \bibinfo{year}{2021}\natexlab{}.


\bibitem[\protect\citeauthoryear{{VESA}}{{VESA}}{[n. d.]}]%
        {edp_psr}
\bibfield{author}{\bibinfo{person}{{VESA}}.}
\newblock \bibinfo{title}{{{eDP Embedded DisplayPort: The New Generation
  Digital Display Interface for Embedded Applications }}}.
\newblock
\newblock
\newblock
\shownote{https://bit.ly/2TE2Sps.}



\bibitem[\protect\citeauthoryear{{VESA}}{{VESA}}{2021a}]%
        {vesa}
\bibfield{author}{\bibinfo{person}{{VESA}}.}
\newblock \bibinfo{title}{{{VESA publishes Embedded DisplayPort (eDP) Standard
  Version 1.4a }}}.
\newblock
\newblock
\newblock
\shownote{https://bit.ly/2sglj8f.}
 \bibinfo{year}{2021}\natexlab{a}.


\bibitem[\protect\citeauthoryear{{VESA}}{{VESA}}{2021b}]%
        {edp1_4b}
\bibfield{author}{\bibinfo{person}{{VESA}}.}
\newblock \bibinfo{title}{{{Vesa Rolls Out Production-ready Embedded
  Displayport Standard 1.4 For Mobile Personal Computing Devices}}}.
\newblock
\newblock
\newblock
\shownote{https://bit.ly/2KYrk03.}
 \bibinfo{year}{2021}\natexlab{b}.


\bibitem[\protect\citeauthoryear{Wiegand, Sullivan, Bjontegaard, and
  Luthra}{Wiegand et~al\mbox{.}}{2003}]%
        {wiegand2003overview}
\bibfield{author}{\bibinfo{person}{Thomas Wiegand}, \bibinfo{person}{Gary~J
  Sullivan}, \bibinfo{person}{Gisle Bjontegaard}, {and} \bibinfo{person}{Ajay
  Luthra}.}
\newblock \showarticletitle{{Overview of the H. 264/AVC Video Coding
  Standard}}.
\newblock \bibinfo{journal}{\emph{TCSVT}}.
 \bibinfo{year}{2003}\natexlab{}.
\newblock


\bibitem[\protect\citeauthoryear{Wikipedia}{Wikipedia}{2019}]%
        {sunrise_point_skl}
\bibfield{author}{\bibinfo{person}{Wikipedia}.}
\newblock \bibinfo{title}{{Skylake Microarchitecture}}.
\newblock
\newblock
\newblock
\shownote{https://bit.ly/2NG8Hz2.}
 \bibinfo{year}{2019}\natexlab{}.


\bibitem[\protect\citeauthoryear{Wikipedia}{Wikipedia}{2020a}]%
        {cc_wiki}
\bibfield{author}{\bibinfo{person}{Wikipedia}.}
\newblock \bibinfo{title}{{Closed captioning - Online Video Streaming}}.
\newblock
\newblock
\newblock
\shownote{https://bit.ly/30VB96C.}
 \bibinfo{year}{2020}\natexlab{a}.


\bibitem[\protect\citeauthoryear{Wikipedia}{Wikipedia}{2020b}]%
        {msft_surface}
\bibfield{author}{\bibinfo{person}{Wikipedia}.}
\newblock \bibinfo{title}{{Microsoft Surface}}.
\newblock
\newblock
\newblock
\shownote{https://bit.ly/38N8vrS.}
 \bibinfo{year}{2020}\natexlab{b}.


\bibitem[\protect\citeauthoryear{Yabe, Aimoto, Motomura, Takizawa, Miyamoto,
  Iwasaki, et~al\mbox{.}}{Yabe et~al\mbox{.}}{1998}]%
        {yabe1998compression}
\bibfield{author}{\bibinfo{person}{Y Yabe}, \bibinfo{person}{Y Aimoto},
  \bibinfo{person}{M Motomura}, \bibinfo{person}{T Takizawa},
  \bibinfo{person}{T Miyamoto}, \bibinfo{person}{T Iwasaki}, {et~al\mbox{.}}}
\newblock \showarticletitle{{Compression/Decompression DRAM for Unified Memory
  Systems: a 16Mb, 200MHz, 90\% to 50\% Graphics-Bandwidth Reduction
  Prototype}}. In \bibinfo{booktitle}{\emph{ISSCC}}.
 \bibinfo{year}{1998}\natexlab{}.
\newblock


\bibitem[\protect\citeauthoryear{Yang, Nieh, Selsky, and Tiwari}{Yang
  et~al\mbox{.}}{2002}]%
        {yang2002performance}
\bibfield{author}{\bibinfo{person}{S~Jae Yang}, \bibinfo{person}{Jason Nieh},
  \bibinfo{person}{Matt Selsky}, {and} \bibinfo{person}{Nikhil Tiwari}.}
\newblock \showarticletitle{{The Performance of Remote Display Mechanisms for
  Thin-Client Computing.}}. In \bibinfo{booktitle}{\emph{USENIX ATC}}.
 \bibinfo{year}{2002}\natexlab{}.
\newblock


\bibitem[\protect\citeauthoryear{Yasin, Haj-Yahya, Ben-Asher, and
  Mendelson}{Yasin et~al\mbox{.}}{2019}]%
        {yasin2019metric}
\bibfield{author}{\bibinfo{person}{Ahmad Yasin}, \bibinfo{person}{Jawad
  Haj-Yahya}, \bibinfo{person}{Yosi Ben-Asher}, {and} \bibinfo{person}{Avi
  Mendelson}.}
\newblock \showarticletitle{{A Metric-guided Method for Discovering Impactful
  Features and Architectural Insights for Skylake-based Processors}}.
\newblock \bibinfo{journal}{\emph{TACO}}.
 \bibinfo{year}{2019}\natexlab{}.
\newblock


\bibitem[\protect\citeauthoryear{Yedlapalli, Nachiappan, Soundararajan,
  Sivasubramaniam, Kandemir, and Das}{Yedlapalli et~al\mbox{.}}{2014}]%
        {yedlapalli2014short}
\bibfield{author}{\bibinfo{person}{Praveen Yedlapalli},
  \bibinfo{person}{Nachiappan~Chidambaram Nachiappan},
  \bibinfo{person}{Niranjan Soundararajan}, \bibinfo{person}{Anand
  Sivasubramaniam}, \bibinfo{person}{Mahmut~T Kandemir}, {and}
  \bibinfo{person}{Chita~R Das}.}
\newblock \showarticletitle{{Short-Circuiting Memory Traffic in Handheld
  Platforms}}. In \bibinfo{booktitle}{\emph{MICRO}}.
 \bibinfo{year}{2014}\natexlab{}.
\newblock


\bibitem[\protect\citeauthoryear{Zahir, Ewert, and Seshadri}{Zahir
  et~al\mbox{.}}{2013}]%
        {zahir2013medfield}
\bibfield{author}{\bibinfo{person}{Rumi Zahir}, \bibinfo{person}{Mark Ewert},
  {and} \bibinfo{person}{Hari Seshadri}.}
\newblock \showarticletitle{{The Medfield Smartphone: Intel Architecture in a
  Handheld Form Factor}}.
\newblock \bibinfo{journal}{\emph{IEEE Micro}}.
 \bibinfo{year}{2013}\natexlab{}.
\newblock


\bibitem[\protect\citeauthoryear{Zhang, Rengasamy, Zhao, Nachiappan,
  Sivasubramaniam, Kandemir, et~al\mbox{.}}{Zhang et~al\mbox{.}}{2017}]%
        {rts}
\bibfield{author}{\bibinfo{person}{Haibo Zhang},
  \bibinfo{person}{Prasanna~Venkatesh Rengasamy}, \bibinfo{person}{Shulin
  Zhao}, \bibinfo{person}{Nachiappan~Chidambaram Nachiappan},
  \bibinfo{person}{Anand Sivasubramaniam}, \bibinfo{person}{Mahmut~T Kandemir},
  {et~al\mbox{.}}}
\newblock \showarticletitle{{Race-to-sleep+ Content Caching+ Display Caching: A
  Recipe for Energy-efficient Video Streaming on Handhelds}}. In
  \bibinfo{booktitle}{\emph{MICRO}}.
 \bibinfo{year}{2017}\natexlab{}.
\newblock


\bibitem[\protect\citeauthoryear{Zhao, Zhang, Bhuyan, Mishra, Ying, Kandemir,
  et~al\mbox{.}}{Zhao et~al\mbox{.}}{2020}]%
        {zhao2020deja}
\bibfield{author}{\bibinfo{person}{Shulin Zhao}, \bibinfo{person}{Haibo Zhang},
  \bibinfo{person}{Sandeepa Bhuyan}, \bibinfo{person}{Cyan~Subhra Mishra},
  \bibinfo{person}{Ziyu Ying}, \bibinfo{person}{Mahmut~T Kandemir},
  {et~al\mbox{.}}}
\newblock \showarticletitle{{D{\'e}j{\`a} View: Spatio-Temporal Compute Reuse
  for ‘Energy-Efficient 360° VR Video Streaming}}. In
  \bibinfo{booktitle}{\emph{ISCA}}.
 \bibinfo{year}{2020}\natexlab{}.
\newblock


\bibitem[\protect\citeauthoryear{Zyda}{Zyda}{2005}]%
        {zyda2005visual}
\bibfield{author}{\bibinfo{person}{Michael Zyda}.}
\newblock \showarticletitle{{From Visual Simulation to Virtual Reality to
  Games}}.
\newblock \bibinfo{journal}{\emph{IEEE Computer Society}}.
 \bibinfo{year}{2005}\natexlab{}.
\newblock


\end{thebibliography}










\end{document}